\documentclass[fleqn,usenatbib]{mnras}

\usepackage[T1]{fontenc}

\DeclareRobustCommand{\VAN}[3]{#2}
\let\VANthebibliography\thebibliography
\def\thebibliography{\DeclareRobustCommand{\VAN}[3]{##3}\VANthebibliography}

\usepackage{graphicx}	
\usepackage{amsmath}	
\usepackage{amssymb}	
\usepackage{natbib}
\usepackage{multicol}
\usepackage{subfig}
\usepackage{appendix}
\usepackage{multirow}
\usepackage{pdflscape}	
\usepackage{footnote}
\usepackage{threeparttable}
\usepackage{hyperref}
\usepackage{newtxtext,newtxmath}

\title[Env. effects in USS1558 at z=2.53]{Enhanced star formation and metallicity deficit in the USS\,1558--003 forming protocluster at z=2.53}

\author[J. M. Perez-Martinez et al.]{J. M. P\'erez-Mart\'inez,$^{1}$\thanks{E-mail: jm.perez@astr.tohoku.ac.jp}
T. Kodama,$^{1}$
Y. Koyama,$^{2}$
R. Shimakawa,$^{3,4}$
T. L. Suzuki,$^{5}$
K. Daikuhara,$^{1}$
K. Adachi,$^{1}$
\newauthor M. Onodera,$^{2,6}$
I. Tanaka$^{2}$
\\
$^{1}$ Astronomical Institute, Tohoku University, 6-3, Aramaki, Aoba, Sendai, Miyagi, 980-8578, Japan.\\
$^{2}$ Subaru Telescope, National Astronomical Observatory of Japan, National Institutes of Natural Sciences (NINS), 650 North A’ohoku Place, Hilo, HI 96720, USA.\\
$^{3}$ Waseda Institute for Advanced Study (WIAS), Waseda University, 1-21-1, Nishi-Waseda, Shinjuku, Tokyo 169-0051, Japan. \\
$^{4}$ National Astronomical Observatory of Japan, 2-21-1, Osawa, Mitaka, Tokyo, 181-8588, Japan.\\
$^{5}$ Kavli Institute for the Physics and Mathematics of the Universe (WPI), University of Tokyo, Kashiwa, Chiba, 277-8583, Japan.\\
$^{6}$ Department of Astronomical Science, SOKENDAI (The Graduate University for Advanced Studies), Osawa 2-21-1, Mitaka, Tokyo,
181-8588, Japan\\
}

\date{Accepted XXX. Received YYY; in original form ZZZ}

\pubyear{2022}

\begin{document}
\label{firstpage}
\pagerange{\pageref{firstpage}--\pageref{lastpage}}
\maketitle

\begin{abstract}
{We use K-band multi-object near-infrared spectroscopy with Keck/MOSFIRE to search for environmental imprints on the gas properties of 27 narrow-band selected H$\alpha$ emitters (HAEs) across the three major clumps of the assembling USS1558--003 protocluster at $z=2.53$. We target the H$\alpha$ and [N{\sc{ii}}]$\lambda$6584 emission lines to obtain star-formation rates (SFR) and gas-phase oxygen abundances for our sources, confirming the membership of 23 objects. HAEs belonging to this protocluster display enhanced SFRs with respect to the main sequence of star formation at the same cosmic epoch. This effect is more prominent for low-mass galaxies ($\mathrm{\log M_*/M_\odot<10.0}$), which may be experiencing a vigorous phase of mass assembly shortly after they were formed. We compute the individual and stacked gas-phase metallicities for our sources finding a metallicity deficit for low-mass objects when compared against the field mass-metallicity relation and the massive Spiderweb protocluster at $z=2.16$. These results suggest that HAEs within USS1558--003 may be less evolved than those in the Spiderweb protocluster. Finally, we explore the gas metallicity - gas fraction relation for a small sample of five galaxies with CO(3-2) molecular gas information. Assuming our objects are in equilibrium, we obtain a relatively wide range of mass loading factors ($\mathrm{\lambda=0.5-2}$) matching field samples at the cosmic noon but in contrast with our previous results in the Spiderweb protocluster. We speculate that these discrepancies between protoclusters may be (partly) driven by differences in their current dynamical and mass assembly stages, hinting at the co-evolution of protoclusters and their galaxy populations at $2<z<3$.}

\end{abstract}




\section{Introduction}
\label{S:Intro}

In the framework of the hierarchical growth of large-scale structures in the Universe, mass is assembled inhomogeneously along walls and filaments creating a "cosmic web" (\citealt{Bond96}). Galaxy clusters are formed at intersections of such filaments and, in the local universe, they are dominated by old quiescent massive early-type galaxies (with $>10$ Gyrs stellar ages) indicating that the majority of their stellar components were formed in the early universe ($z>2$, \citealt{Thomas10}) when clusters were still being assembled (i.e., protoclusters, see \citealt{Overzier16} and \citealt{Alberts22} for a review). Thus, while clusters represent an important component of the total stellar mass content of the universe (5-10\%), protoclusters are thought to vigorously contribute to the star formation rate density (SFRD) at high redshifts (\citealt{Chiang17}), e.g., 20-50\% at $z>2$. Simulations predict that cold gas streams along the filaments of the cosmic web supply the required fuel to galaxies and drive their elevated SFRs at high-z (e.g., \citealt{Keres05}; \citealt{Dekel06}; \citealt{Genel08}). It is expected that the channeling of cold gas streams towards the center of galaxies, the effect of dynamical friction, and the onset of environmental effects contribute to compressing/altering the gas distribution (\citealt{Tacchella16}) of protoclusters members, possibly triggering central starburst (\citealt{Gomez-Guijarro19}) and supporting their inside out mass growth (\citealt{vanDokkum15}). Thus, galaxies populating high-redshift protoclusters are expected to be sites of exceptionally vigorous (though short-lived) star formation (\citealt{Hickox12}). At times, this star formation may be detectable in the rest-frame UV/optical, but during intense starburst ($\mathrm{SFR\approx10^2-10^3\,M_\odot/yr}$) it can be heavily obscured by dust, emitting mainly in the rest-frame infrared (\citealt{Riechers13}). Therefore, unlike galaxy clusters in the local universe which are dominated by their passive population, young forming high-z protoclusters are observed as strong overdensities of gas-rich dusty starbursts (e.g., \citealt{Dannerbauer14}; \citealt{Popesso15}; \citealt{Casey16,Casey17}; \citealt{Chiang17}; \citealt{Oteo18}; \citealt{Long20}; \citealt{Zhang22}) which may be triggered by mergers and are believed to be the predecessors of the red ellipticals that dominate the clusters cores by $z=0$ (e.g., \citealt{Ivison13}; \citealt{Smail14}). It is for these reasons that the epoch between $2<z<3$ represents not only the peak epoch of star-formation and black hole activity (\citealt{Madau14}) but also a crucial stage in the evolutionary pathway of protocluster galaxy populations, setting the (pre-)conditions for the start of environmental effects both as a consequence of changes in the gas accretion due to their rapid dark matter halo growth (\citealt{Dekel09a}; \citealt{Keres09}) as well as due to enhanced galaxy-galaxy interactions (e.g., \citealt{Gottlober01}; \citealt{Genel14}; \citealt{Hine16}; \citealt{Coogan18}; \citealt{Watson19}).

As such, numerous studies have been carried out during the last decade to determine the interplay between the protocluster environment and the accelerated evolution of the galaxy populations therein in terms of star-formation activity, metal enrichment, and molecular gas content (see \citealt{Alberts22} for a review). However, a unified picture depicting how the cluster assembly process affects the early evolution of galaxies is still missing, with contradicting results between protoclusters at similar cosmic epochs. Some works suggest the enhancement of star formation relative to the field in high-z dense environments both as a result of higher gas accretion and merging rate (\citealt{Alberts14}; \citealt{Shimakawa18a}; \citealt{Lemaux22}; \citealt{Monson21}; \citealt{Shi23}), which would contribute to an accelerated mass assembly compared to their field counterparts. However, others report no such differences (e.g. \citealt{Toshikawa14}; \citealt{Cucciati14}; \citealt{Shi21}; \citealt{Sattari21}; \citealt{Koyama21}; \citealt{Polletta21}; \citealt{PerezMartinez23}). A similar situation is found regarding the metal enrichment of galaxies in protoclusters. Some authors claim that the early development of an intracluster medium (ICM) in massive protoclusters (e.g., \citealt{Willis20}; \citealt{Tozzi22b}; \citealt{DiMascolo23}) would contribute to detach infalling galaxies from the cosmic web, progressively hampering their gas accretion and forcing them to enrich their ISM via strangulation (\citealt{Maier19b}) and by recycling of already processed gas due to the external pressure of an overdense IGM (e.g. \citealt{Kulas13}; \citealt{Shimakawa15}; \citealt{PerezMartinez23}). On the other hand, various levels of metallicity deficiency have been reported in other protoclusters at similar redshifts (\citealt{Valentino15}; \citealt{Chartab21}; \citealt{Sattari21}). Paradoxically, the proposed physical scenario to explain such results involves the powerful accretion of cold gas in very young and not massive ($\mathrm{\log\,M_{cl}/M_\odot<13.5}$) high-z protoclusters, which would contribute to both fuel star formation and dilute the metal content with respect to the general field. Conversely, some other studies have not observed significant environmental dependence of the mass-metallicity relation during this epoch (e.g. \citealt{Tran15}; \citealt{Kacprzak15}; \citealt{Namiki19}). 

Furthermore, the molecular gas properties of protocluster galaxies at $z>2$, which are key to understanding the gas feeding and consumption processes fueling star formation, have been traced by a significant number of studies albeit over relatively small sample sizes (\citealt{Dannerbauer17}; \citealt{Coogan18}; \citealt{Wang18}; \citealt{Tadaki19}; \citealt{Zavala19}; \citealt{Champagne21}; \citealt{Jin21}; \citealt{Aoyama22}; \citealt{Ikeda22}). In \cite{PerezMartinez23} the authors proposed a co-evolution scenario where the total mass of different protoclusters could play a key role in the physical properties of the galaxies inhabiting them due to the predicted changes on their accretion mode (cold vs hot, \citealt{Dekel09a}) as a consequence of the protocluster dark matter halo growth and early virialization (\citealt{Overzier08}; \citealt{Toshikawa14}). However, this scenario requires further observational support and a number of potential biases may also play a role in explaining some of the aforementioned contradictions. For example, the different selection criteria for the parent samples of protocluster galaxies under scrutiny, the lack of statistics due to small sample size, the use of a diverse set of star-formation, metal enrichment, and molecular gas tracers, the difficulties in precisely defining environmental variables in this forming structures, and the presence of AGNs and the relevance of their fraction (e.g., \citealt{Macuga19}; \citealt{Monson21}; \citealt{Tozzi22a}) in the recent past and future protocluster evolution.

Among the many attempts to establish an unbiased and systematic census of star-formation activities in high-z protoclusters, the MAHALO-Subaru project (MApping H-Alpha and Lines of Oxygen with Subaru) led by \citealt{Kodama13} stands as one of the most successful and prolific environmental surveys over the last decade. It has carried out a series of narrow-band observations targeting H$\alpha$ and Oxygen emission lines tracing the star formation activity within the cores of 10 overdensities at $0.5<z<2.5$ (e.g. \citealt{Hayashi10,Hayashi11,Hayashi12,Hayashi16}; \citealt{Koyama10,Koyama13,Koyama14}; \citealt{Tanaka11}; \citealt{Tadaki12}; \citealt{Shimakawa18a,Shimakawa18b}). In this work, we spectroscopically follow up one of these structures, the USS\,1558--003 protocluster at $z=2.53$ (hereafter USS1558), to investigate the effect of the environment over the star formation activity, gas-phase metallicity, and molecular gas properties of narrow-band selected H$\alpha$ emitters (HAEs). USS1558 was first discovered as an overdensity of distant red galaxies (DRGs) around the radio galaxy 4C-00.62 (\citealt{Kajisawa06}; \citealt{Kodama07}; \citealt{Galametz12}; \citealt{Hayashi12}). The MAHALO-Deep NB imaging observations targeting H$\alpha$ emission at $z=2.53$ identified $>100$ H$\alpha$ emitters (HAEs) in an area of $\mathrm{3\times2\,pMpc^{2}}$ around the radio galaxy (\citealt{Hayashi12,Hayashi16}; \citealt{Shimakawa18a}; \citealt{Suzuki19}). The membership of a significant number of them have been spectroscopically confirmed by rest-frame optical emission lines and ALMA combined CO and dust continuum observations (\citealt{Shimakawa14,Shimakawa15}; \citealt{Tadaki19}; \citealt{Aoyama22}), with most sources being distributed in a filamentary structure made of three dense clumps (\citealt{Hayashi12}) dominated by relatively low mass galaxies ($\mathrm{M_*<10^{10.5}M_\odot}$, \citealt{Shimakawa18a}), yielding a total protocluster mass of $\mathrm{M_{USS1558}\lesssim10^{14}M_\odot}$ (\citealt{Shimakawa14}). Given the group-like structure, and the lack of a diffuse X-ray emission detection in this field (\citealt{Macuga19}), USS1558 is considered to be an immature system at the stage of vigorous assembly and thus the ideal test site to search for environmental effects in a rapidly growing protocluster at the cosmic noon.

In this manuscript we will investigate the star-forming activity, metal enrichment, and gas reservoir of HAEs within USS1558 and compare our results with our recent findings in the Spiderweb protocluster (hereafter PKS1138), which is a more mature coeval system (\citealt{PerezMartinez23}), following the same approach. The manuscript is structured in the following way: $\text{Sect.}$ \ref{S:Data} describes our Keck/MOSFIRE spectroscopic campaign and the archival data available within this field. $\text{Sect.}$ \ref{S:Methods} outlines the methods used to analyze the star formation, metal enrichment, and gas reservoir of the USS1558 protocluster members as well as the parametrization we use to define the environment. $\text{Sect.}$ \ref{S:Results} and \ref{S:Discussion} present our main results and the discussion of their physical interpretation in the context of galaxy evolution respectively. Finally, $\text{Sect.}$ \ref{S:Conclusions} outlines the major conclusions of this study. Throughout this article, we assume a \citet{Chabrier03} initial mass function (IMF), and adopt a flat cosmology with $\Omega_{\Lambda}$=0.7, $\Omega_{m}$=0.3, and $H_{0}$=70 km\,s$^{-1}$Mpc$^{-1}$. All magnitudes quoted in this paper are in the AB system (\citealt{Oke83}).


\section{Observations}
\label{S:Data}

\subsection{Keck/MOSFIRE spectroscopy}
\label{SS:KMOS} 

\begin{figure}
\centering
\includegraphics[width=\linewidth]{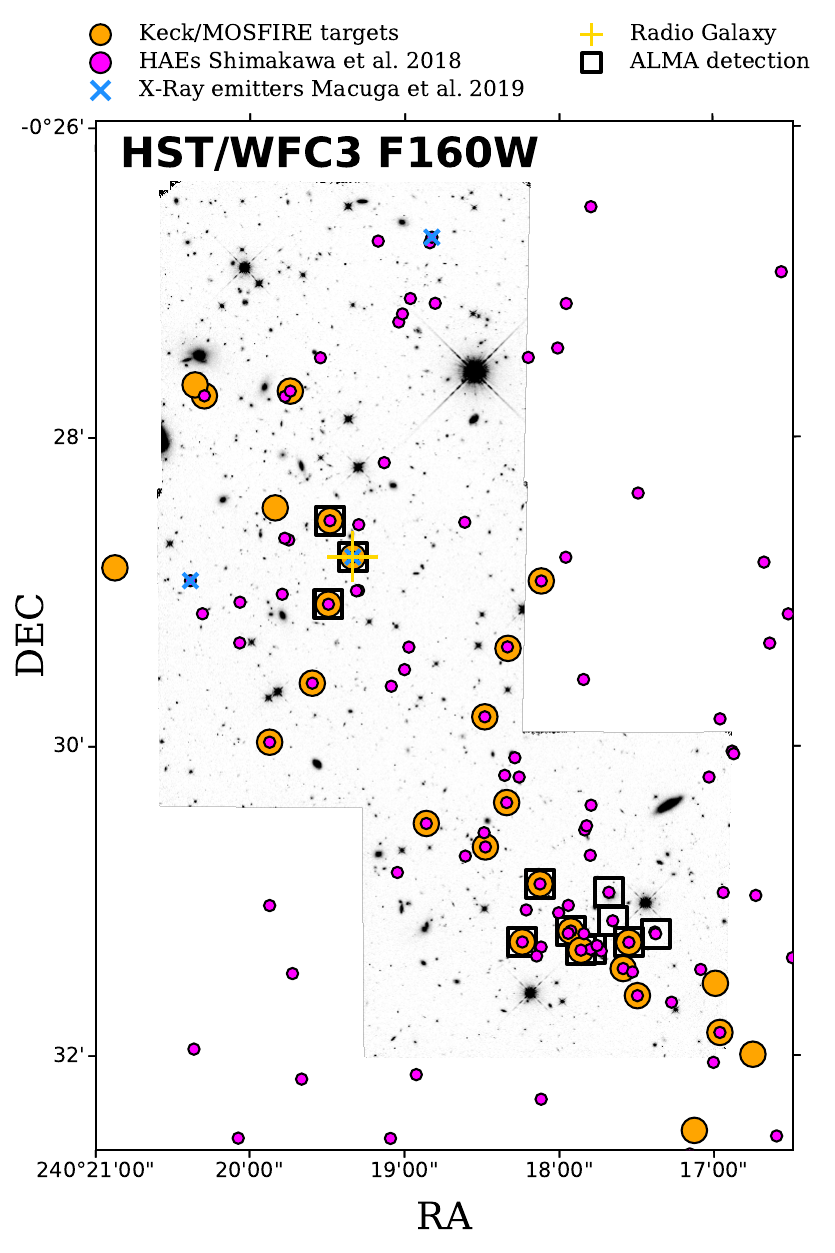}\par
\caption{Distribution of narrow-band HAEs in the field of the USS1558 protocluster. Orange circles show the sample of Keck/MOSFIRE targets studied in this work. Magenta small circles depict the parent sample of confirmed narrow-band HAEs from \protect\cite{Shimakawa18a}. Empty squares highlight a small overlapping sample of ALMA CO(3-2) sources from \protect\cite{Tadaki19} and dust continuum emitters from \protect\citealt{Aoyama22}. Blue crosses display the position of X-ray emitters from \protect\cite{Macuga19}. The big yellow cross shows the position of the radio galaxy 4C-00.62 (\protect\citealt{Kajisawa06}). All targets are overlaid over the HST/WFC3 F160W mosaic.}
\label{F:Map}
\end{figure}

\begin{figure}
\centering
\includegraphics[width=\linewidth]{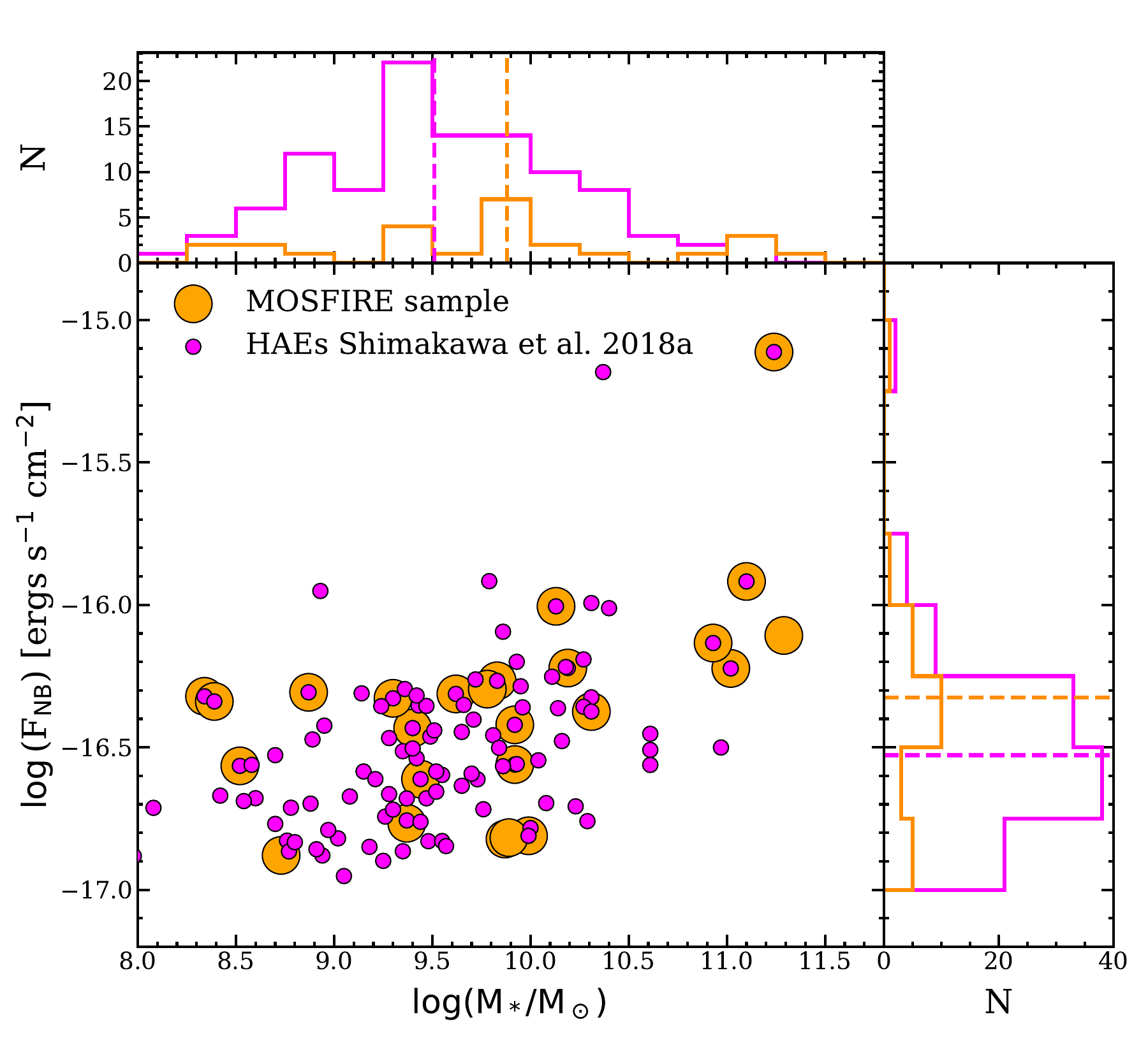}
\caption{Main diagram: Stellar-mass and narrow-band flux excess properties of our 27 Keck/MOSFIRE targets (orange circles) and their parent sample of 107 confirmed HAEs (magenta circles) from \protect\cite{Shimakawa18a}. Side panels: Distribution (histograms) and median values (dashed lines) of the MOSFIRE and parent samples. Colors follow the same scheme applied in the main diagram.}
\label{F:SEL}
\end{figure}

We carried out multi-object spectroscopy observations of a sample of narrow-band selected HAEs in the USS1558 protocluster (\citealt{Hayashi12,Hayashi16}; \citealt{Shimakawa18a}) to obtain H$\alpha$ and [N{\sc{ii}}]$\lambda6584$ emission line fluxes and study the galaxies' star-formation activity, gas-phase metallicities and their relation with the surrounding environment. Our spectroscopic targets encompass the three densest clumps across the large scale structure of USS1558 (Fig.\,\ref{F:Map}) and were selected to have narrow-band flux excess above $\mathrm{F_{NB}\geq1.5\times10^{-17}}$ ergs\,s$^{-1}$\,cm$^{-2}$ (Fig.\,\ref{F:SEL}). The observations were carried out using the NIR multi-object spectrograph MOSFIRE (\citealt{McLean10,McLean12}) installed at the Cassegrain focus of the Keck I telescope in Mauna Kea, Hawaii (Program ID: S20A-075; PI: T. Kodama) using $0.7\arcsec$ wide slits. We used a single mask configuration targetting 27 objects previously identified by \cite{Hayashi12,Hayashi16} and \cite{Shimakawa18a}, plus one star to monitor the sky conditions during the night. Out of these 27 targets, \cite{Shimakawa18a} classified 20 of them as confirmed HAEs (magenta circles in Fig.\,\ref{F:Map}) and constitute our primary sample. The remaining 7 objects were labeled as candidate HAEs by the same authors due to the lack of color-color diagram (e.g., $\mathrm{r'JK_s\,and\,Br'K_s}$) or spectroscopic confirmation. These targets were included in our mask design either as fillers or as companion galaxies within the same slit assigned to one of our primary targets. Finally, eight objects within our USS1558 sample have ALMA molecular gas information available either through CO(3-2) emission (\citealt{Tadaki19}) or from dust continuum observations (\citealt{Aoyama22}). 

The observing run was carried out during the night of June 25th 2021 following an ABA'B' dithering pattern of individual exposures consisting of 180s each in K-band ($19540-23970$\,\AA). This configuration yields a nominal spectral resolution of R$\sim$3600 around the central wavelength of the grating enabling us to resolve H$\alpha$ and [N{\sc{ii}}]$\lambda6584$ at $z=2.53$, and a spatial sampling of $0.2\arcsec$ per pixel. We integrated for 174 min under average $0.6-0.7\arcsec$ seeing conditions (FWHM) and airmass value of 1.5. The spectra were reduced using the MOSFIRE Data Reduction Pipeline (MOSFIRE-DRP, \citealt{Steidel14}) following standard reduction procedures (i.e., bias subtraction, flat-field normalization, sky subtraction, wavelength calibration) producing a final 1D science and error spectra for each target using the \cite{Horne86} algorithm. We carry out the flux calibration of our sample by using the slit star (RA=16:01:19.26, DEC=-00:30:35.49) within our mask. First, we identify the spectral type of the star based on the G-band absolute magnitude and effective temperature ($\mathrm{t_{eff}}$) reported by the Gaia DR3 survey (\citealt{Gaia16,GaiaDR3}). This yields $\mathrm{t_{eff}=5096^{+35}_{-42}\,K}$ and $\mathrm{G\approx6.8\,mag}$ which correspond to a main sequence K1 star. The continuum emission of K stars is affected by strong features in the optical regime, usually compromising their utility as flux calibrators. However, these features are relatively weak in the K-band. Thus, we retrieve a model of such spectral type (\citealt{Pickles1998}) and re-scale it to match the 2MASS NIR broad-band photometry of the star (\citealt{2MASS}), thus modeling the intrinsic spectrum. By dividing the modeled and observed stellar spectra we obtain the transmission function which includes the conversion from raw units to physical flux, the slit loss correction, and all instrumental and atmospheric effects present during our observing run. Finally, we apply the same transmission function to all our targets. 

\subsection{Archival Photometry}
\label{SS:Photometry} 

Over the last decade, the USS1558 protocluster field has been surveyed by several optical to NIR photometric campaigns (\citealt{Hayashi12,Hayashi16}. These observations include the Subaru/Suprime-Cam B, r', z' bands and the Subaru/MOIRCS J, H, K$_s$ bands and the NB2315 narrow-band filter (Program IDs: S10B-028 and S15A-047, PI: T. Kodama). In addition, this field is almost completely covered by a Hubble Space Telescope (HST) WFC3 program in H-band (i.e, F160W; Program ID: GO-13291, PI: M. Hayashi). In this work, we make use of the reprocessed mosaics in each band (\citealt{Shimakawa18a}) to compute the stellar mass and dust extinction of our main targets. Even though the characteristics of these observations have been extensively described in detail in previous publications (\citealt{Hayashi16}; \citealt{Shimakawa18a}), we provide a brief summary of their main features in Table\,\ref{T:Imaging}. 

\begin{table*}
\caption{Optical to NIR photometric bands available for the USS1558 protocluster field. Integration times, FWHM, pixel size, and galactic extinction reproduced from \protect\cite{Shimakawa18a}. Ellipsis are placed to avoid repetitions in the instrument and reference columns.}
\centering
\begin{tabular}{llcccccc}
\hline
\noalign{\vskip 0.1cm}
Instrument &  Filter   & Int. Time &  FWHM & Pixel-size & $\mathrm{A_\lambda}$ & Reference  \\ 
           &           & (min)       &   ($\arcsec$) & ($\arcsec$) & (mag) &  \\ \hline 
\noalign{\vskip 0.1cm}
Suprime-Cam/Subaru & B        &  80  &  0.70 & 0.200 & 0.55 & \citealt{Hayashi12} \\
\ldots        & r'            &  90  &  0.67 & 0.200 & 0.35 & \ldots \\
\ldots        & z'            &  55  &  0.67 & 0.200 & 0.19 & \ldots \\
MOIRCS/Subaru & J             &  191 &  0.67 & 0.116 & 0.11 & \citealt{Hayashi12,Hayashi16}\\
\ldots        & H             &  45  &  0.67 & 0.116 & 0.07 & \ldots\\
\ldots        & K$_s$         &  207 &  0.67 & 0.116 & 0.05 & \ldots \\
\ldots        & NB2315        &  583 &  0.67 & 0.116 & 0.05 & \ldots \\ 
WFC3/HST      & F160W         &  87  &  0.21 & 0.130 & 0.08 & \ldots \\ 
\hline
\end{tabular}
\label{T:Imaging}
\end{table*}

\section{Methods}
\label{S:Methods}

\subsection{Emission line fitting}
\label{SS:EL}

Our Keck/MOSFIRE program aims at obtaining information from the H$\alpha$ and [N{\sc{ii}}]$\lambda6584$ emission lines for HAEs at $z\approx2.5$. We will carry out the emission line fitting by making use of a self-written code written in {\sc{python}} (Astropy library, \citealt{Astropy13, Astropy18}) which was previously utilized to fit the same spectral features in HAEs belonging to PKS1138 at similar redshift (\citealt{PerezMartinez23}). We present a brief summary of its most important features in the following paragraph.

First, we visually inspect the flux-calibrated spectrum of each HAE searching for emission lines within the width of the NB2315 narrow-band filter (i.e., $\sim23000-23300$\,\AA). Our sample is initially selected to be made of 27 confirmed and candidate HAEs at $z\approx2.5$ from \cite{Shimakawa18a}, based both on narrow-band excess and color-color selection throughout the r'JK$_s$ and Br'K$_s$ diagrams, which separate interlopers at lower and higher redshift respectively. Only two objects (IDs 9 and 84) displayed no detectable emission within the inspected wavelength range. ID 84 was classified as a non-emitter by the parent sample and it is a filler in our spectroscopic campaign. On the other hand ID 9 was identified as HAE candidate by \cite{Shimakawa18a} albeit it was not possible to discard it as a background galaxy due to its dim colors in the Br'K$_s$ diagram. However, no emission could be detected for this object. In addition, two [O{\sc{iii}}]$\lambda5007$ emitters at $z\approx3.6$ were spectroscopically confirmed in our observing run (IDs 25 and 151). These objects were classified as HAE candidates in \cite{Shimakawa18a} due to their narrow-band flux excess although their interloper nature could not be discarded for the same reasons applied to ID 9. This represents contamination of $\sim12\%$ within the emitting candidates (i.e. 3 out of 26 objects) or a $\sim88\%$ success rate in the spectroscopic identification of narrow-band selected emitters at $z\approx2.5$. This rate is in line with previous narrow-band studies suggesting a contamination rate by fore- and background sources of $\sim10\%$ (e.g., \citealt{Sobral13}, \citealt{PerezMartinez23}). These four objects have been removed from subsequent analyses yielding a final sample composed of 23 confirmed HAEs. 

Once an emission line is identified, we proceed to fit and subtract a local continuum consisting of several windows free from skyline contamination and encompassing a few hundred angstroms left and right of the detected line. Then, we perform a simultaneous triple Gaussian fit for the H$\alpha$ line and [N{\sc{ii}}]$\lambda\lambda6548, 6584$ doublet while tying the center of the [N{\sc{ii}}] lines to their expected value based on the H$\alpha$ fit and fixing the flux ratio $F_{\lambda6584}/F_{\lambda6548}=3$ (\citealt{Storey00}). Furthermore, we also couple the H$\alpha$ and [N{\sc{ii}}] line widths since they originate in the same star-forming region while their amplitude remains a free parameter. The radio galaxy (ID 138) displays an extremely broad H$\alpha$ profile ($\sigma>1000$ km/s) characteristic of type-1 AGN activity and it is the only X-ray emitter from \cite{Macuga19} within our sample. This object is flagged in all subsequent analyses due to the difficulty to separate its AGN and star-forming components. Another target (ID 38) also displays a broad emission line component ($\sigma\approx700$ km/s) suggesting the presence of an AGN in spite of not being detected in X-rays. In these two cases, we use an additional free gaussian component to account for the broad H$\alpha$ emission in addition to the three narrower H$\alpha$ + [N{\sc{ii}}]$\lambda\lambda6548, 6584$ components (see Fig.\,\ref{F:Spec}).

In total, we spectroscopically confirm the protocluster membership of 23 H$\alpha$ emitters (including the two broad-component objects) with 7 of them being new sources unknown to previous spectroscopic studies in this protocluster (\citealt{Shimakawa14,Shimakawa15}; \citealt{Tadaki19}). Finally, we estimate the uncertainties over the fitting procedure by using the Montecarlo approach. The fitting procedure outlined above is repeated 1000 times allowing for random gaussian variations on every spectral data point with a maximum amplitude equal to the value found in the 1D noise spectra. The mean fitted values are then used to compute the total flux for the H$\alpha$ and [N{\sc{ii}}] emission lines. We use the standard deviation over the 1000 realizations as the formal uncertainty of our measurements. Finally, we consider a line to be detected if the ratio between the total flux and the flux error measured in the way described above is higher than two. This constraint applies both to H$\alpha$ and [N{\sc{ii}}]. In total, we identify 10 dual H$\alpha$ + [N{\sc{ii}}]$\lambda6584$ detections at $>2\sigma$ level. Flux upper limits can be derived for 3 extra cases while the remaining 10 objects without [N{\sc{ii}}] estimates will be only used for the stacking analysis. Figure\,\ref{F:Spec} displays the spectral fitting results for every object within our sample. 

\subsection{Stellar masses and dust extinction}
\label{SS:SED}

The USS1558 protocluster field has extensive optical to NIR photometry (see Table\,\ref{T:Imaging}) that can be used to obtain reliable stellar mass values for our targets via SED fitting. To this end, we first homogenized the pixel size and FWHM of our available bands to that of the narrow-band filter. Then, we use the NB2315 image for source detection while running \textsc{SExtractor} (\citealt{Bertin96}) in dual-image mode thus obtaining the observed magnitudes (MAG\_AUTO) for each object in every other band available. This approach ensures that the targets' fluxes are measured consistently over the same area at different wavelengths. We carry out the SED fitting over the multi-band photometric measurements (excluding NB2315) we have gathered for every object with the Code Investigating GALaxy Emission (CIGALE, \citealt{Boquien19}). This code incorporates the stellar, nebular, and dust attenuation and re-emission into a grid of models while conserving the energy balance between its different components. CIGALE then applies Bayesian statistics to compute the physical properties of each object based on the probability distribution function (PDF) of models with respect to the best fit. Thus, the physical properties are statistically derived as the likelihood-weighted mean of the PDF and standard deviations represent the uncertainties taking into account both the goodness of the fit as well as the intrinsic degeneracies between physical parameters.

CIGALE offers great flexibility to customize the grid of models used during the SED fitting through its different modules. In this work, we assume the stellar population synthesis models of Bruzual \& Charlot (\citealt{BC03}) applied to an exponentially delayed star formation history with the following functional form:
\begin{equation}
\mathrm{SFR(t)\propto\frac{t}{\tau^2}\times\exp{(-t/\tau)}}
\label{EQ:SFR}
\end{equation}
with e-folding times ($\tau$) ranging between 0.1 and 8 Gyr, stellar population ages constraint to be younger than the age of the Universe at $z=2.5$ (i.e., $\sim2.5$ Gyr), and allowing for the possibility of a recent minor star-forming burst accounting for up to 1, 5, or 10\% of the mass fraction with age not older than 300 Myrs. The reason behind this choice lies in the higher interaction rate expected during the formation and maturing phases of young protoclusters, which could increase the frequency of starburst events within their members. In addition, we assume a Chabrier IMF (\citealt{Chabrier03}) and subsolar metallicity (i.e $Z = 0.004$) for the stellar component. We constrain the ionization parameter ($U$) between $-2.4<\log(U)<-2.8$ for the nebular emission, which encompasses the typical values for star-forming galaxies at $z\sim2$ (\citealt{Cullen16}). Finally, the dust extinction is modeled throughout Calzetti’s attenuation law (\citealt{Calzetti2000}) with reddening values ranging $\mathrm{E(B-V)=0-1\,mag}$ in steps of 0.1 mag. Most objects display reasonable fits with average reduced $\chi^2=1.7$ within a range of $0.2\leq\chi^2\leq7.9$. The average accuracy of the derived stellar masses is 0.2 dex. All fitted SEDs as well as their reduced $\chi^2$ values are shown in the Appendix. This combination of parameters was previously used by \cite{PerezMartinez23} to describe a sample of HAEs in PKS1138 at $z=2.16$, except for the maximum age of stellar populations which matched the age of the universe at that cosmic epoch (i.e., 3 Gyrs). We decided to follow exactly the same approach for consistency in our environmental analysis (Sect.\,\ref{S:Results}).

\begin{figure*}
 \centering
 \begin{multicols}{4}
      \includegraphics[width=\linewidth]{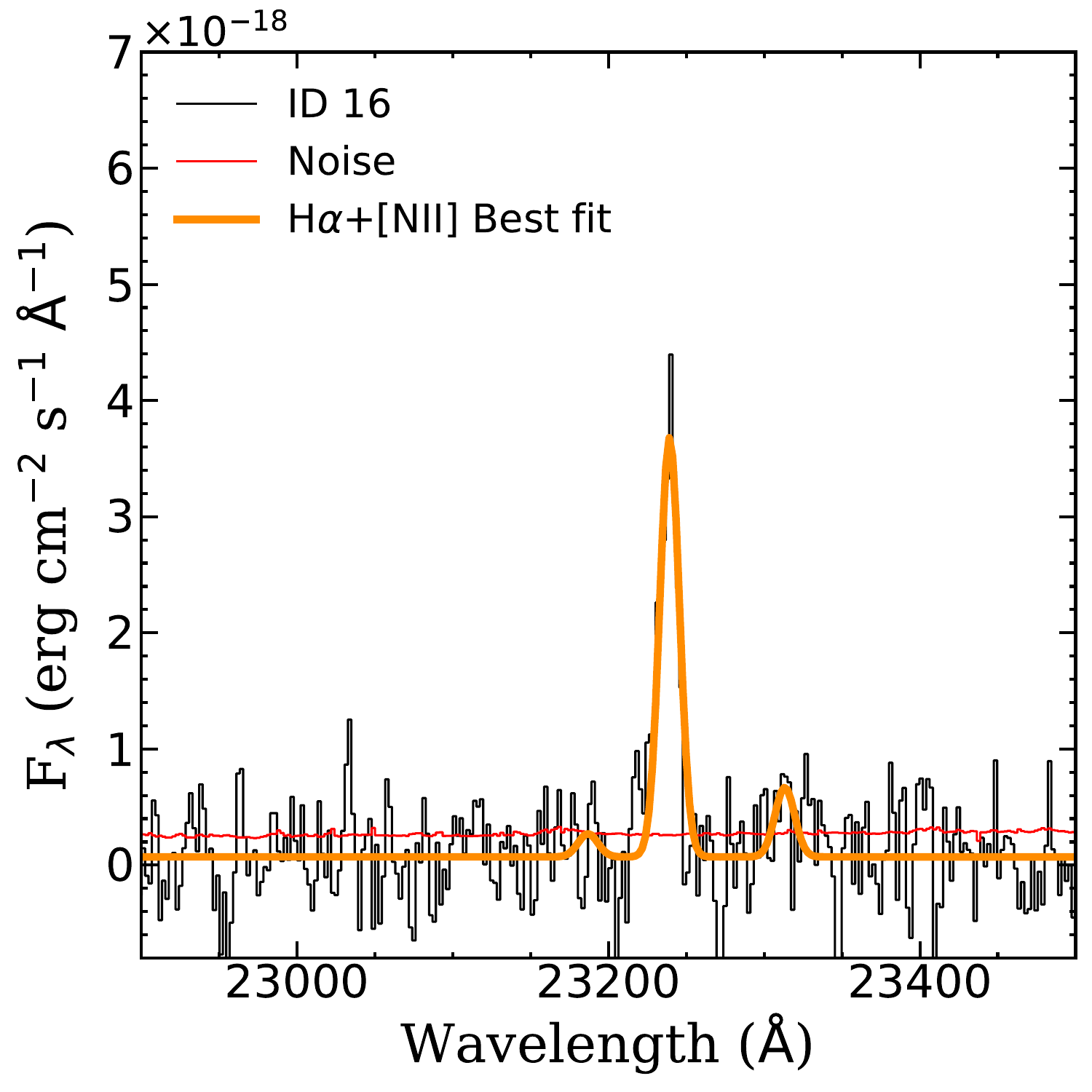}\par
      \includegraphics[width=\linewidth]{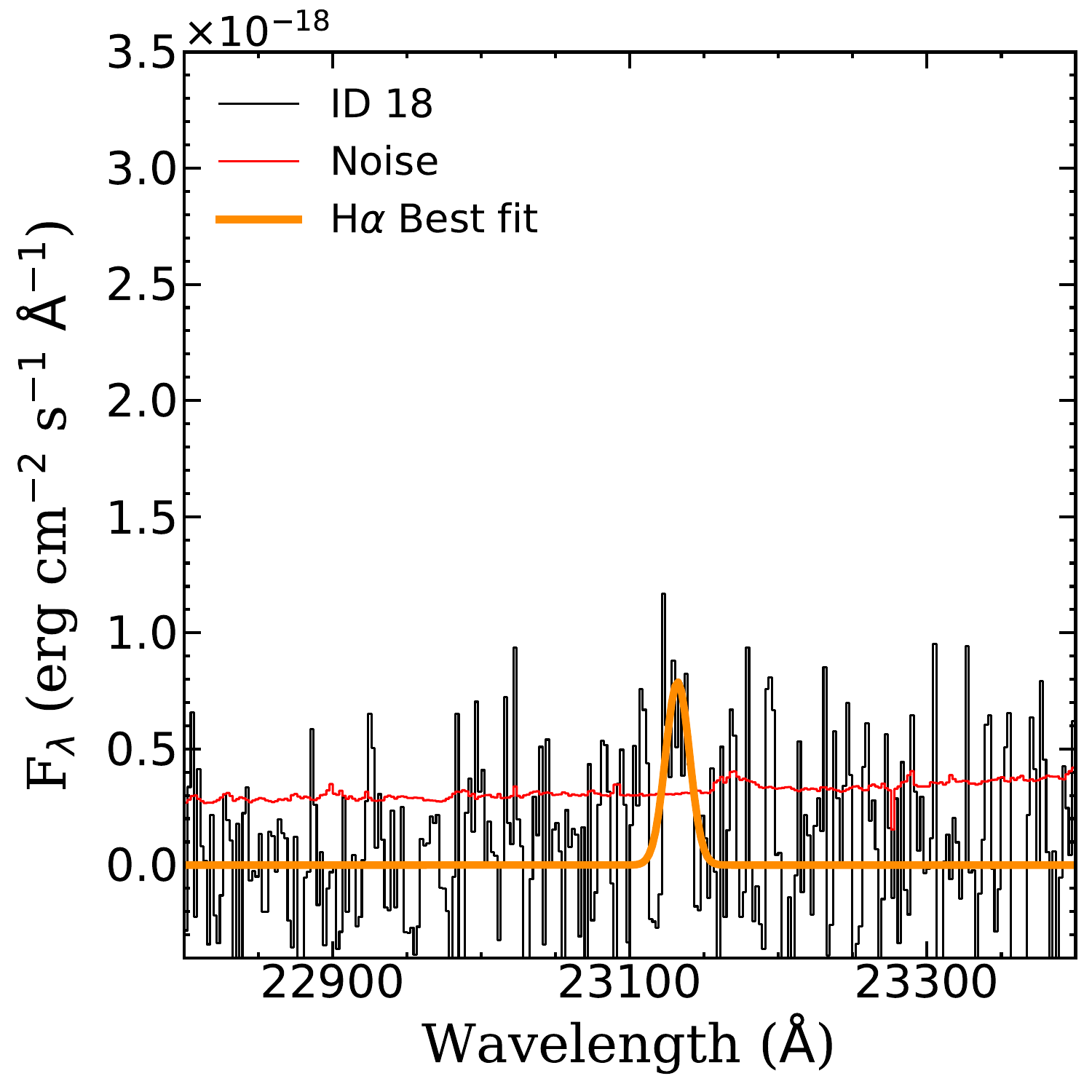}\par
      \includegraphics[width=\linewidth]{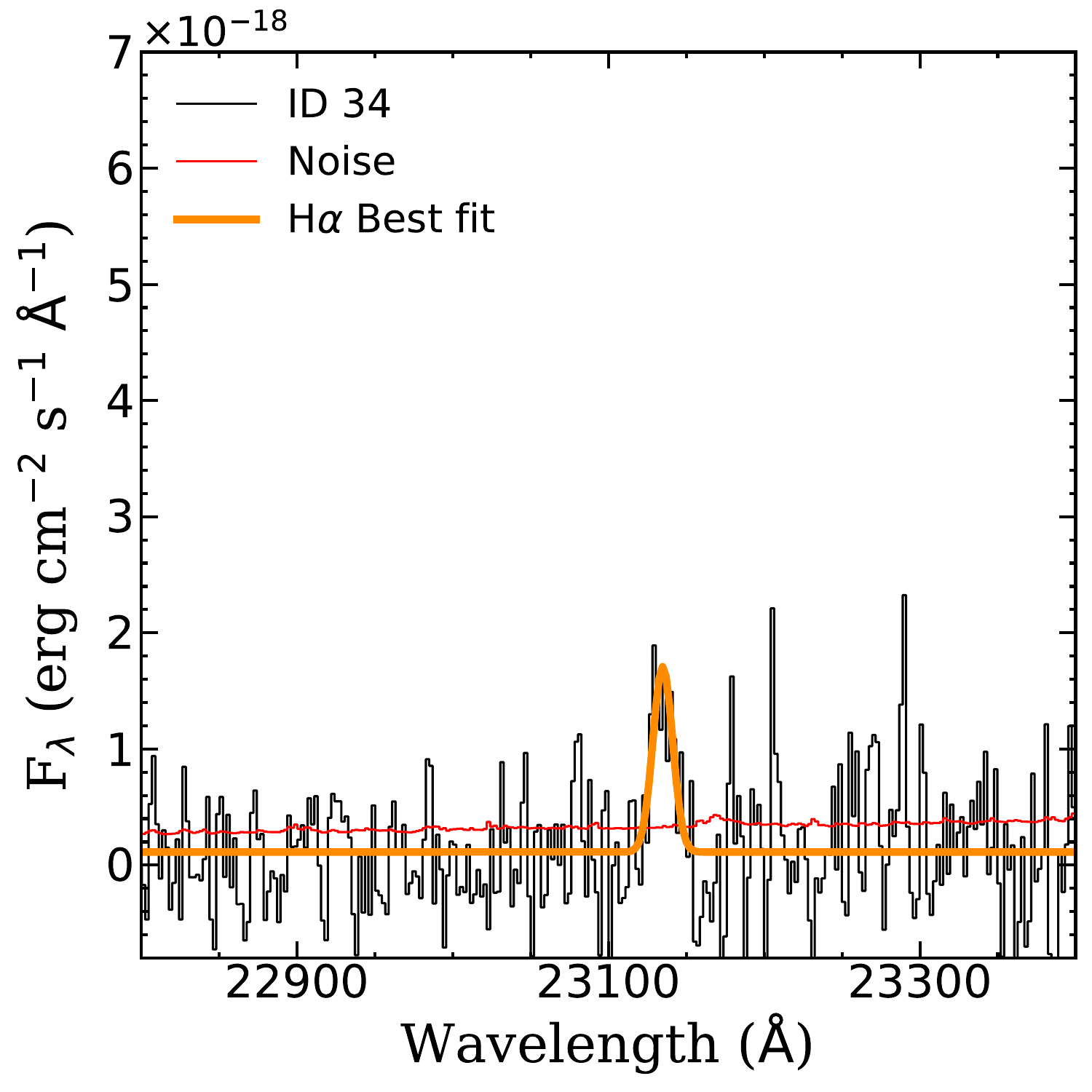}\par
      \includegraphics[width=\linewidth]{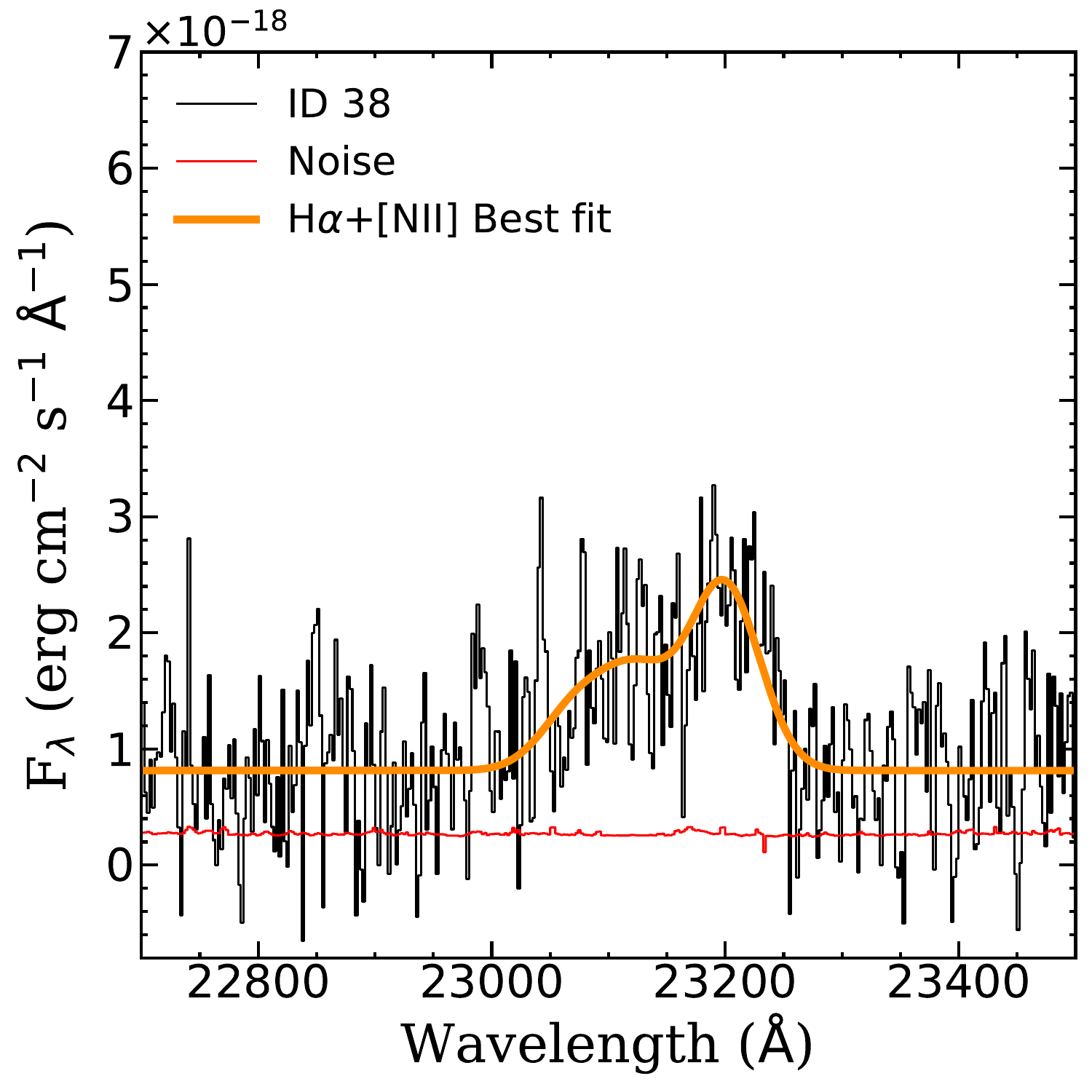}\par
      \includegraphics[width=\linewidth]{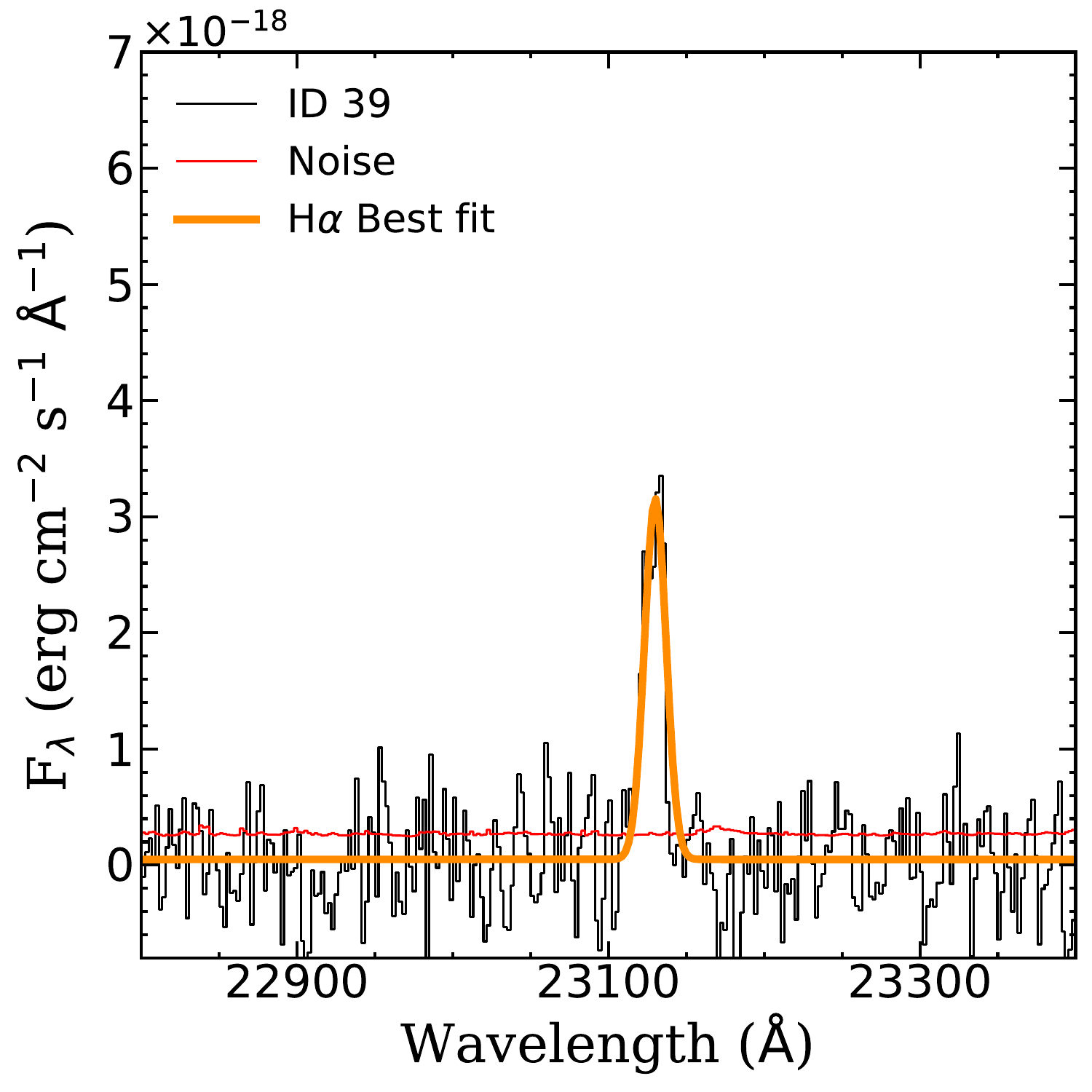}\par
      \includegraphics[width=\linewidth]{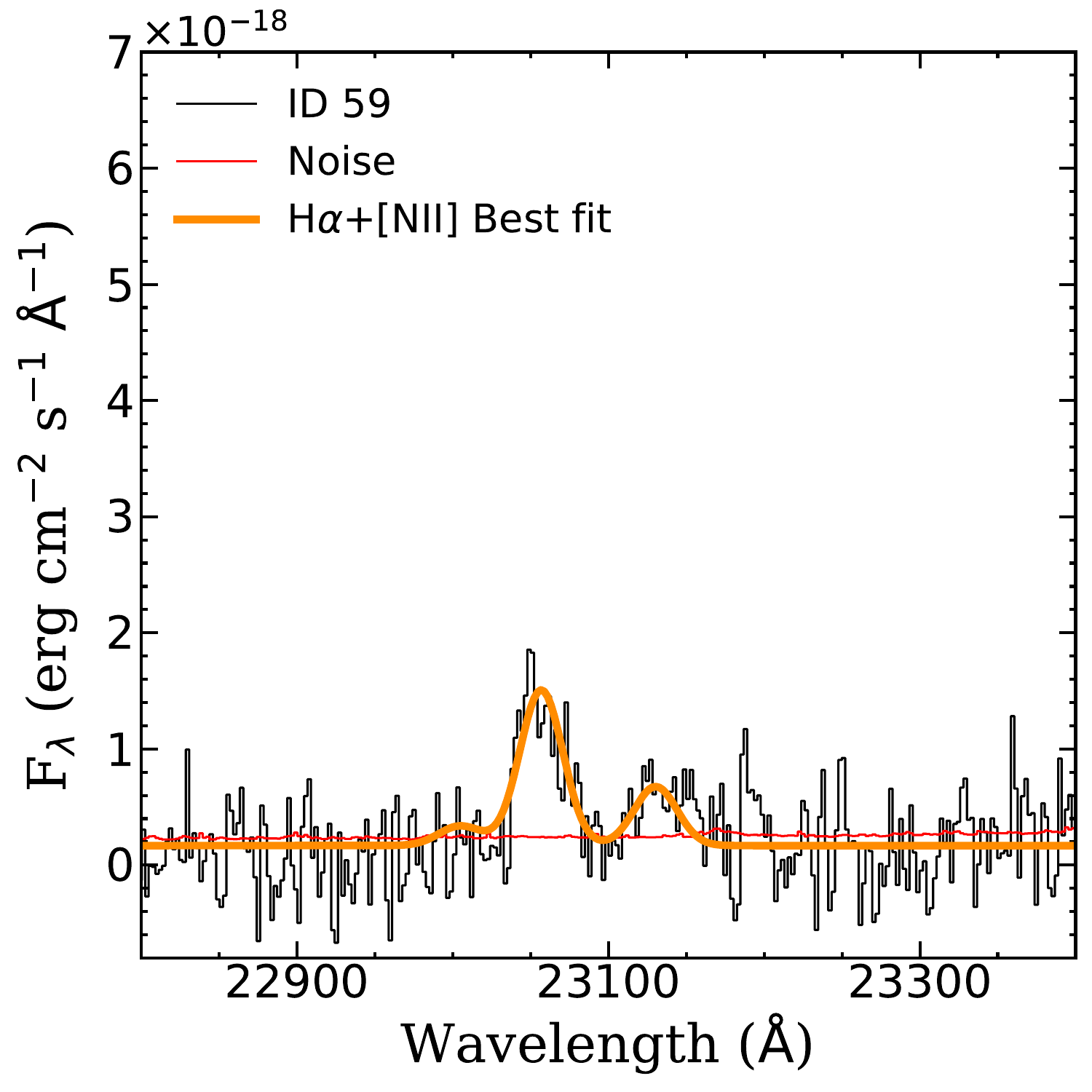}\par
      \includegraphics[width=\linewidth]{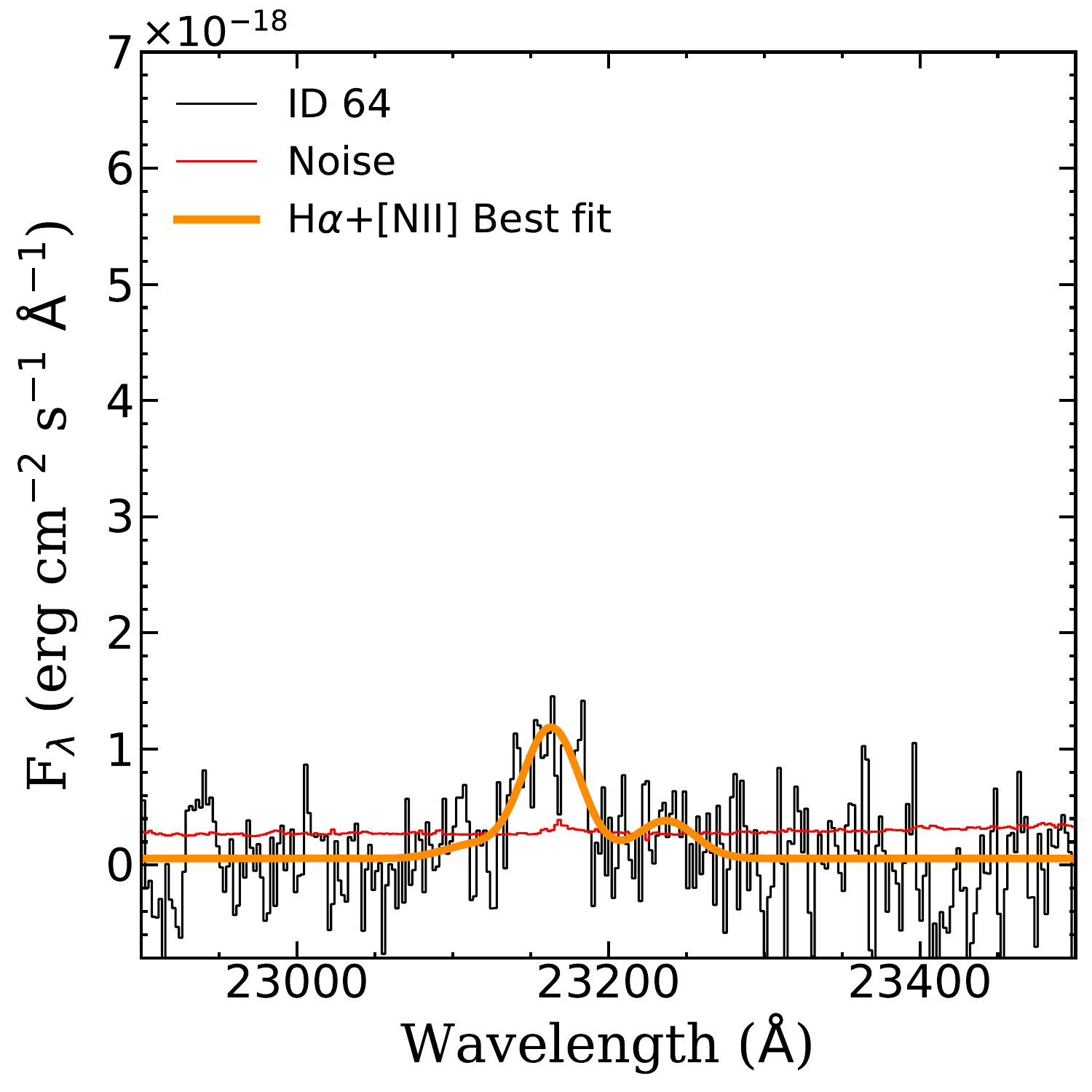}\par
      \includegraphics[width=\linewidth]{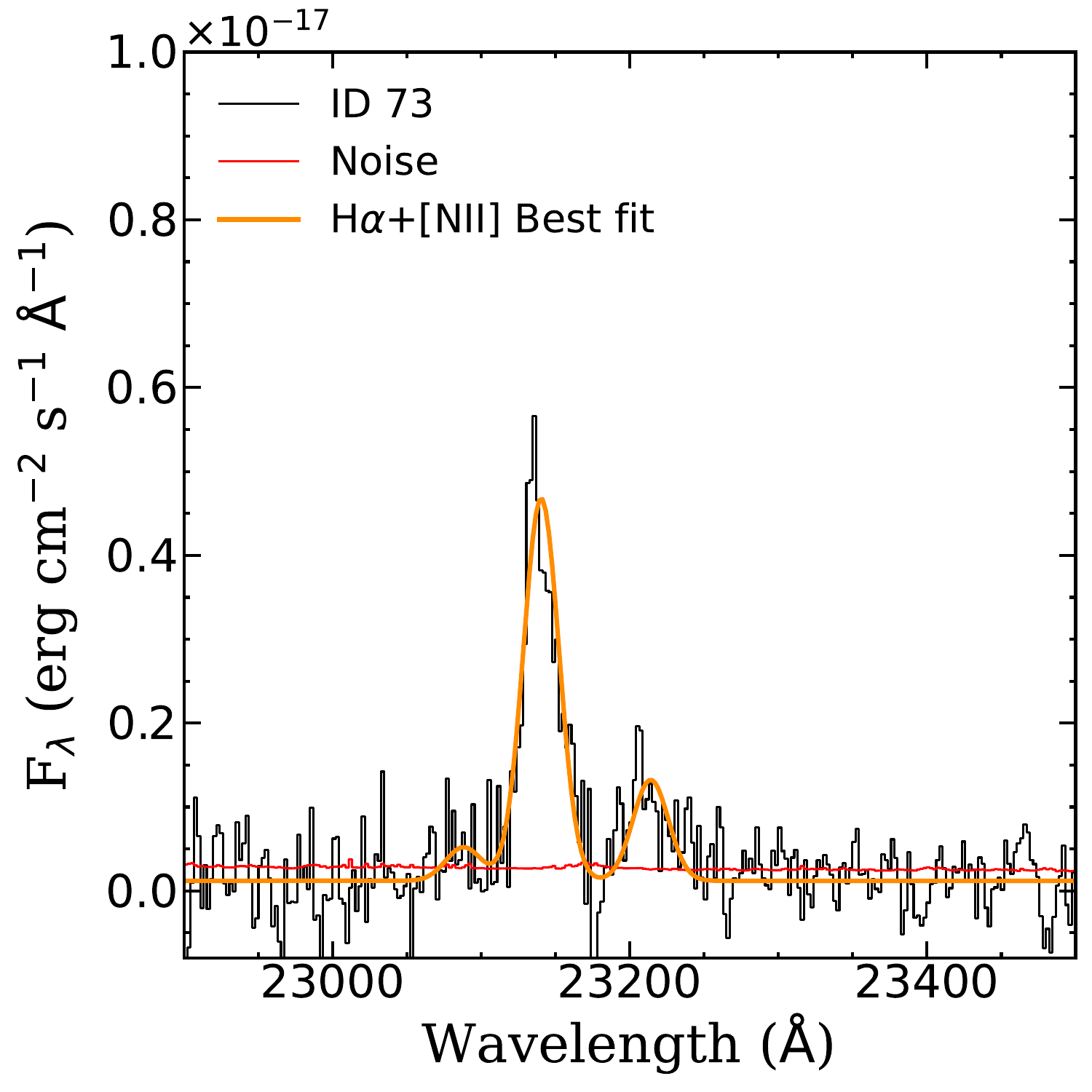}\par
      \includegraphics[width=\linewidth]{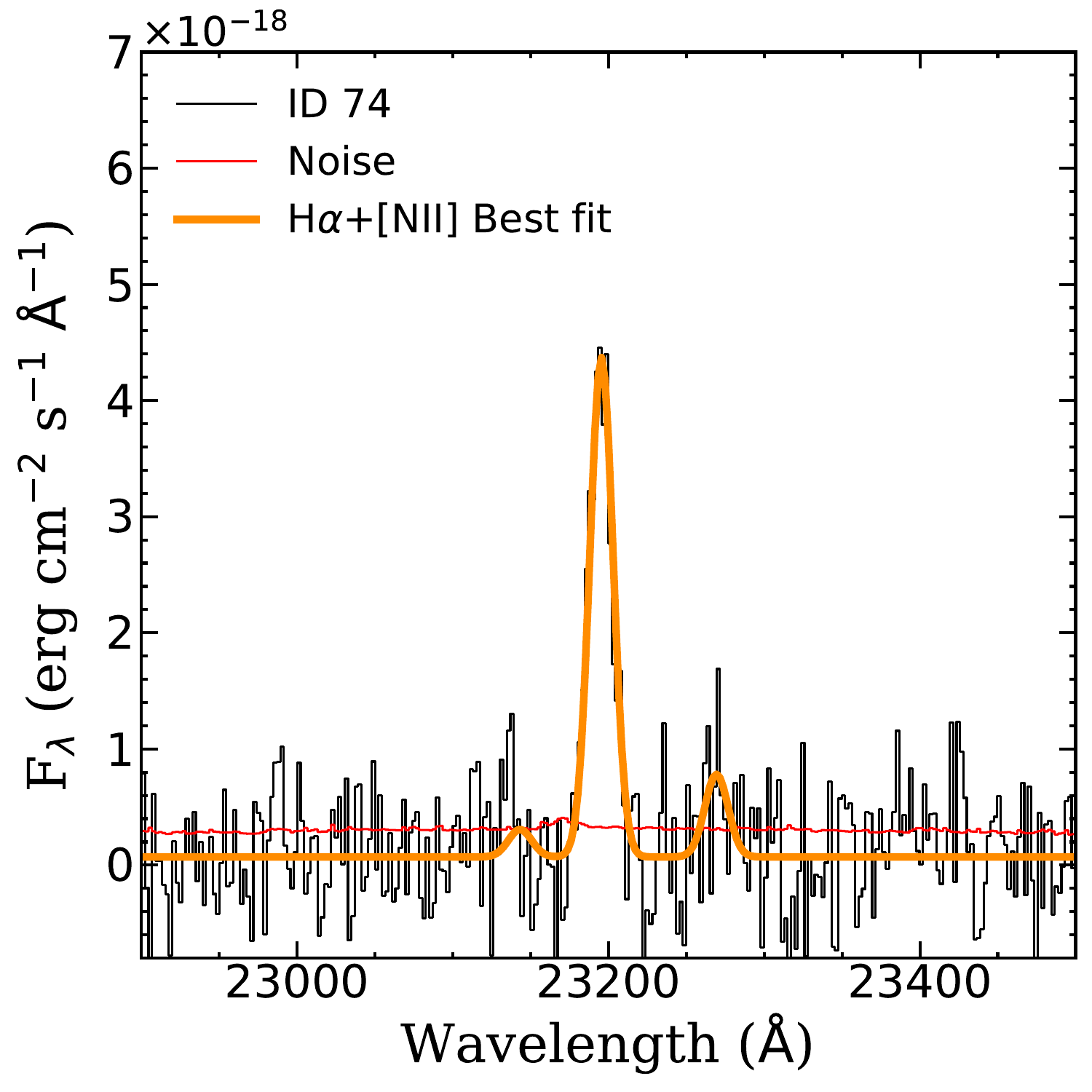}\par
      \includegraphics[width=\linewidth]{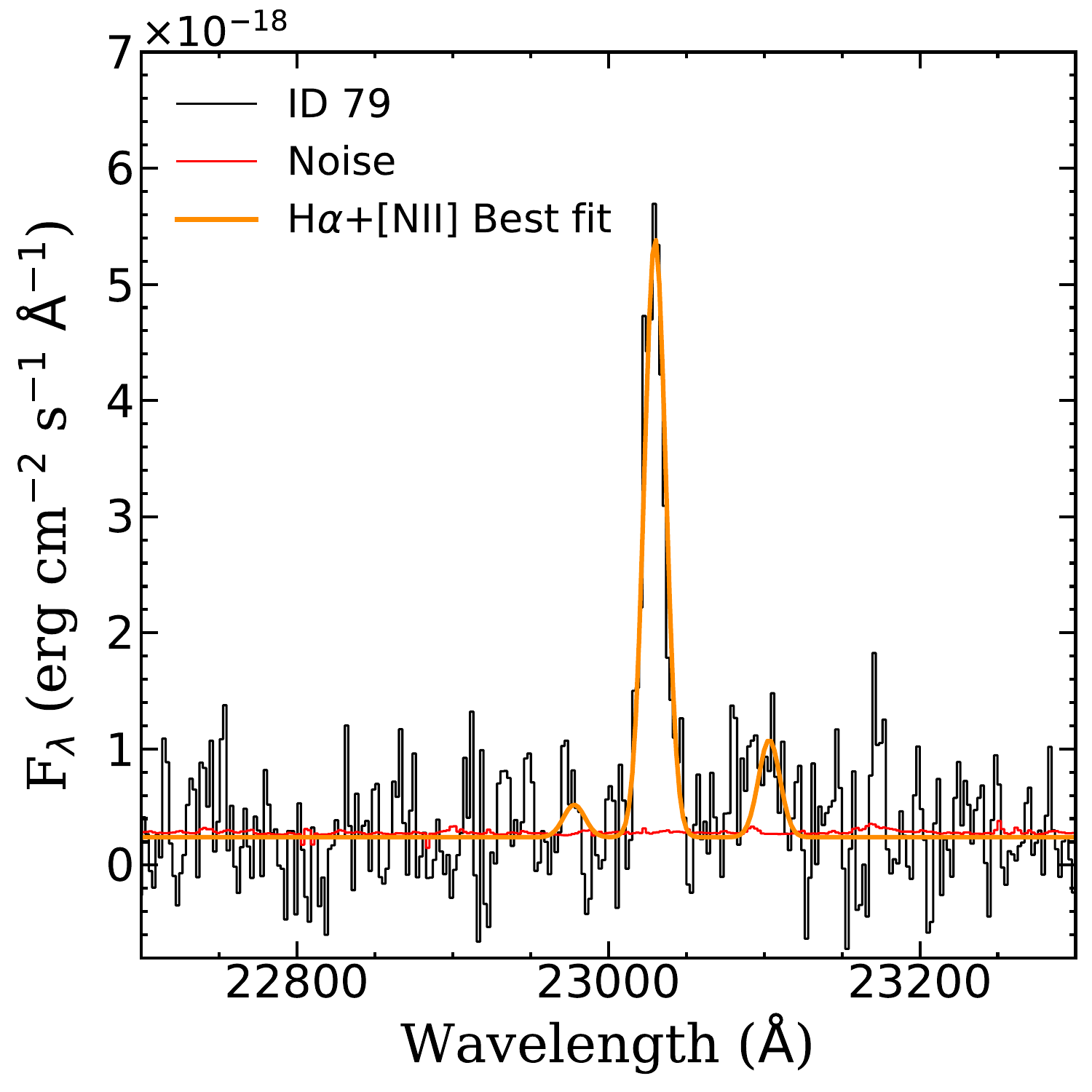}\par
      \includegraphics[width=\linewidth]{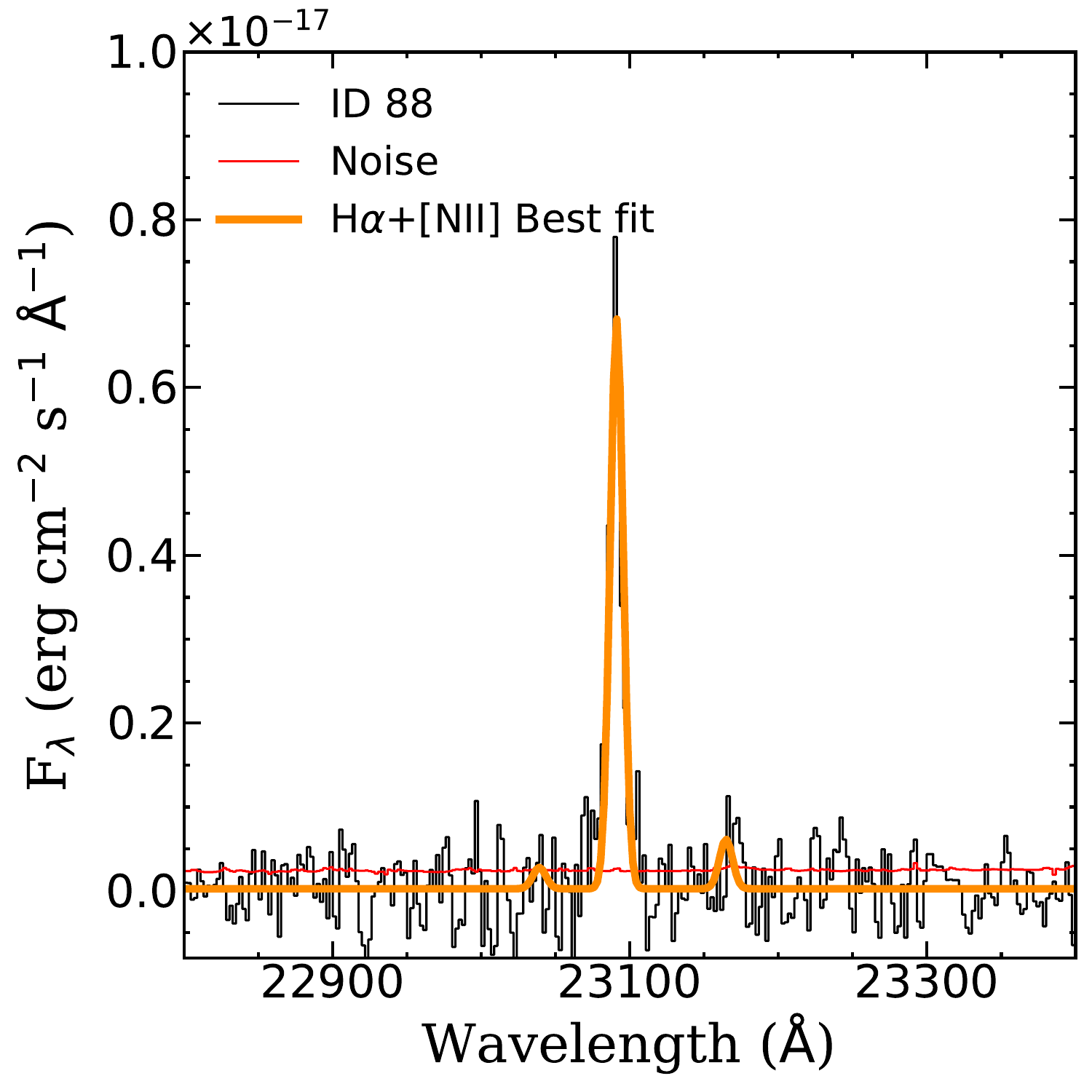}\par
      \includegraphics[width=\linewidth]{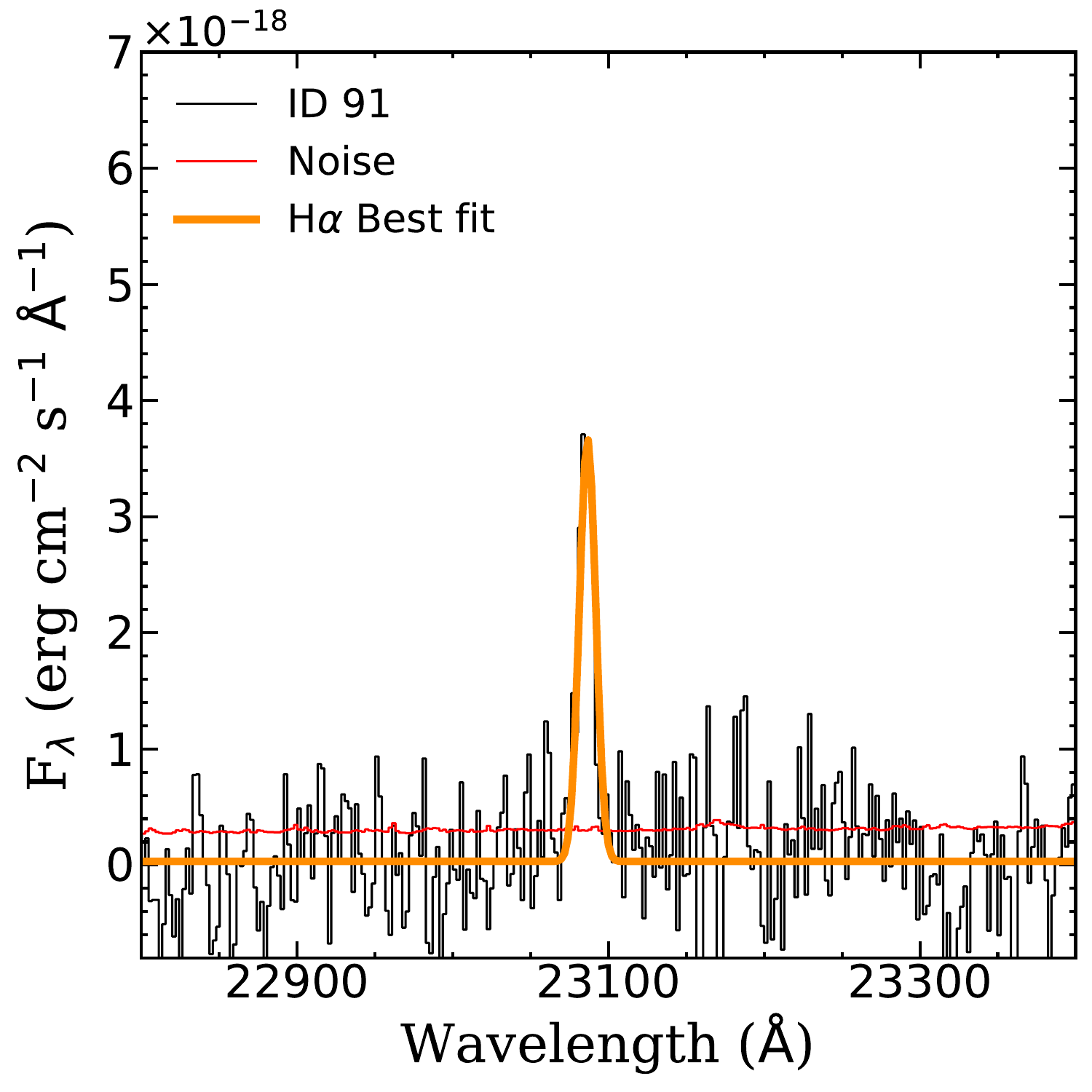}\par
      \includegraphics[width=\linewidth]{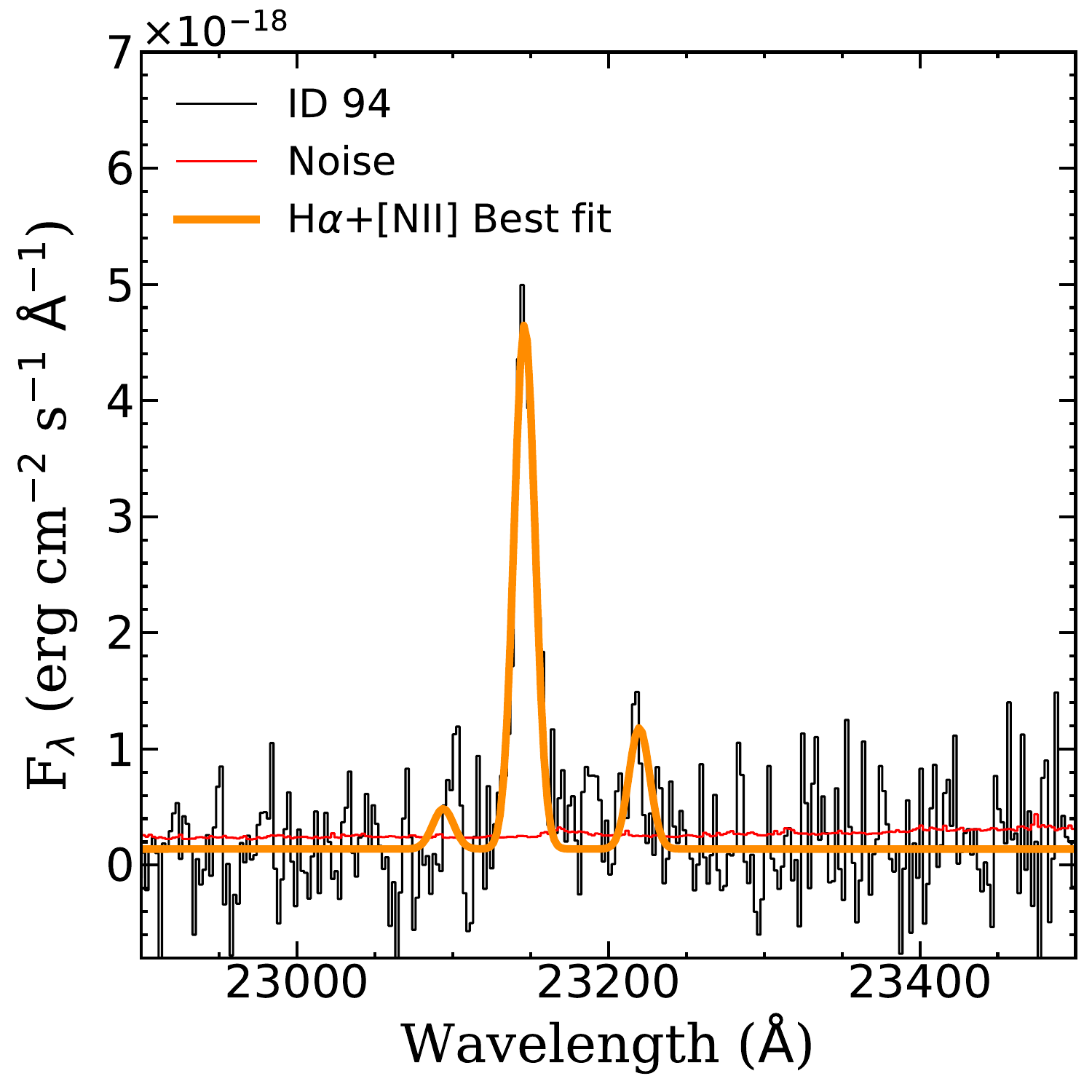}\par
      \includegraphics[width=\linewidth]{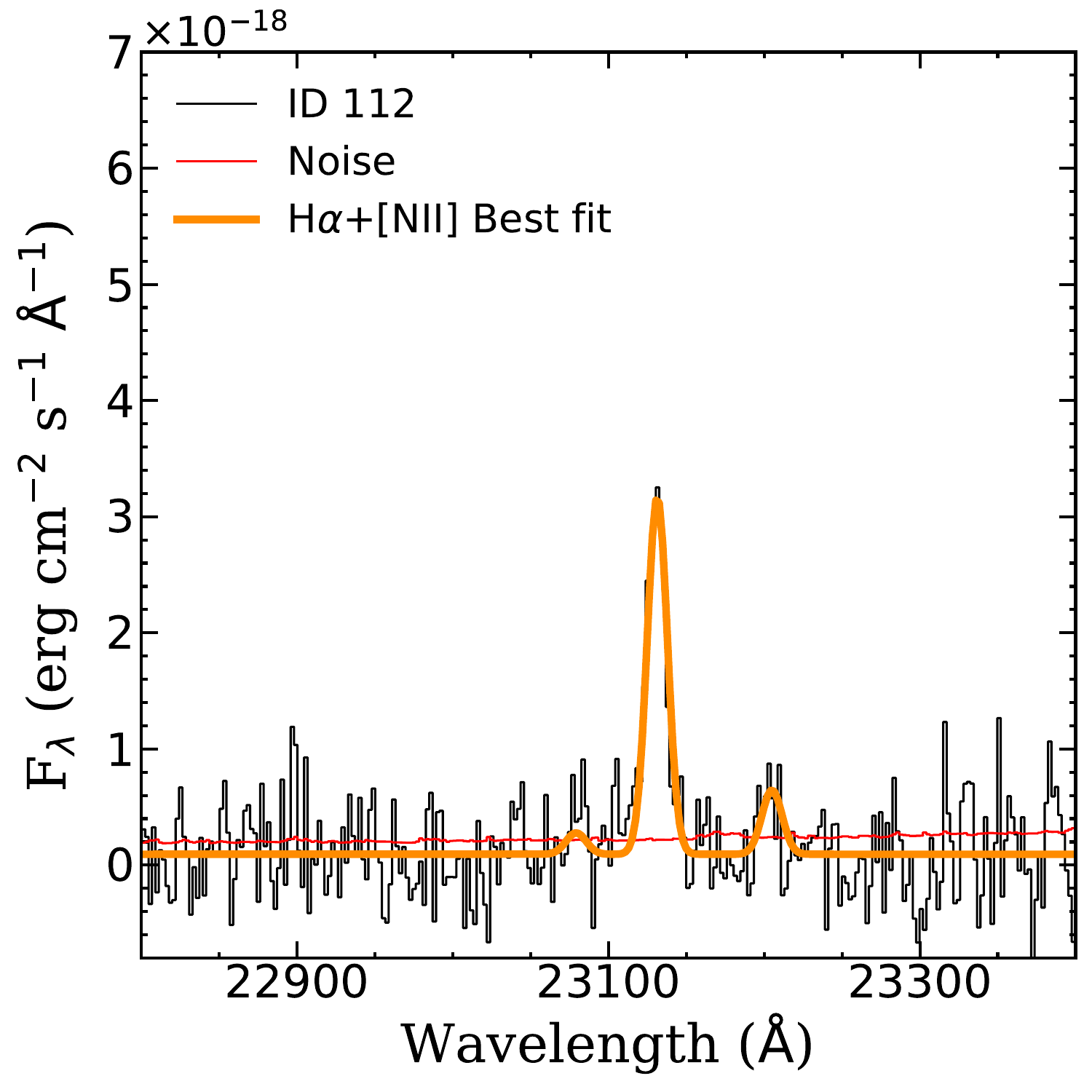}\par
      \includegraphics[width=\linewidth]{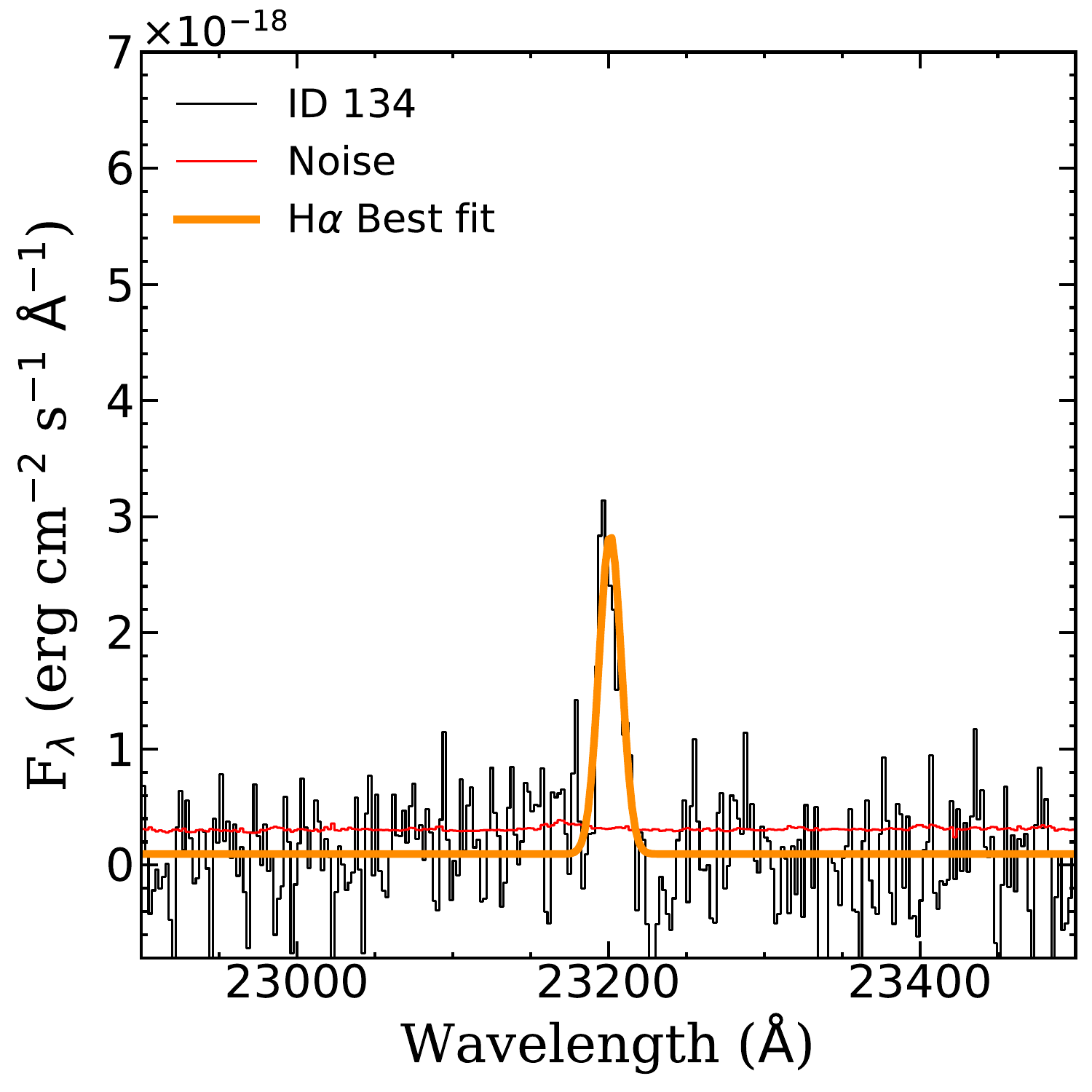}\par
      \includegraphics[width=\linewidth]{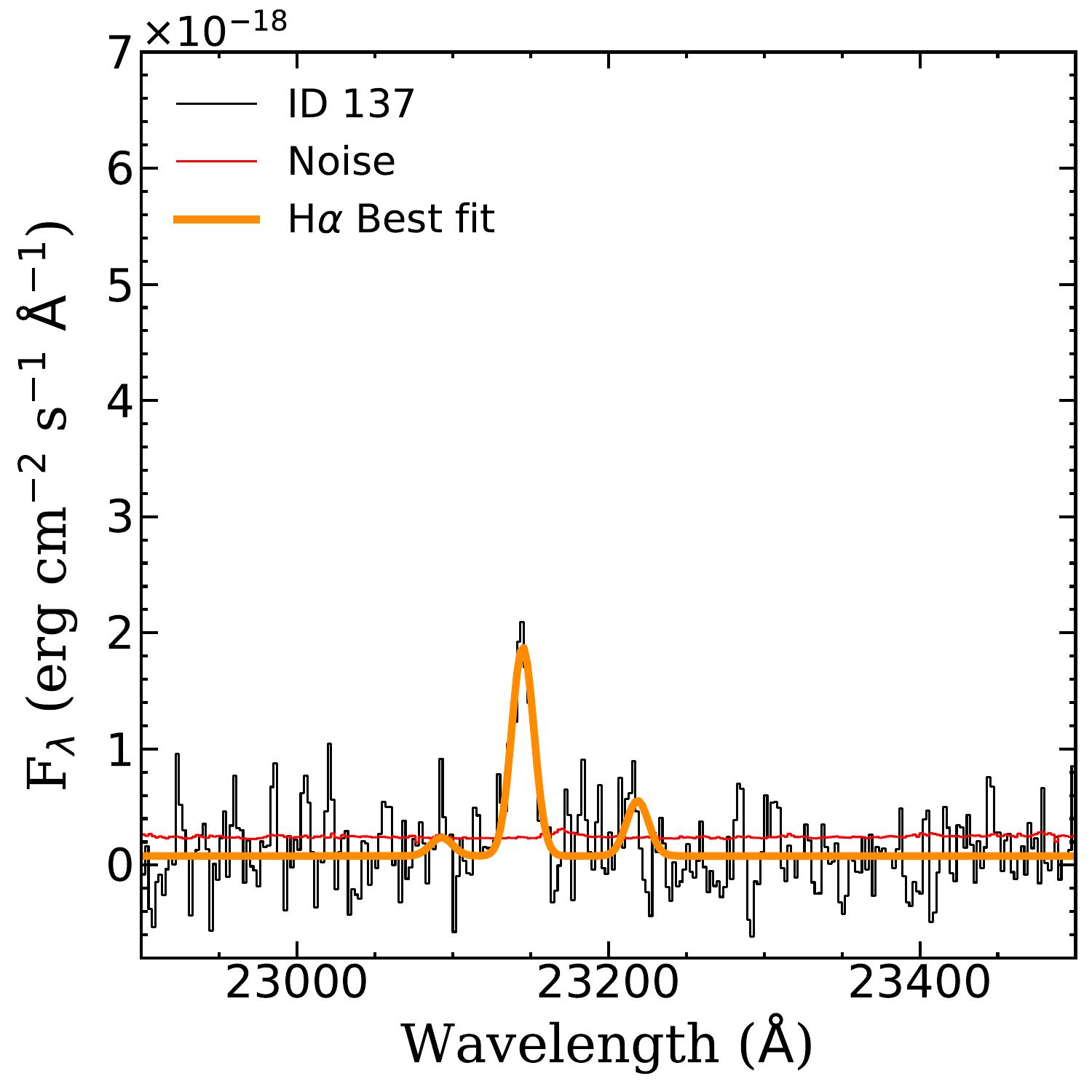}\par
      \includegraphics[width=\linewidth]{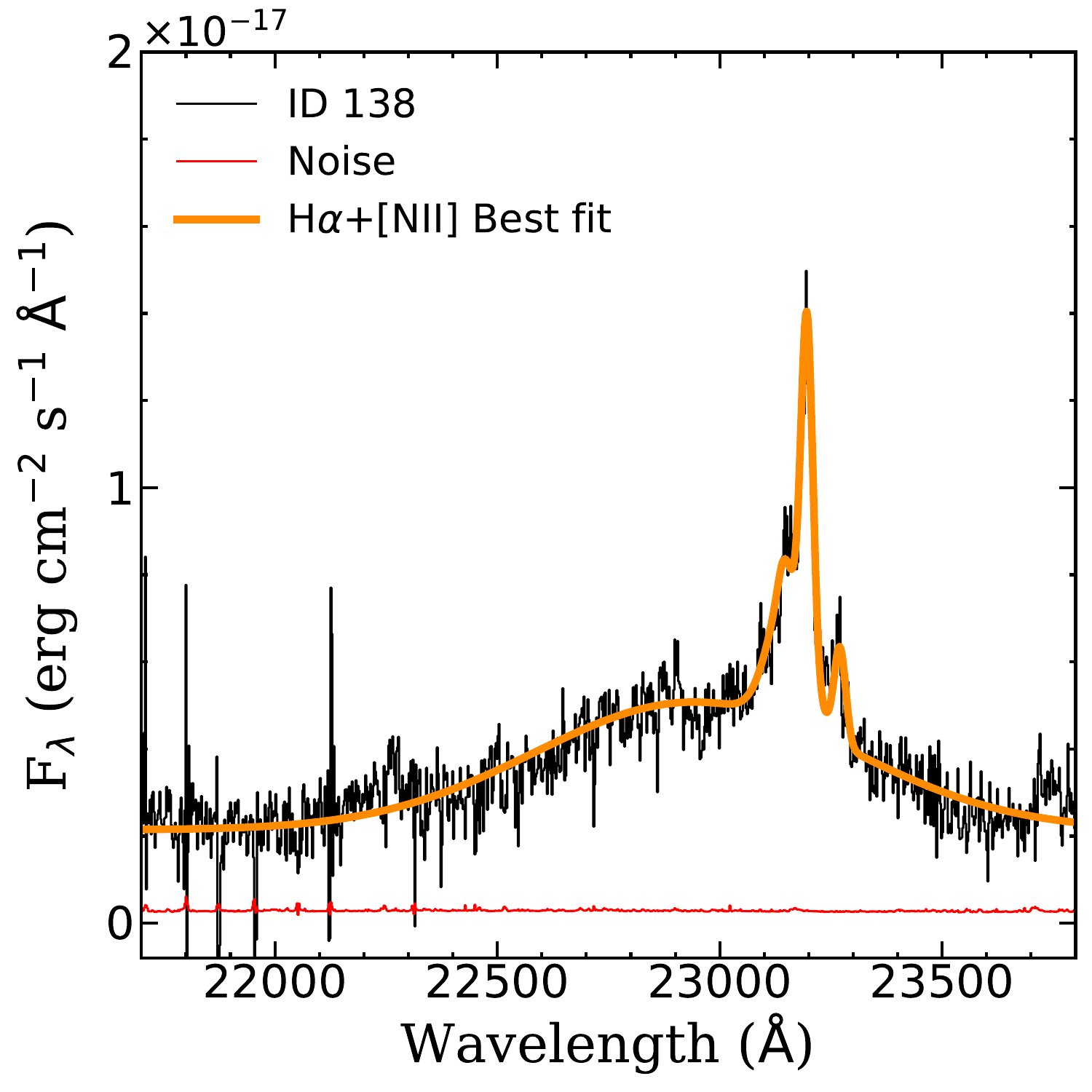}\par
      \includegraphics[width=\linewidth]{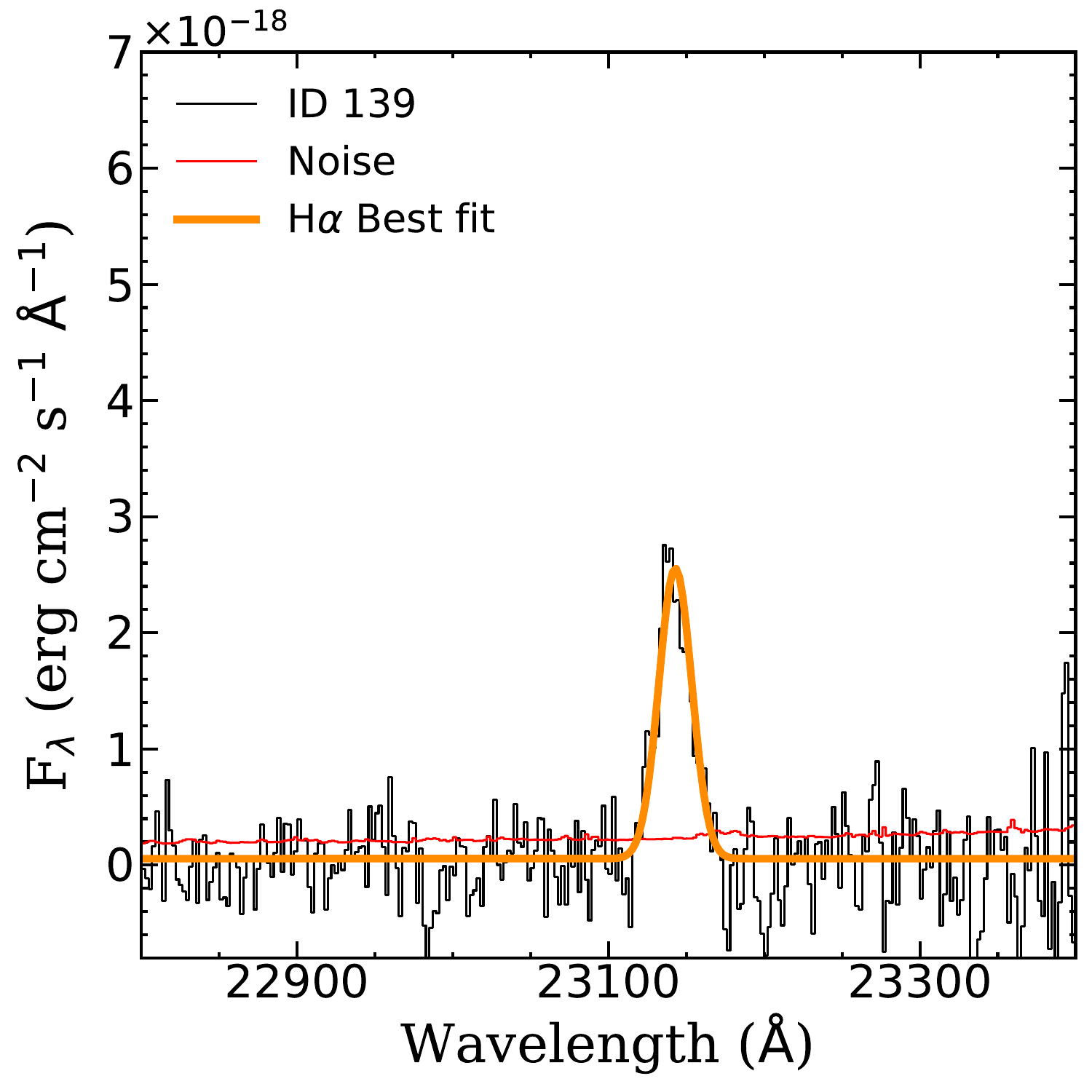}\par
      \includegraphics[width=\linewidth]{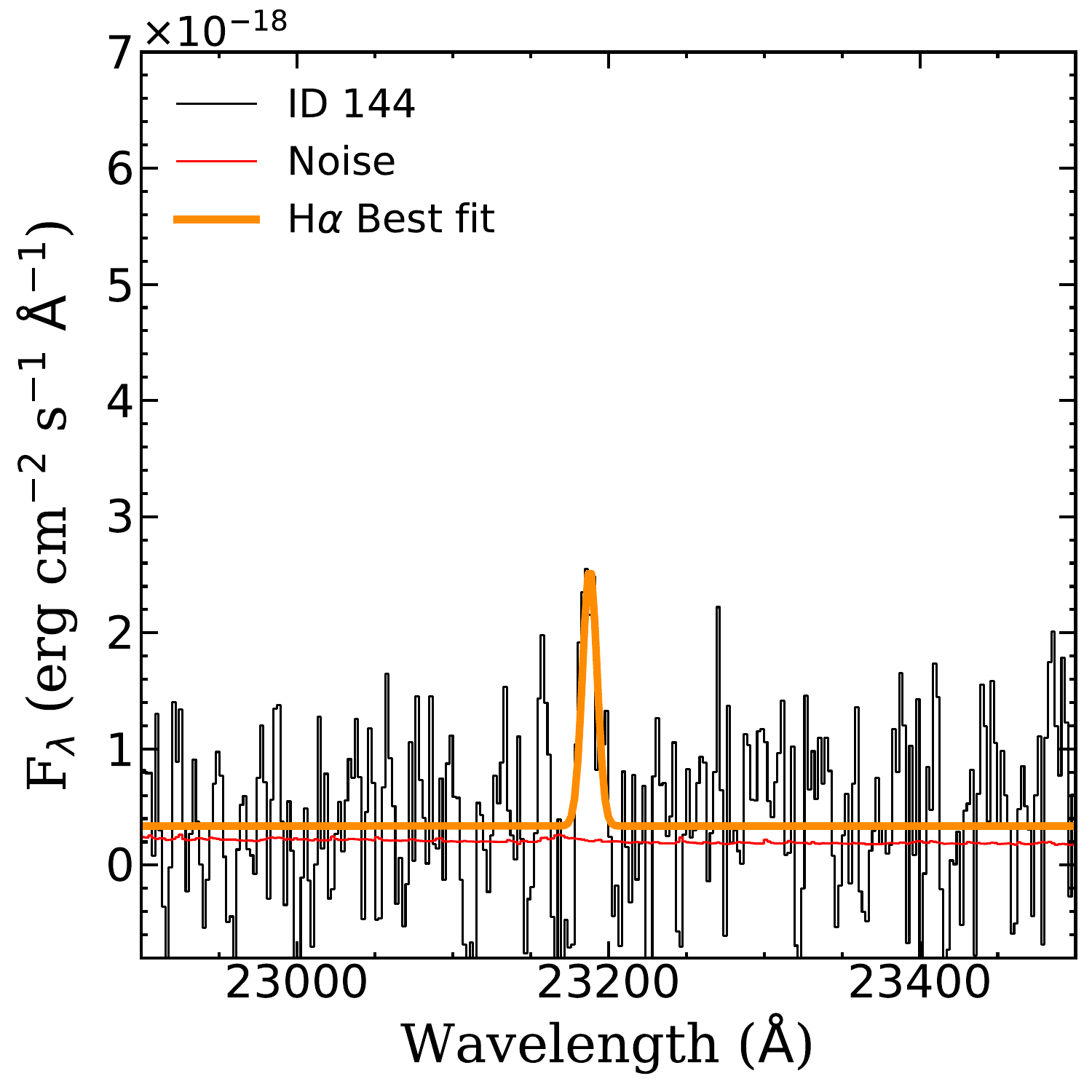}\par
      \includegraphics[width=\linewidth]{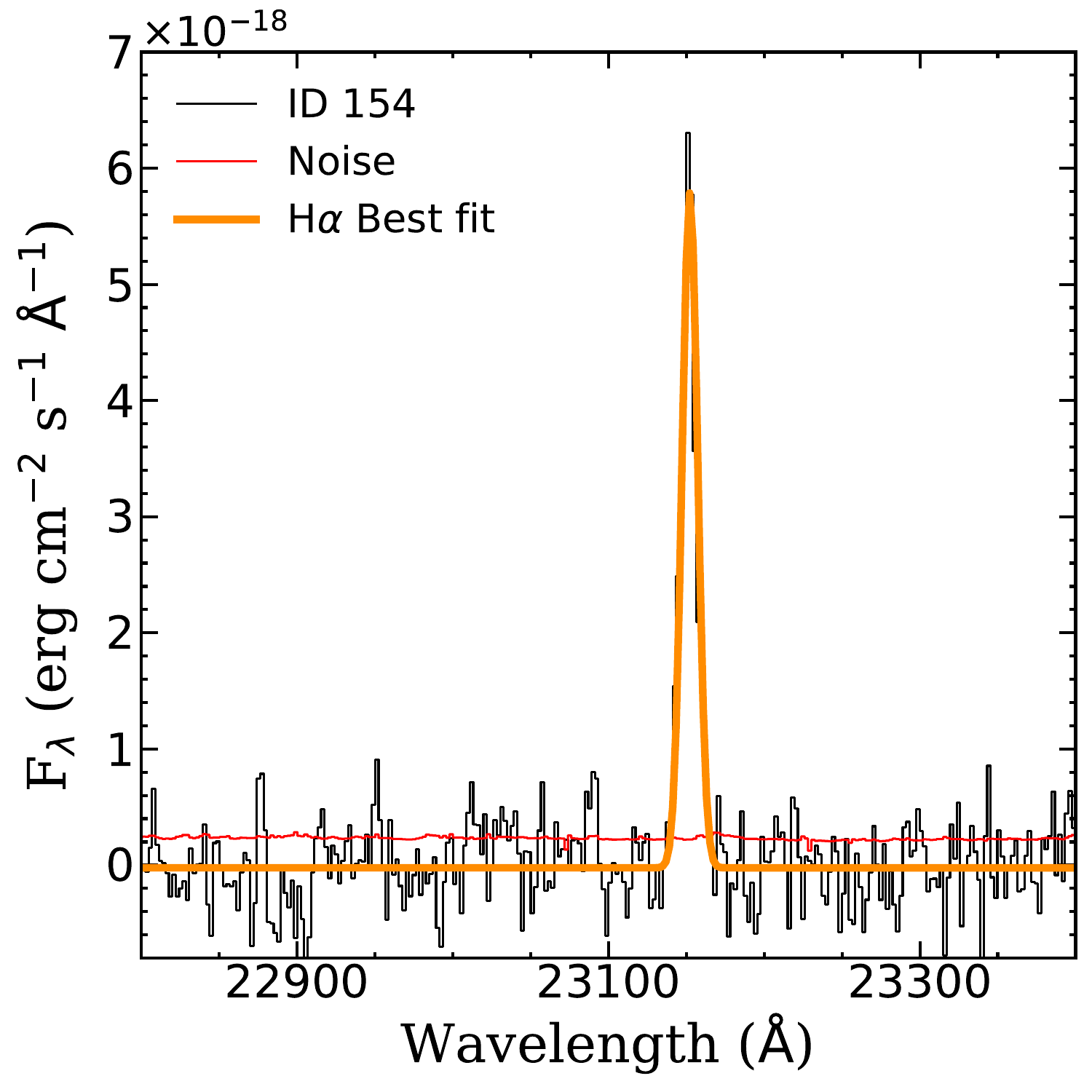}\par
      \end{multicols}
      \caption{Keck/MOSFIRE 1D reduced spectra and emission line fitting modeling. We use a black solid line for the aperture extracted signal spectrum, a red solid line for the noise, and an orange solid line for the H$\alpha$ and [N{\sc{ii}}] fit. The 23 spectroscopically confirmed protocluster objects are ordered in four columns following the ID numbering given in Table\,\ref{T:BigTable}.}
      \label{F:Spec}
\end{figure*}

\begin{figure*}
 \centering
 \begin{multicols}{4}
      \includegraphics[width=\linewidth]{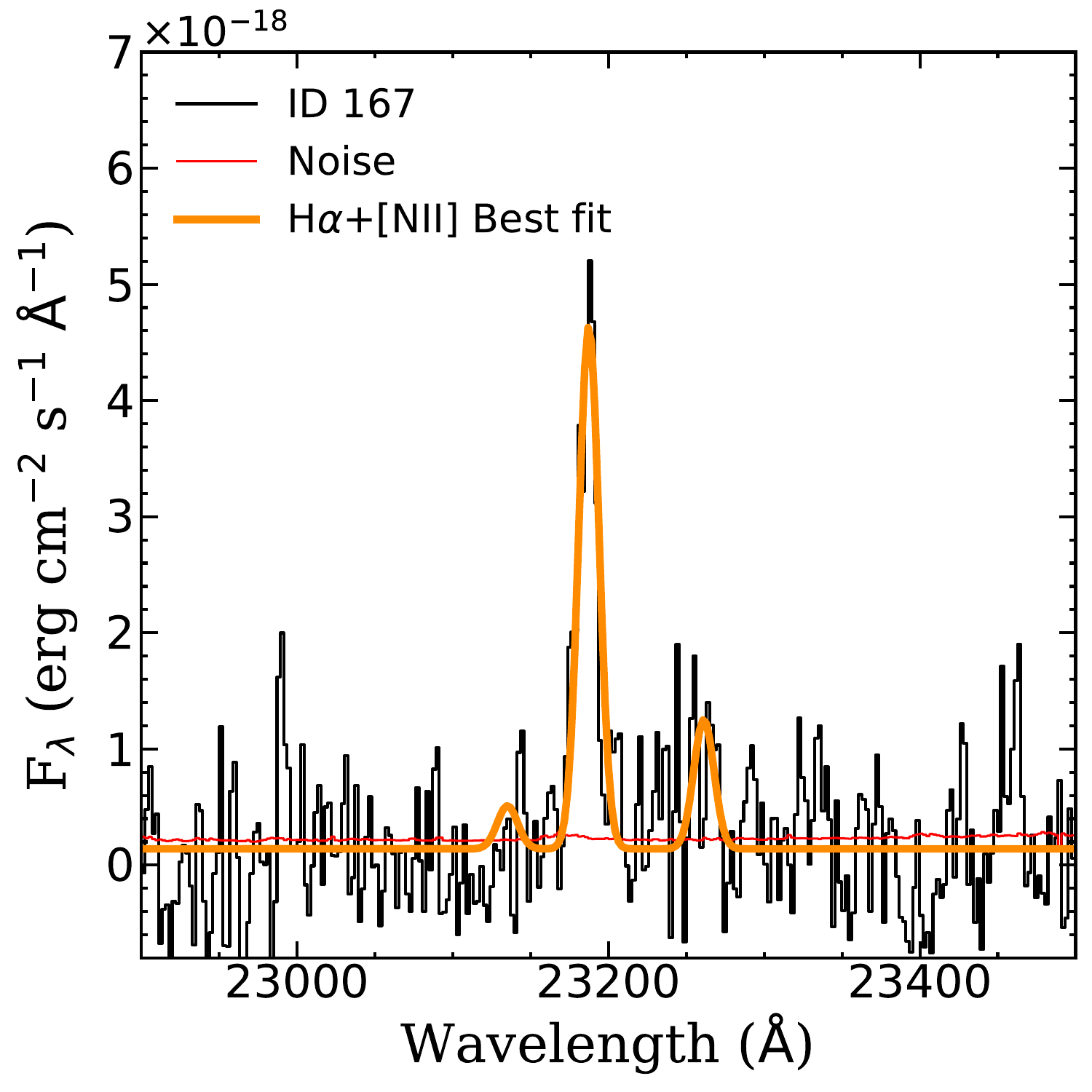}\par
      \includegraphics[width=\linewidth]{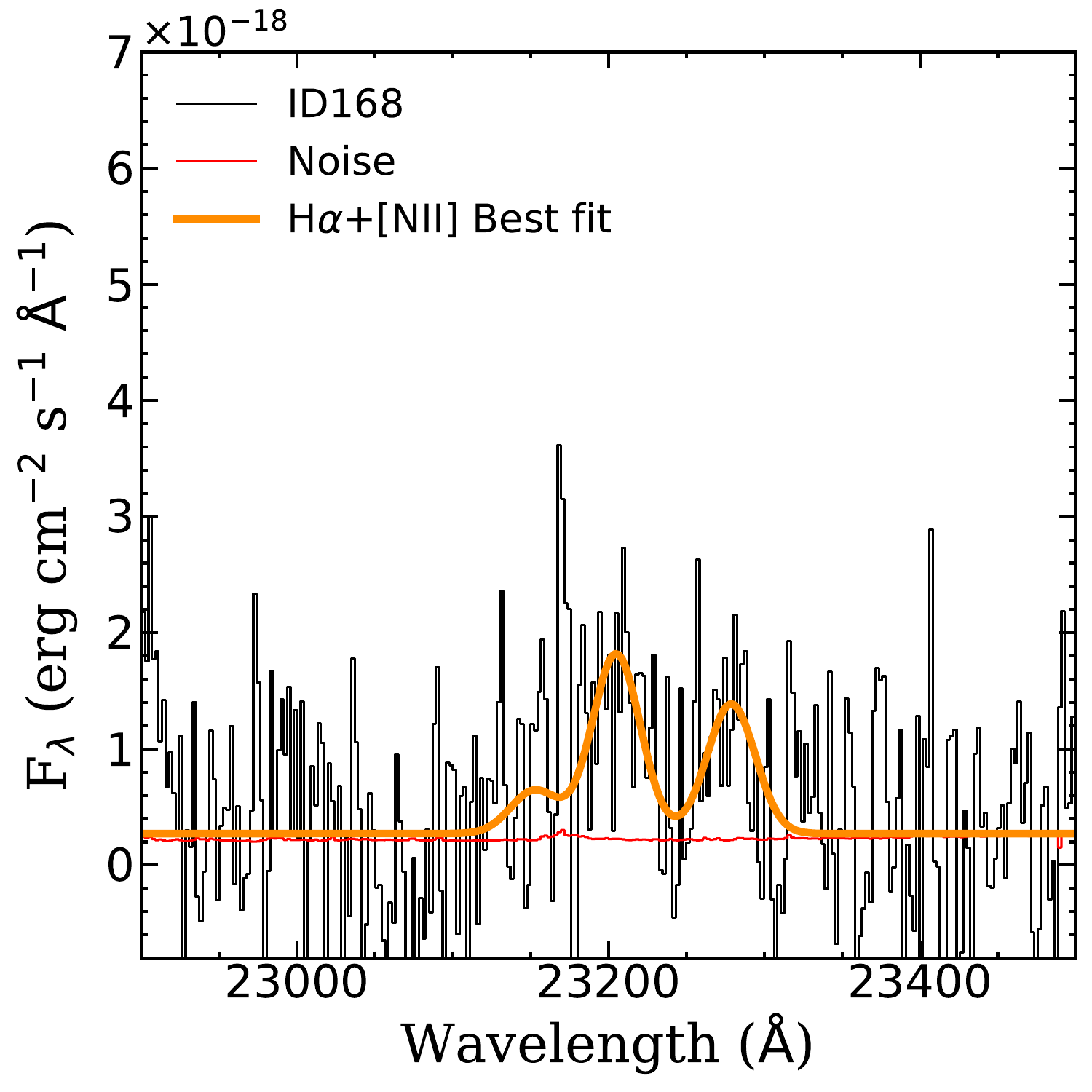}\par
      \includegraphics[width=\linewidth]{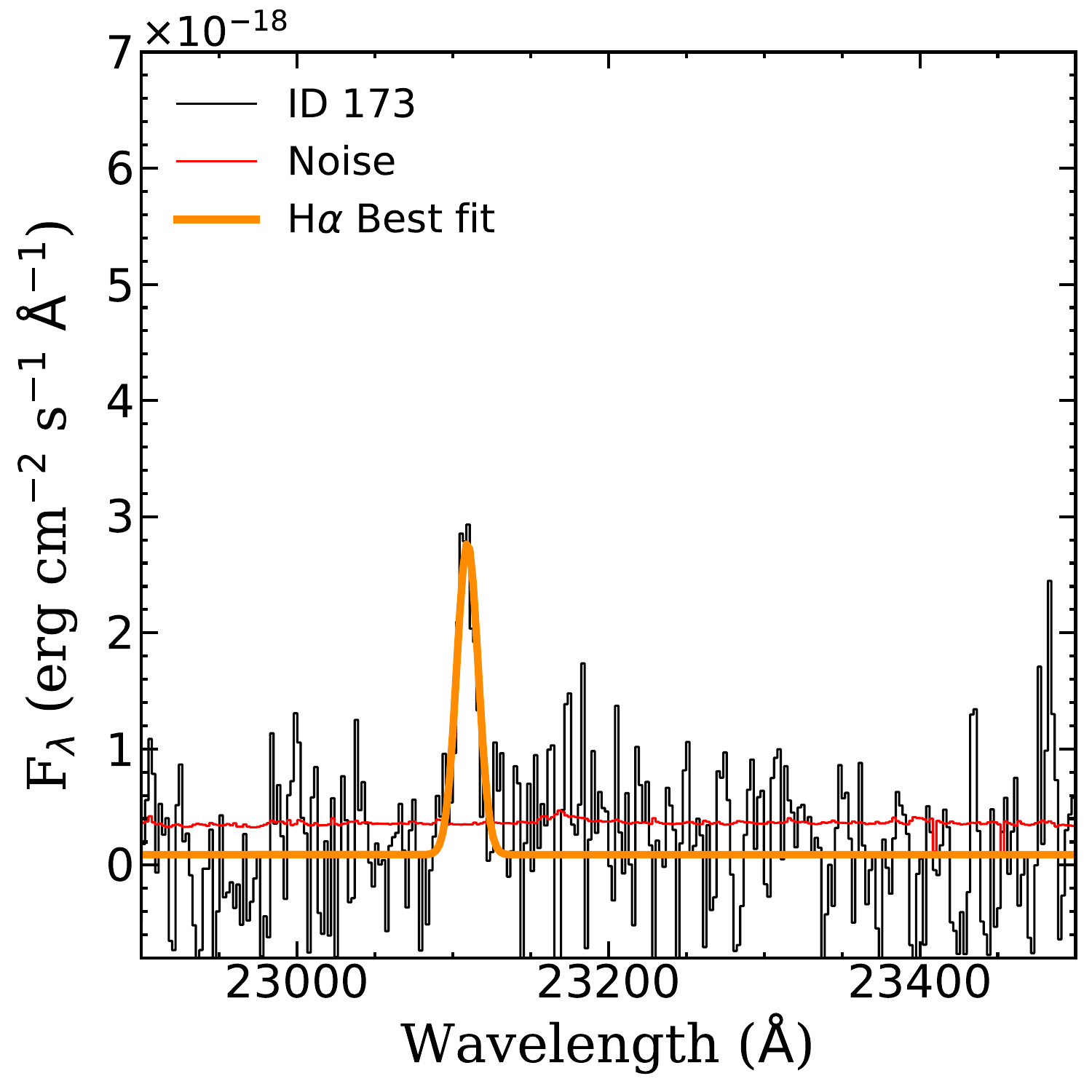}\par
      \end{multicols}
      \contcaption{}
\end{figure*}

\subsection{Star-formation activity}  
\label{SS:SFR_method}

Our Keck/MOSFIRE spectroscopic campaign allows us to trace the H$\alpha$ emission line of protocluster members at $z\approx2.53$. We follow a standard approach to compute the star-formation rate (SFR) of our sources by applying the calibration developed by \cite{Kennicutt98} modified for a Chabrier IMF (Eq.\,\ref{EQ:SFR}). This calibration has proven to be one of the most reliable both at local and high redshift (e.g. \citealt{Moustakas06}, \citealt{Wisnioski19}):
\begin{equation}
\mathrm{SFR(H\alpha)\ [M_{\odot}/{yr}]=4.65\times10^{-42}L(H\alpha)\ [ergs\ s^{-1}]}
\label{EQ:SFR}
\end{equation}
where $\mathrm{L(H\alpha)}$ is the luminosity of the H$\alpha$ emission-line. We estimate the value of this quantity by measuring the H$\alpha$ spectroscopic fluxes of our targets and assuming a Calzetti's extinction law ($\mathrm{R_v}$=4.05, \citealt{Calzetti2000}) to account for the dust attenuation. SED extinction ($\mathrm{A_v}$) values are used to correct for the diffuse dust attenuation in the galaxy's continuum following: $\mathrm{A_{cont}=0.82A_{v,SED}}$ (\citealt{Wuyts13}; \citealt{Wisnioski19}). However, the nebular contribution produced in the active star-forming regions of the galaxy remains unaccounted for at this stage. To tackle this problem, \cite{Wuyts13} add an extra extinction term ($\mathrm{A_{extra}}$) that can be parametrized as $\mathrm{A_{extra}=0.9A_{cont}-0.15A_{cont}^{2}}$ and is in good agreement with the previous extinction estimates made by \cite{Calzetti2000} in the local universe. Thus, the total extinction applied to the measured H$\alpha$ fluxes is $\mathrm{A(H\alpha)=A_{cont}+A_{extra}}$. This method is also applied by \cite{PerezMartinez23} in PKS1138 and by the KMOS3D team in \cite{Wisnioski19}, which we will use as our main field comparison sample in Sect.\,\ref{S:Results}. Two objects (IDs 38 and 138) display very broad H$\alpha$ line widths suggesting the presence of type 1 AGNs. In these two cases, the SFR values were computed by using only the H$\alpha$ narrow component in Eq. \ref{EQ:SFR}. The final SFR and reddening values of our protocluster sample can be found at the end of this work in Table \ref{T:BigTable}.

\subsection{Gas phase metallicities}  
\label{SS:metallicity}

The ISM metal enrichment of galaxies is frequently computed as a function of the Oxygen to Hydrogen ratio. However, the difficulties to obtain a simultaneous and direct measurement of auroral and nebular Oxygen line intensities, such as [O{\sc{iii}}]$\lambda$4363/$\lambda$5007 Oxygen species in the optical rest-frame (i.e., direct or $t_e$ method, \citealt{Aller84}; \citealt{Stasinska05}; \citealt{Izotov06}) led to the development of empirical calibrations that uses either a reduced set of these Oxygen lines or different species such as Nitrogen and Sulfur to indirectly trace the Oxygen enrichment based on stellar evolution physics (see \citealt{Kewley08} for a review of different methods). At $z>1$ the access to these diagnostics shifts to the NIR and thus empirical calibrations are the most common tracer of metallicities due to the difficulty to obtain multi-band spectroscopy. In this work, we apply the N2 calibration developed by \cite{Pettini04} which involves the [N{\sc{ii}}]$\lambda$6584/H$\alpha$ ratio: 

\begin{equation}
\mathrm{12+log(O/H)=8.90+0.57\times N2}
\label{EQ:N2}
\end{equation}
where N2 is equivalent to $\log$(F{[N{\sc{ii}}]}/F{(H$\alpha$})). This approach has been successfully applied for field galaxies up to $z\sim3$ (\citealt{Erb06}, \citealt{Wuyts16}, \citealt{Sanders21}) albeit it was calibrated locally. However, the accuracy of its absolute values at high redshift is still under debate (\citealt{Steidel14}) with an estimated systematic scatter of $<0.1$\,dex (\citealt{Curti17}). The aim of this work, however, is to compare the relative metal enrichment of USS1558 protocluster members with those belonging to PKS1138 (\citealt{PerezMartinez23}) and the coeval field. Thus calibration uncertainties do not play a role in our analysis as long as we consistently apply the same empirical calibration at a fixed cosmic time. Furthermore, we consider the effect of AGN contamination by splitting our sample in objects below and above $\log$([N{\sc{ii}}]/H$\alpha$)$\geq-0.35$. This threshold effectively separates the purely star-forming from the composite and AGN regions in the BPT diagram (\citealt{Baldwin81}) even at relatively low [O{\sc{iii}}]/H$\beta$ ratios (e.g. $\log$([O{\sc{iii}}]/H$\beta$)$\,\leq-0.2$, \citealt{Agostino21}). Thus, we use $\log$([N{\sc{ii}}]/H$\alpha$)$\geq-0.35$ as an upper limit beyond which the ionization source of most objects is either composite or dominated by the AGN contribution and label such objects as AGN candidates in our subsequent analyses.

\begin{figure*}
    \begin{multicols}{2}
      \includegraphics[width=\linewidth]{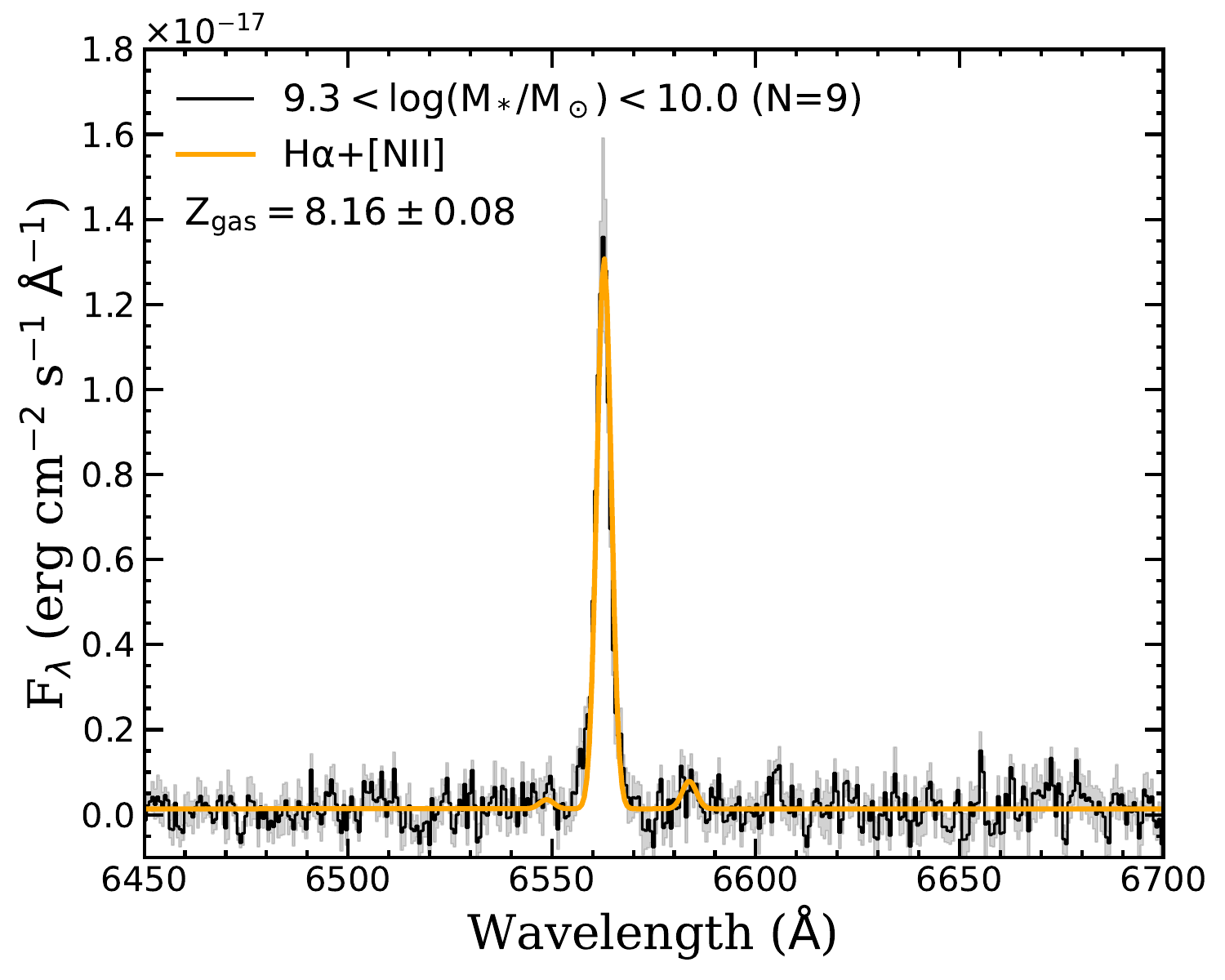}\par 
      \includegraphics[width=\linewidth]{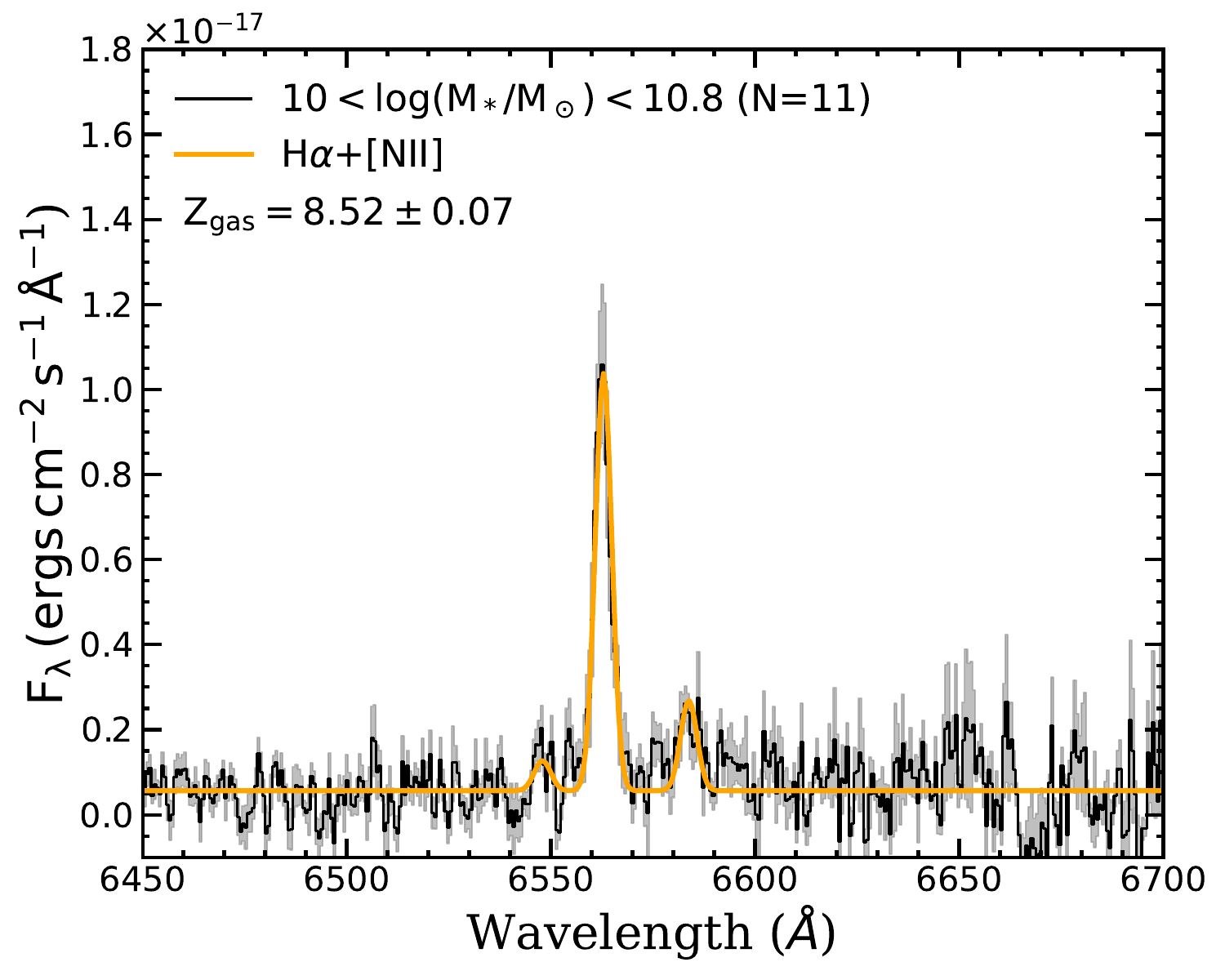}\par
      \end{multicols}
      \caption{Mean bootstrapped rest-frame spectra (black), 1$\sigma$ confidence interval (grey), and fit around the H$\alpha$ and [N{\sc{ii}}]$\mathrm{\lambda\lambda}$6548,\,6584 emission lines (orange) for the two inspected stellar-mass bins.}
      \label{F:Stacked}
      \end{figure*}

Only 10 out of the 23 inspected objects yielded [N{\sc{ii}}]$\lambda$6584 fluxes above the $2\sigma$ level and thus, we resort to the stacking analysis (excluding AGN candidates) to trace the mean metal enrichment of our sample in two mass bins ($\mathrm{9.3<\log M_*/M_\odot<10.0\,and\,10.0<\log M_*/M_\odot<10.8}$). Only three galaxies have stellar masses above this limit and are thus excluded from the stacking analysis: IDs 38 and 138 (radio galaxy) which display broad emission line components and ID 59 which lies just above the limit with $\mathrm{\log M_*/M_\odot=10.88}$. The spectral stacking is performed following the approach outlined in \cite{Shimakawa15}: 

\begin{equation}
\mathrm{F_{stack}=\sum_{i}^{n}\frac{F_{i}(\lambda)}{\sigma_{i}(\lambda)^2}\Bigg/\sum_{i}^{n} \frac{1}{\sigma_{i}(\lambda)^2}}
\label{EQ:stack}
\end{equation}
where $F_{i}(\lambda)$ is the flux density of an individual spectrum and $\sigma_{i}(\lambda)$ is the noise as a function of wavelength. Within a given bin, the galaxies' spectrum is summed up using a median sigma clipping algorithm to reduce the contamination by sky residuals. In addition, we use bootstrapping, thus resampling the number of spectra within our stellar mass bins over a thousand realizations, to limit the impact of emission line flux variability between individual sources during the stacking process. After this, we take the mean bootstrapped spectrum as the final one for each bin, and one standard deviation as the new spectral error. These two quantities are used to fit and measure the N2 ratio following the same procedure applied to every individual spectrum (see Sect.\,\ref{SS:EL}). Finally, based on the values reported in Table\,\ref{T:BigTable} we estimate the median uncertainties for our gas-phase abundances to be of the order of 0.10\,dex for the individual measurements across our stellar mas range, while we report $<0.10$ dex for the stacking analysis measurements after bootstrapping. Figure\,\ref{F:Stacked} displays the stacked spectrum for each bin and their best fit.

\subsection{Molecular gas masses}
\label{SS:MOL}

USS1558 has been subject to deep multiwavelength observations including the submillimeter/radio regime with ALMA (e.g., CO(3-2) from \citealt{Tadaki19} and dust continuum at $\sim 1.1$\,mm from \citealt{Aoyama22}). In this work, we will make use CO(3-2) measurements to estimate the molecular gas mass of a small subsample of our targets. In order to minimize the systematics between different molecular gas tracers we will focus solely on the available CO(3-2) measurements published by \cite{Tadaki19}. In total, we share 6 overlapping targets with 4 of them being detected in CO at more than $5\sigma$ level and two upper limits. In order to convert the CO(3-2) luminosity, $L'_{\mathrm{CO(3-2)}}$, to a molecular gas mass estimate we assume a conversion factor, $\alpha_{\mathrm{CO(3-2)}}$, between the CO(3-2) and the CO(1-0) luminosity, which better trace the amount of molecular gas from optically thick virialized clouds (\citealt{Dickman86}, \citealt{Solomon87}):
\begin{equation}
M_{\mathrm{mol,CO}}=\alpha_{\mathrm{CO(3-2)}}\times L'_{\mathrm{CO(3-2)}}
\label{EQ:mol}
\end{equation}

\begin{equation}
\alpha_{\mathrm{CO(3-2)}}=\alpha_{\mathrm{CO(1-0)}}\times \frac{1}{r_\mathrm{31}}\times \mathrm{f_{corr}}
\label{EQ:rJ}
\end{equation}
However, the estimates of $r_\mathrm{31}$ for high-z star-forming galaxies reported in the literature show large variations with some works suggesting values close to those reported in the local universe ($r_\mathrm{31}\approx0.5$, \citealt{Genzel12}; \citealt{Aravena14}; \citealt{Daddi15}), while others suggest higher values as a consequence of the expected high star formation activity at this cosmic epoch ($r_\mathrm{31}\approx0.8-0.9$, \citealt{Sharon16}; \citealt{Riechers20}). Interestingly, \cite{Aoyama22} found that the molecular gas masses computed for USS1558 objects both from CO(3-2) and from dust continuum observations achieve a higher degree of agreement when the \cite{Riechers20} prescription ($r_\mathrm{31}\approx0.9\pm0.14$) is assumed. However, we will assume $r_\mathrm{31}\approx0.56\pm0.14$ (\citealt{Genzel12}) for our own molecular gas computation in order to be consistent with the calibrations of our field comparison samples (e.g., \citealt{Sanders23}). In addition, $\alpha_{\mathrm{CO(1-0)}}$ is found to be approximately constant for nearby galaxies with solar gas-phase metallicities ($\alpha_{\mathrm{CO(1-0)}}=4.36\pm0.90$, \citealt{Bolatto13}). At lower metallicities, the conversion factor inversely correlates with metallicity. Therefore, we compute our molecular gas masses by assuming the metallicity dependent conversion factor, $\mathrm{f_{corr}}$, outlined in \cite{Genzel15} and later revisited by \cite{Tacconi18}:
\begin{equation}
\mathrm{f_{corr}=\sqrt{0.67\times\exp{(0.36\times 10^{-(Z_{gas}-8.67)}\times 10^{-1.27\times(Z_{gas}-8.67)})}}}
\label{EQ:alpha}
\end{equation}
where $\mathrm{Z_{gas}=12+log(O/H)}$ assuming the \cite{Pettini04} metallicity calibration. This conversion factor is based on the geometric sum of the metallicity corrections proposed by \cite{Genzel12} and \cite{Bolatto13}. However, both methods strongly diverge in the low metallicity regime. Thus, we constrain the range of application of this prescription to galaxies with $\mathrm{12+log(O/H)}>8.44$, which corresponds to a maximum offset between methods of 0.2\,dex. Given the distribution of our galaxies in the mass-metallicity relation (see Sect. \ref{SS:MZR}), this approximation allows us to constrain the molecular gas fraction ($\mathrm{f_{gas}=M_{Mol}/(M_*+M_{Mol})}$) of 6 targets with CO(3-2) emission although only 5 of them have individual metallicity estimates in this work. 

\subsection{Environment quantification}
\label{SS:Environment}

The accurate quantification of the environment around a given object is a recurrent problem in galaxy evolution studies. Two main paths have been developed to answer this question although any of them can only offer an incomplete picture due to projection effects and the difficulties to trace all galaxy populations within a given region. First, global environmental indicators such as clustercentric distance or phase-space variables require the pre-identification and physical characterization (in terms of $M_{200}$, $R_{200}$, and $\sigma_{cl}$, \citealt{Carlberg97}) of strong overdensities within the field of view studied. This method is usually applied to massive virialized clusters but it is not appropriate when the structure under scrutiny is made up of several clumps or groups lacking virialization, as is the case of USS1558.

In addition, the environment can be measured locally, as a number density of objects within a given projected area (e.g. \citealt{Dressler80}). This is a very common environmental tracer because it does not require the pre-identification of any kind of overdensity (group, cluster, or protocluster) in a given area to be applied, and its accuracy is only limited by the number and type of objects identified within a sufficiently narrow redshift range. Its main weakness resides in the simultaneous identification of different galaxy populations (e.g., star-forming vs passive) at a given cosmic epoch so that we can obtain a complete picture of the density field. Protoclusters such as USS1558 at $z=2.53$ are overwhelmingly populated by star-forming galaxies (e.g., \citealt{Overzier16}; \citealt{Chiang17}; \citealt{Remus23}; \citealt{Alberts22}) which are easily identified by their emission lines using deep NB surveys such as the MAHALO-Subaru project (\citealt{Kodama13}). This approach, however, may miss a population of heavily obscured star-forming galaxies whose main emission is concentrated in the far infrared and (sub-)millimeter. In the case of USS1558, this has been ruled out by the ALMA CO(3-2) and dust continuum studies of \cite{Tadaki19} and \cite{Aoyama22} respectively, who found that most ALMA sources are also identified as narrow-band HAEs. Similarly, the NB approach misses potential protocluster members without detectable emission lines (i.e., quenched objects). Even though this population is expected to be a minority within protoclusters, previous observational works in this field have identified an overdensity of distant red objects (DRGs) close to the radio galaxy and to the main clump in the bottom right corner of Fig.\,\ref{F:Map} using color diagnostics (\citealt{Kodama07}; \citealt{Galametz12}; \citealt{Hayashi12}). This means that the distribution of DRGs is largely consistent with the overdensities depicted by the HAEs alone. It is for these reasons that we decided to use the catalog of Subaru/MOIRCS narrow-band confirmed HAEs in USS1558 published by \cite{Shimakawa18a} to compute the local density of every object. This catalog contains 107 sources and represents the parent sample of our spectroscopic campaign. The spectroscopic confirmation of narrow-band selected HAES in this and previous works related to MAHALO-Subaru yields a $>90\%$ success rate (e.g., \citealt{Shimakawa14,Shimakawa15}; \citealt{PerezMartinez23}), which allow us to assume that all remaining HAEs without spectroscopic confirmation have a high chance of belonging to USS1558. The position of every HAE is projected into a plane where distances between objects are measured by their sky angular separation and transformed to the physical scale using a fixed cosmology with the scale factor ($a$) corresponding with the redshift of the cluster (e.g. $a=8.050$ kpc/\arcsec at $z=2.53$). This way we define a number surface density in the following way:
\begin{equation}
   \Sigma_N=\frac{N}{\pi r^2_{N-1}}
\label{LocalDensity}
\end{equation}
For consistency with previous works in this protocluster (\citealt{Shimakawa18a}), we will use $\Sigma_5$, i.e. the local density defined by the minimum radius enclosing five neighboring galaxies. The determination of galaxy positions is carried out during the source detection (Sect.\,\ref{SS:SED}) by fitting a 2D surface brightness profile to each object. Typical uncertainties are of $\pm1$ pixel, which results on $\mathrm{\sigma_{\Sigma_{5}}<0.01\,Mpc^{-2}}$. The values of $\Sigma_5$ for each member of our sample can be found in Table\,\ref{T:BigTable}. This choice enables us to trace the local density gradient inside each of the three densest clumps as well as identify additional small compact galaxy groups representing local density peaks outside the main structures.

\begin{figure*}
 \centering
 \begin{multicols}{2}
      \includegraphics[width=\linewidth]{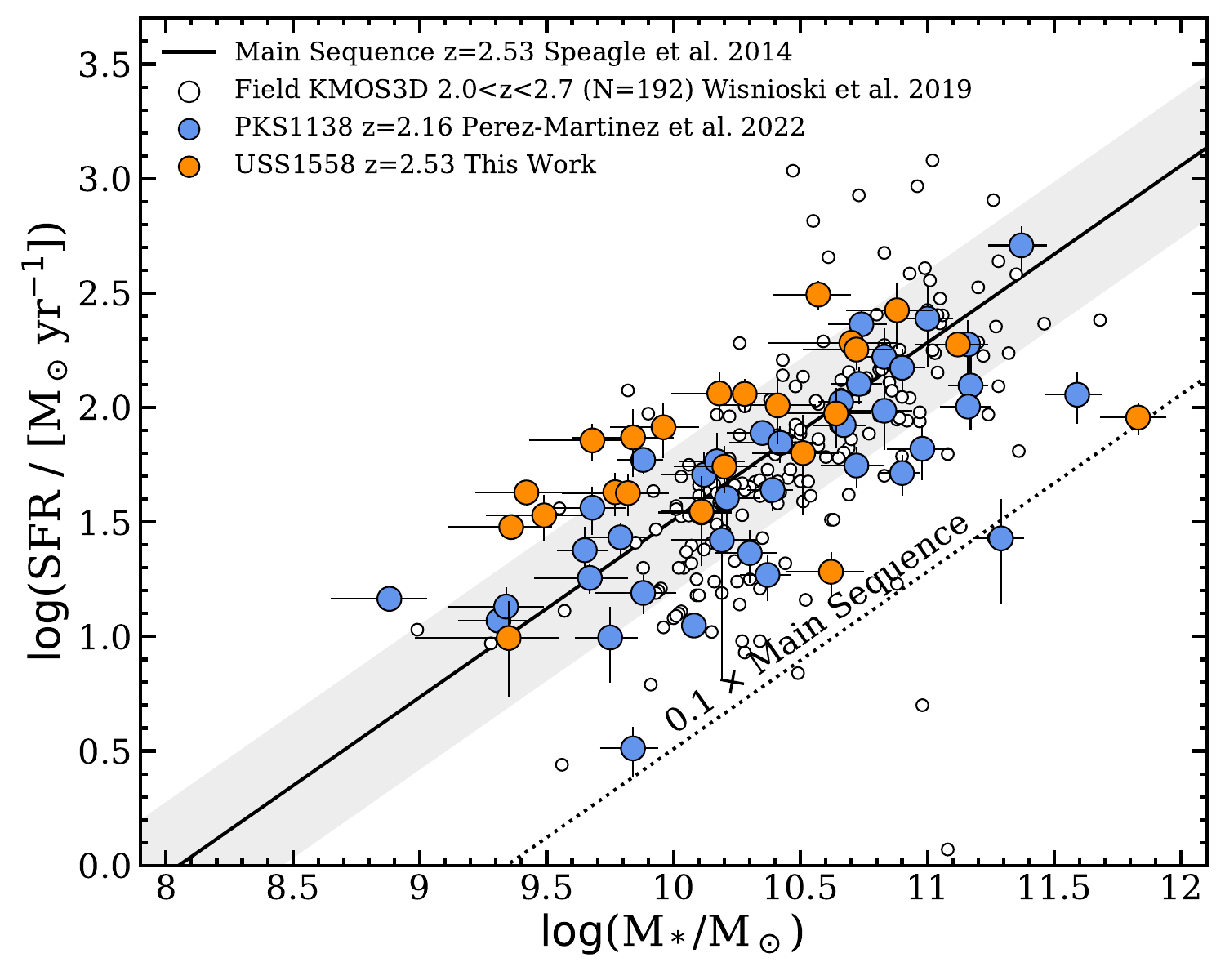}\par
      \includegraphics[width=\linewidth]{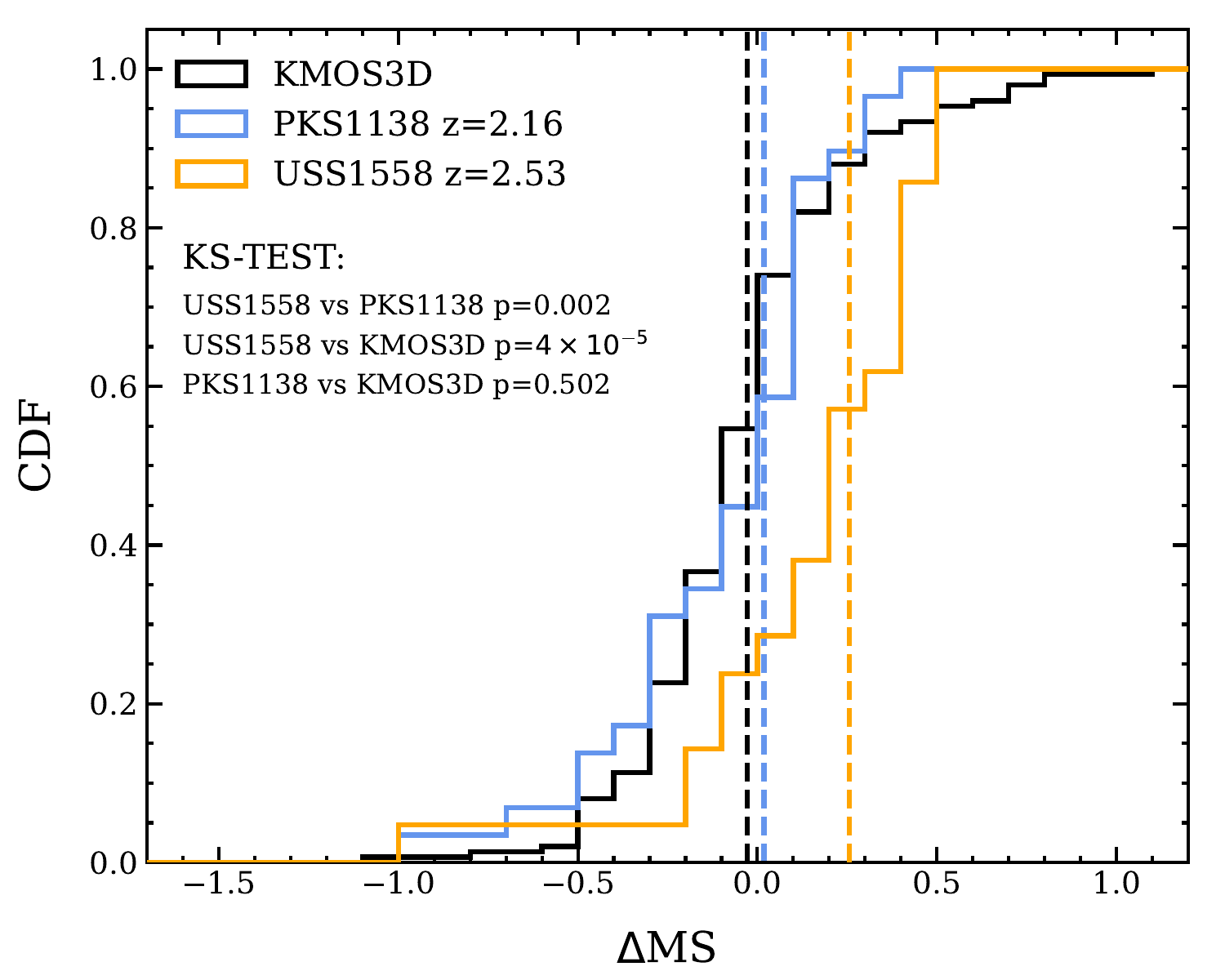}\par
      \end{multicols}
      \caption{Left: Star-forming main sequence diagram. Orange circles show the distribution of our Keck/MOSFIRE sample in USS1558. Blue circles display protocluster HAEs within PKS1138 (\protect\citealt{PerezMartinez23}). Empty circles depict the field comparison sample from the KMOS3D survey (\protect\citealt{Wisnioski19}). This dataset consists of 192 field galaxies at a similar mean redshift to the USS1558 sample. The black solid line represents the star-forming main sequence parametrized for $z=2.2$ (\protect\citealt{Speagle14}), while the grey shaded region marks the 3$\mathrm{\sigma}$ scatter around it. Right: Cumulative distribution function of the offsets with respect to the main sequence for the three inspected samples after matching their minimum SFR to $\mathrm{25 M_\odot/yr}$}. Vertical dashed lines display the median offset for each dataset. The colors remain the same as in the left-hand panel.
         \label{F:SFR}
\end{figure*}
 \begin{figure}
      \includegraphics[width=\linewidth]{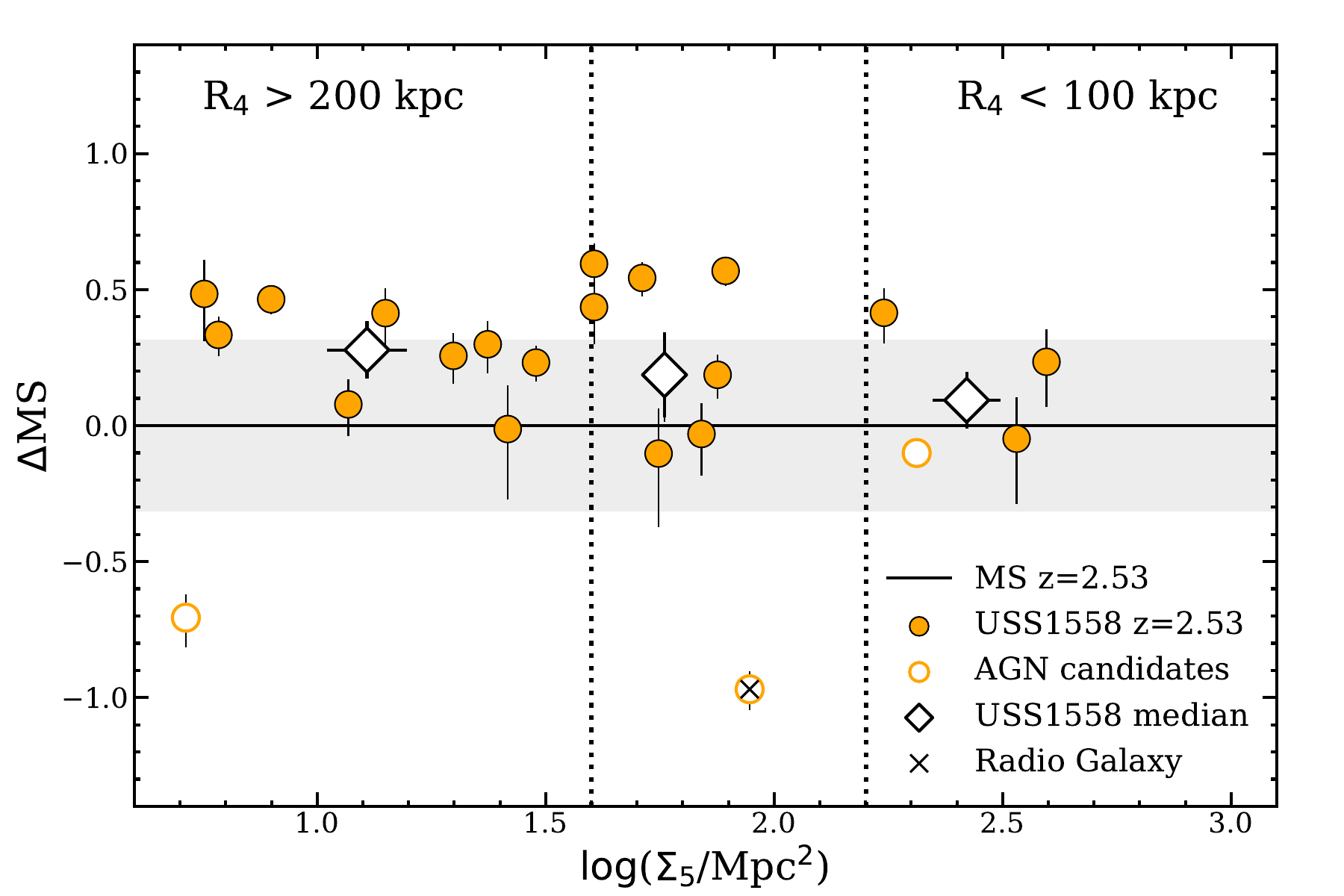}\par 
      \caption{Residuals from the main sequence of galaxies as a function of the local number density of objects, which is defined as the minimum area to enclose five neighboring galaxies. The horizontal solid line and grey area depict the position and 3$\mathrm{\sigma}$ scatter of the main sequence (\protect\citealt{Speagle14}). Vertical dashed lines separate the three density regimes outlined in Sect. \ref{SS:Environment}. The radio galaxy (cross) and AGN candidates (empty circles) are flagged using different symbols.}
         \label{F:Density_SFR}
      \end{figure}

\section{Results}
\label{S:Results}

\subsection{Star formation in the assembling USS1558 protocluster}
\label{SS:SFR}
In this section, we trace the star formation activity of our sample of HAEs belonging to the USS1558 protocluster and compare our results with previous studies in massive protoclusters (\citealt{PerezMartinez23}) and the coeval field (\citealt{Speagle14}; \citealt{Wisnioski19}). Figure\,\ref{F:SFR} (left) shows the distribution of our targets within the SFR-$\mathrm{M_*}$ and along the so-called "Main Sequence" of star formation at $z=2.53$. USS1558 galaxies are preferentially located above the main sequence, with at least one-third of the sample displaying an SFR enhancement of $\mathrm{\sim0.3\,dex}$. This effect becomes ever clearer when inspecting the low-mass end of the sample. Eight out of the nine objects at $\mathrm{\log M_*/M_\odot<10.0}$ lie above the $3\sigma$ scatter of the main sequence suggesting that low-mass galaxies are going through a rapid phase of mass build-up. This effect was previously reported by \cite{Hayashi16} who analyzed a larger sample of HAEs in this very same structure relying only on narrow-band photometry for computing SFRs. On the other hand, a recent study by \cite{PerezMartinez23} (blue in Fig.\,\ref{F:SFR}) found that HAEs belonging to the massive PKS1138 protocluster at $z=2.16$ agree well with the main sequence prescription parametrized for the protocluster redshift. 

However, Fig.\,\ref{F:SFR} also shows a clear mismatch between the minimum typical SFR between USS1558 and PKS1138. In fact, $>90\%$ of the USS1558 sample lies at $\mathrm{\log SFR >1.4}$ (i.e., $\mathrm{SFR>25 M_\odot/yr}$) while PKS1138 objects trace the star formation activity down to $\mathrm{SFR>10 M_\odot/yr}$. This may be the result of different flux cuts applied over their respective narrow-band parent samples coupled with the effects of spectroscopic sampling during our observations. For these reasons, we decide to compute the cumulative distribution function of the offsets with respect to the main sequence (i.e, $\Delta\mathrm{MS=SFR-SFR_{MS}}$) for the three samples previously discussed after imposing an $\mathrm{SFR>25 M_\odot/yr}$ cut on them (right panel of Fig.\,\ref{F:SFR})
Our results show that the USS1558 (orange) distribution is shifted towards higher offsets in comparison with both the PKS1138 (blue) and the KMOS3D (black) samples, which are consistent with their respective main sequences. In the case of PKS1138, we have computed its $\Delta\mathrm{MS}$ by parametrizing the main sequence of \cite{Speagle14} to $z=2.16$, thus avoiding possible biases due to the expected evolution of star formation with cosmic time. The median values for each sample are represented by vertical dashed lines confirming the SFR enhancement of the USS1558 sample ($\mathrm{\Delta\mathrm{MS}\approx0.26\pm0.07\,dex}$). In addition, we have performed a Kolmogorov-Smirnov test between our samples to investigate if these differences are statistically significant. We choose $\mathrm{p<0.05}$ as the threshold to test if a given distribution is significantly different with respect to the reference sample, which in this case is USS1558. The results of the KS-test suggest that the differences found between USS558 and PKS1138 are statistically significant ($\mathrm{p=2\times10^{-3}}$) as well as those with respect to the KMOS3D sample ($\mathrm{p=4\times10^{-5}}$). On the other hand, the KMOS3D sample and PKS1138 are statistically indistinguishable ($\mathrm{p=0.502}$). These results suggest that HAEs from different protoclusters at approximately the same cosmic epoch may be experiencing different phases of their stellar-mass build-up which could potentially be related to the evolutionary stage of the large-scale structure they belong to. 

Finally, we investigate the influence of the protocluster environment over the star formation activity of USS1558 HAEs. We decide to trace the local density across the protocluster structure by computing the minimum area to enclose five neighboring galaxies (i.e., $\Sigma_5$) from the position of each of our targets. In Figure\,\ref{F:Density_SFR} we show the distribution of main sequence offsets (i.e., $\Delta\mathrm{MS}$) as a function of $\Sigma_5$ and compute the median offset within each density bin. The SFR enhancement detected for the USS1558 sample is approximately constant ($\mathrm{\approx0.25\,dex}$) across different density regimes. However, given the limited size of our sample, we can not discard that particularly overdense regions within the protocluster play a role in boosting star formation, as suggested by \cite{Shimakawa18a}.

\subsection{Gas-phase metallicities}
\label{SS:MZR}

\begin{figure*}    
\centering
\begin{multicols}{2}
\includegraphics[width=\linewidth]{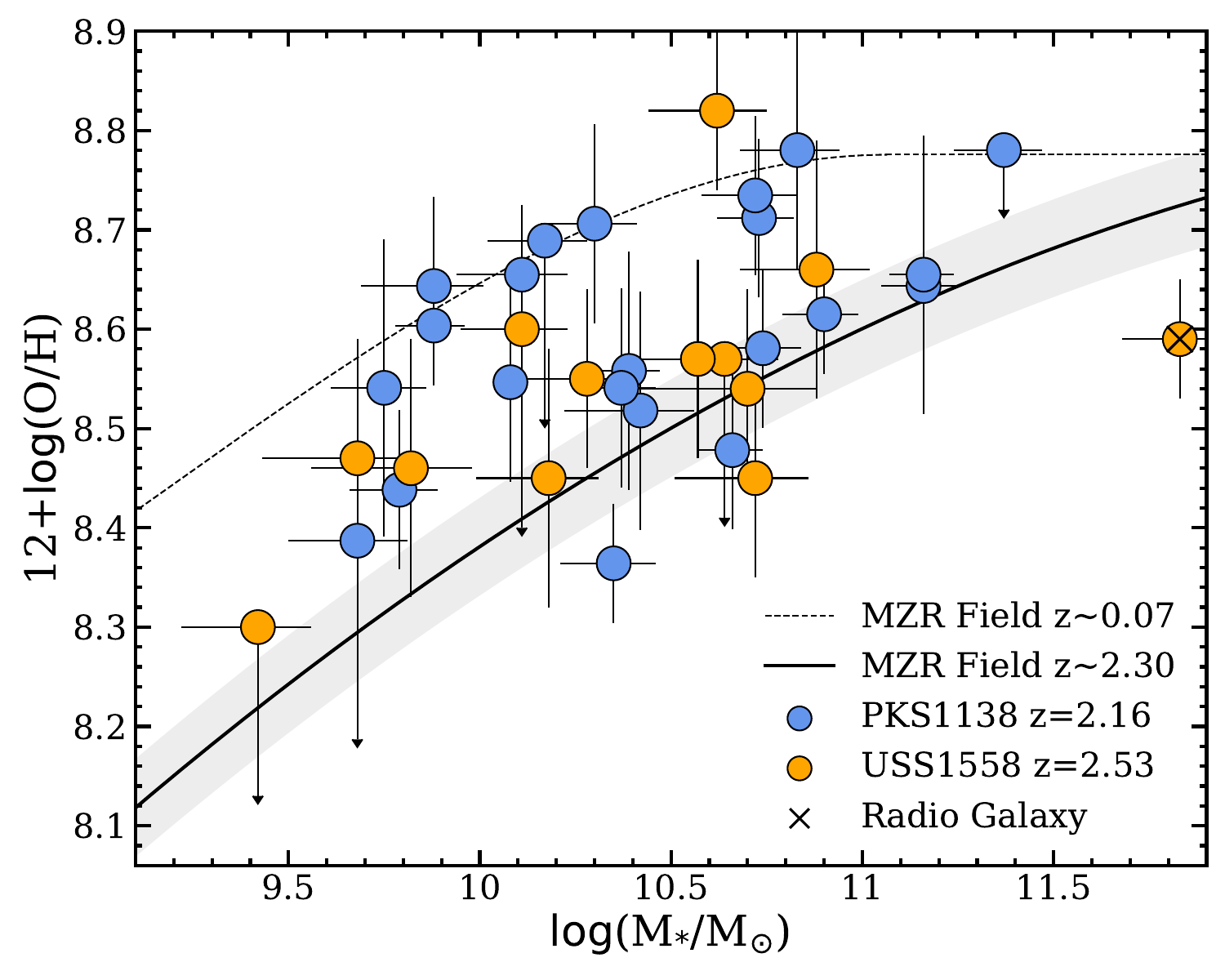}\par
\includegraphics[width=\linewidth]{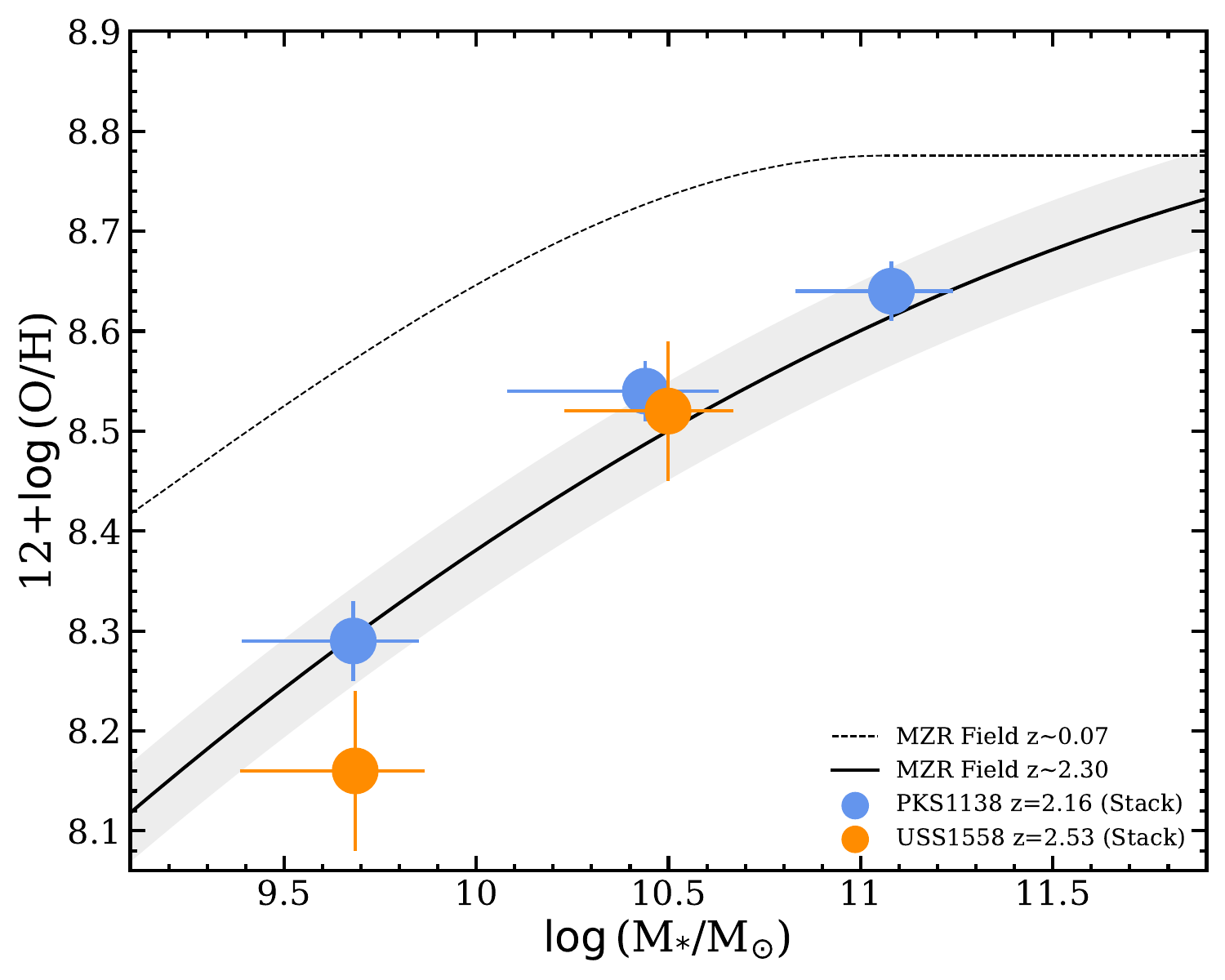}\par
\end{multicols}
\caption{Mass-metallicity relation (MZR) diagrams. Left: Individual gas-phase metallicity measurements for USS1558 (this work) and PKS1138 (\protect\citealt{PerezMartinez23}) displayed by orange and blue circles respectively. Right: Stacked metallicity results for USS1558 and PKS1138. Horizontal error bars display the stellar mass standard deviation within each bin. In both panels, the solid line depicts the field MZR using a second-order polynomial fit over the combined results of the \protect\cite{Erb06} sample, the KMOS3D sample of \protect\cite{Wuyts16} and the MOSDEF sample of \protect\cite{Sanders21} at $z\sim2.3$. The grey area shows the mean metallicity uncertainty for the samples used to define the field MZR at $z\sim2.3$. The dashed line marks the position of the local MZR (\protect\citealt{Kewley08}) assuming the \protect\cite{Pettini04} metallicity calibration but adjusted to the \protect\cite{Chabrier03} IMF.}
\label{F:MOH}
\end{figure*} 

\begin{figure}
      \includegraphics[width=\linewidth]{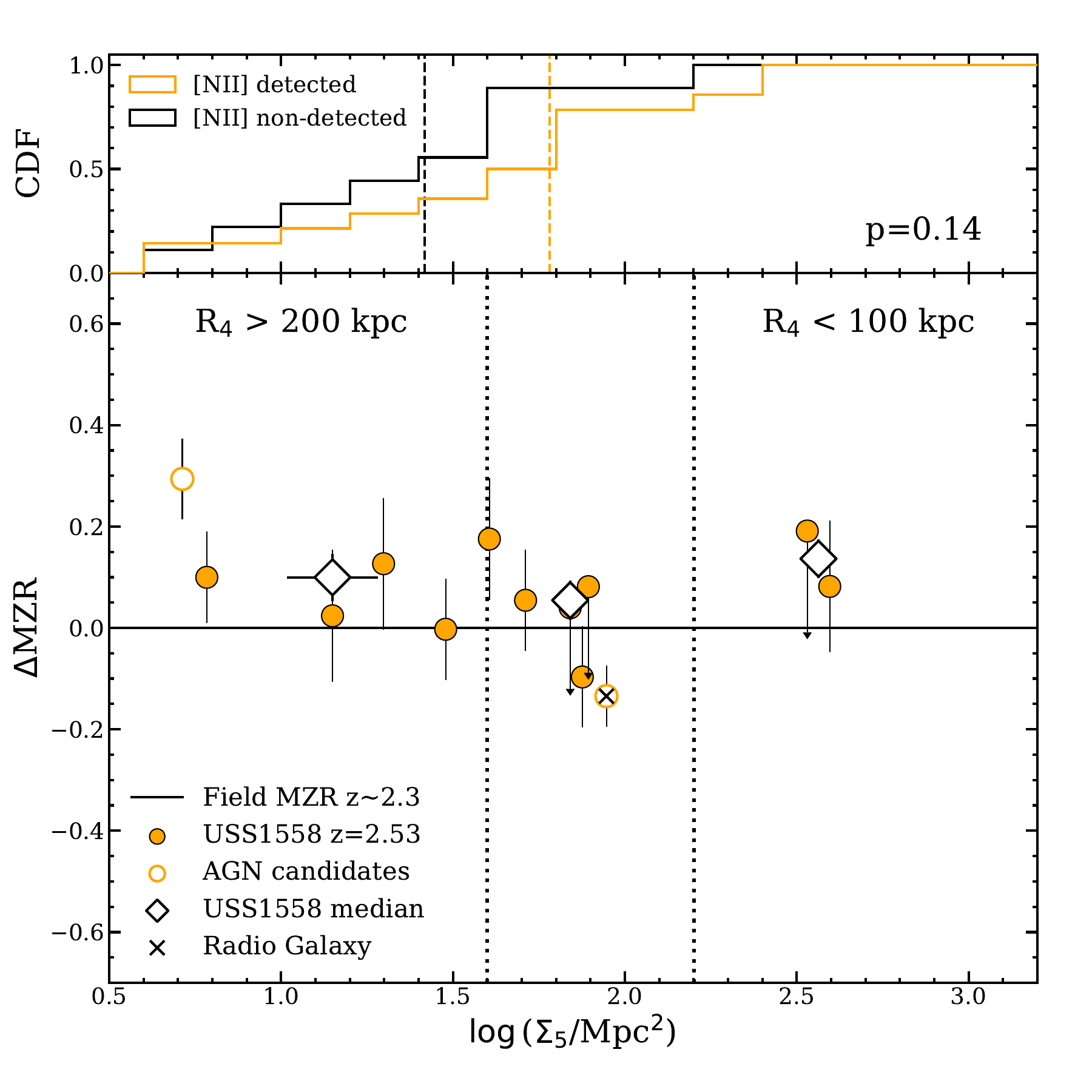}\par 
      \caption{Main panel: Offsets from the field MZR as a function of local density values, which is defined as the minimum area to enclose five neighboring galaxies. The horizontal dashed line shows the position of the field MZR from Fig.\,\ref{F:MOH}. Vertical dashed lines separate the density regimes outlined in Sect. \ref{SS:Environment}. Upper panel: Local density cumulative distribution functions for the [N{\sc{ii}}] detected (orange) and non-detected (black) samples. Dashed vertical lines display the median values for each sample.}
         \label{F:Density_OH}
\end{figure}

This section investigates the mass-metallicity relation (MZR) for HAEs belonging to USS1558 and the impact of the environment on their ISM enrichment. We derive gas-phase metallicities from the H$\alpha$ and [N{\sc{ii}}]$\lambda6584$ emission lines based on the N2 calibration developed by \cite{Pettini04}. We extract individual metallicity measurements for 10 sources (plus 3 upper limits) which are represented by orange circles in Figure\,\ref{F:MOH} (left). We compare our results with those obtained in PKS1138 by \cite{PerezMartinez23} (blue circles) and with a coeval field MZR constructed by combining the $2<z<3$ samples of \cite{Erb06}, \cite{Wuyts16} (KMOS3D) and \cite{Sanders21} (MOSDEF) as it is explained in Sect.\,\ref{SS:metallicity}. Prior to the comparison, we have homogenized all samples to the same IMF (\citealt{Chabrier03}) and the same metallicity diagnostic (N2 following Sect.\,\ref{SS:metallicity}) using the empirical calibration transformations of \cite{Curti17} when needed. 

Our results suggest that most of the individual metallicity measurements performed for the USS1558 sample lie on or above the field MZR, albeit displaying lower metallicities than the PKS1138 counterparts. However, the [N{\sc{ii}}] emission line was not detected in approximately half of the sample thus biasing any comparison with the field. Thus our individual measurements only trace the metal-rich end of the distribution of HAEs and the lack of metal-poor galaxies may be just a reflection of our [N{\sc{ii}}] detection limit. Thus, we resort to the stacking analysis to recover all the [N{\sc{ii}}] non-detections objects and present its results in the right diagram of Figure\,\ref{F:MOH}. Both USS1558 and PKS1138 stacked measurements are compatible with each other at $\mathrm{10.0<\log M_*/M_\odot<10.8}$ displaying the following offsets with respect to the field relation: $\mathrm{\Delta MZR(USS1558)=0.02\pm0.07\,dex}$ and $\mathrm{\Delta MZR(PKS1138)=0.06\pm0.03\,dex}$. However, in the low-mass regime, we find a marginally significant offset between these two samples with USS1558 being more metal-poor compared to PKS1138: $\mathrm{\Delta MZR(USS1558)=-0.14\pm0.08\,dex}$ and $\mathrm{\Delta MZR(PKS1138)=0.00\pm0.04\,dex}$. This result and the SFR enhancement detected in Fig.\,\ref{F:SFR} within the same mass regime are qualitatively consistent with the expectations of the fundamental metallicity relation (FMR, \citealt{Mannucci10}) which predicts that the star-forming activity and the metal enrichment are anti-correlated. 

We note, however, that previous work in the same field by part of our team (\citealt{Shimakawa15}) reported a $\mathrm{\sim0.1 dex}$ metallicity enhancement with respect to the field across a similar mass regime even when accounting for the IMF differences. Several factors may explain this discrepancy: First, the lower sensitivity and spectral resolution ($R\sim500$) of their observations with Subaru/MOIRCS hinder the precision of their measurements, yielding only metallicity upper limits for stacked results in the mid-to-low mass regime ($\mathrm{\log M_*/M_\odot<10.5}$). In addition, they derive the stellar mass of their objects relying on the relationship between $J-K_s$ color and the $M_*/L_{K_s}$ using the stellar population synthesis models of \cite{Kodama98,Kodama99}. This approach systematically underestimates the stellar masses by $\mathrm{\sim0.1 dex}$ when compared with the SED fitting carried out in this work, although this difference increases towards lower mass objects. A possible explanation for this effect is that this approach becomes less accurate for fainter objects in these bands (i.e. lower mass galaxies at a fixed redshift), while Ks-band and $J-K_s$ color remain a suitable proxy to estimate the stellar mass of massive galaxies at $z=2.5$. Finally, the field MZR in \cite{Shimakawa15} is based on the UV-selected galaxies alone (\citealt{Erb06}). UV selection tend to be biased towards objects with relatively young stellar populations while excluding some dust and metal rich galaxies (e.g., \citealt{Steidel04}; \citealt{Stott13b}). This combined effect naturally bias the field MZR towards lower metallicities, albeit the contribution of missing dusty objects may be only marginal according to \cite{Shimakawa15} after comparing the metallicity offset of a subsample of SDSS galaxies selected in a similar manner than the HAEs with their measured extinctions. Nevertheless, we highlight that this work in USS1558 and that of \cite{PerezMartinez23} in PKS1138 have applied exactly the same methods to the derivation of stellar masses, SFRs, and gas-phase metallicities and thus, any comparison between protoclusters remains fair within the statistical limitations of the samples studied.

Finally, we explore the relation of diverse local density regimes within the protocluster with the metallicity offsets (i.e., $\mathrm{\Delta\,MZR=12+\log(O/H)-12+\log(O/H)_{MZR}}$) present in the objects lying therein. Figure\,\ref{F:Density_OH} summarizes our results with the main panel displaying the distribution of the individual metallicity measurements as a function of $\Sigma_5$. The median values (empty diamonds) for each local density bin display a flat trend. However, the distribution of [N{\sc{ii}}] non-detections is also of interest for the environmental analysis as this group is representative of objects with lower metallicities (e.g., lower [N{\sc{ii}}] emission line flux) than their counterparts at fixed stellar-mass. The upper panel of Figure\,\ref{F:Density_OH} displays the cumulative distribution function of metallicity offsets for the [N{\sc{ii}}] detected and non-detected samples as a function of local density. Intriguingly, the non-detections (ND) tend to lie at lower density regimes than the [N{\sc{ii}}] detected (D) sample, which traces the metal-rich end of the metallicity distribution, thus suggesting a relation between the denser areas of the protocluster and higher metal enrichment. The median values (dashed lines in the figure) for each sample also support this description with $\mathrm{\log\Sigma_5(ND)=1.42\pm0.21\,Mpc^{-2}}$ and $\mathrm{\log\Sigma_5(D)=1.78\pm0.18\,Mpc^{-2}}$. However, we should take this result with caution as the KS-test dims these two distributions as statistically indistinguishable ($\mathrm{p=0.14}$). Furthermore, more massive galaxies (typically more metal-rich too) tend to reside in the denser areas of overdensities. This behavior is certain within both the [NII] detected and undetected samples, albeit objects within the latter have overall lower stellar masses than the detected sample which could also explain why they lie in less dense environments.

\subsection{Molecular gas fractions}
\label{SS:Molprop}

In the context of the gas regulator model (\citealt{Peng10,Peng14}), galaxies are progressively metal-enriched throughout successive episodes of star formation. Thus, their gas reservoir is gradually depleted (\citealt{Bothwell13a}; \citealt{Hunt15}) due to gas consumption and the contribution of outflows expelling gas from the galaxy disk via supernova or AGN feedback. The lack of cold gas diminishes the efficiency of star formation and eventually quenches the galaxy unless additional pristine gas is accreted from the cosmic web compensating for such consumption (\citealt{Dekel06}). Therefore, the gas regulator model can be parameterized by four simple parameters: the star formation efficiency, the gas inflow rate, the mass-loading factor which controls the balance between star-formation and outflows ($\lambda$=outflows/SFR), and the return mass fraction, which is assumed to be constant. The mass loading factor is a parameter of special relevance for us as it can be described as a function of the gas fraction and gas phase metallicity, as it was shown by \cite{Suzuki21}.

\begin{figure}
      \includegraphics[width=\linewidth]{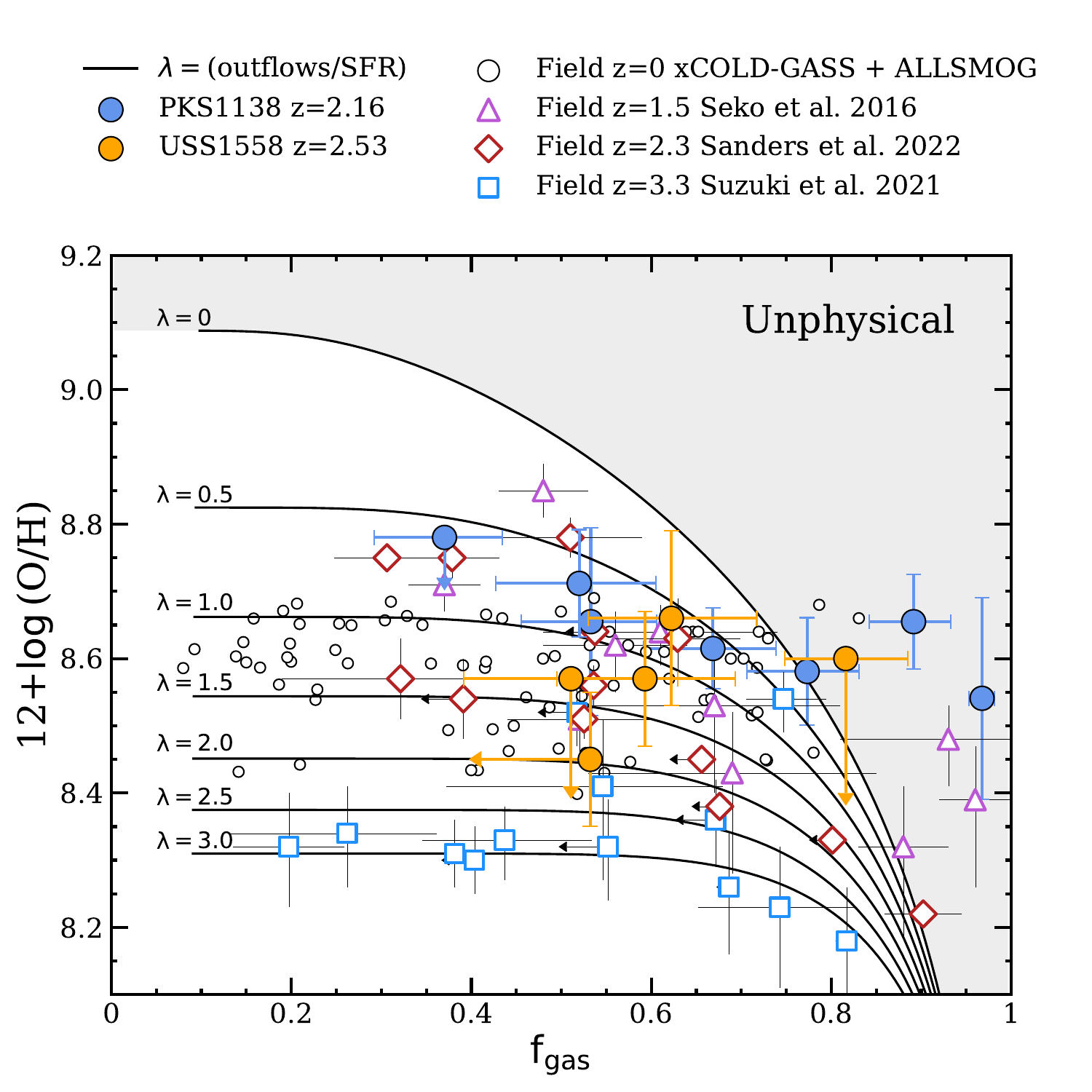}\par 
      \caption{Molecular gas fraction versus gas-phase metallicity. The solid lines show the tracks of constant mass-loading factor (outflows/SFR) assuming equilibrium in the context of the gas regulator model (\protect\citealt{Peng10, Peng14}). Filled orange and blue circles display our subsamples of HAEs with measured molecular gas fractions in the USS1558 (this work) and PKS1138 protoclusters (\protect\citealt{PerezMartinez23}). Blue-edged squares show the distribution of field galaxies at $z=3.3$ from \protect\cite{Suzuki21}. Brown-edged diamons display the field sample from \protect\cite{Sanders23}. Violet-edged triangles depict the field sample of \protect\cite{Seko16} at $z=1.5$. The local field comparison sample is composed of galaxies from the XCOLD GASS (\protect\citealt{Saintonge17}; \protect\citealt{Catinella18}) and the ALLSMOG (\protect\citealt{Bothwell14}) surveys. The grey area would require negative values for the outflow parameter and thus it is labeled as unphysical according to the gas regulator model.}
         \label{F:massload}
      \end{figure}

In this section, we aim to investigate the relation between the molecular gas fraction and the metallicity of five of our spectroscopically confirmed protocluster HAEs for which we have obtained metallicity measurements in this work and CO(3-2) information from \cite{Tadaki19}. We derive molecular gas fractions for these five objects following the procedure outlined in Sect.\,\ref{SS:MOL}. Figure\,\ref{F:massload} displays the distribution of our targets in the gas fraction vs gas metallicity plane in comparison with objects from PKS1138 and four different field samples at $0<z<3.3$. For reference, we overplot the tracks of constant mass-loading factor for $\mathrm{\lambda=0-3}$ (solid black lines) from the origin to the current estimated age of the universe (i.e, 13.6 Gyrs). Star formation proceeds in a more vigorous and short-lived way at high-z than in the local universe (\citealt{Thomas10}) and thus we expect to find higher mass-loading factors during this cosmic epoch as a consequence of enhanced outflows (\citealt{Suzuki21}). As we move towards lower redshifts, star formation proceeds more gradually thus yielding lower mass loading factors as we can see from the \cite{Sanders23} sample at $z=2.3$, the \cite{Seko16} sample at $z=1.5$, and the xCOLD-GASS (\citealt{Saintonge17}; \citealt{Catinella18}) and ALLSMOG (\citealt{Bothwell14}) samples in the local universe. The metallicity calibrations used by the different works presented in Figure\,\ref{F:massload} have been rescaled to the N2 method of \cite{Pettini04} following the empirical prescriptions provided in \cite{Curti17}. This ensures a fair comparison between samples at different cosmic epochs and environments in terms of metallicity. Finally, the grey-shaded area of Figure\,\ref{F:massload} depicts the locus where a negative value for the outflow parameter would be required, and thus it is deemed as unphysical within the gas regulator model. Thus, objects lying within this region may have overestimated values for $\mathrm{M_{mol}}$, or underestimated values for $\mathrm{M_*}$ assuming their metallicities are reliable. The five USS1558 objects scatter around $0.5\lesssim\lambda\lesssim2$ similarly to the $z=2.3$ and at $z=1.5$ field samples but in contrast with higher redshift galaxies of \cite{Suzuki21}. The six objects part of the PKS1138 sample were previously published in \cite{PerezMartinez23} and made use of a sample of CO(1-0) emitters presented in \cite{Jin21} as part of the COALAS project (large program ID: C3181, PI: H. Dannerbauer) with the Australia Telescope Compact Array (ATCA). This dataset was analyzed following the exact same methods quoted in Sect.\,\ref{S:Methods} albeit making use of a lower CO transition. Some objects in both USS1558 and PKS1138 overlap around very low $\lambda$ values albeit the USS1558 sample displays in general lower gas fractions and metallicities. Only one object in the USS1558 sample lies within the unphysical region although its uncertainties in $\mathrm{f_{gas}}$ and gas metallicity make it compatible with several of the mass-loading factor tracks. 

Finally, \cite{PerezMartinez23} argued that the low mass-loading factors displayed by the PKS1138 sample may be (at least partly) caused by the suppression of outflows in those objects, as their SFRs are representative of the Main sequence. This would be the result of the external pressure exerted by an overdense IGM over the gas halo of the protocluster members, preventing the outflows from star formation to leave the gravitational potential of the galaxy and forcing the HAEs to recycle their already enriched gas. This scenario looks less clear in USS1558 given the variety of $\mathrm{\lambda}$ values displayed by this sample, sharing a similar locus with other field samples at similar redshift in Fig. \,\ref{F:massload}.  However, the limited number of objects both in USS1558 and PKS1138 prevent us from drawing firm conclusions at this stage. We will discuss different scenarios of environmentally driven galaxy evolution in Sect.\,\ref{S:Discussion}.

\section{Discussion}
\label{S:Discussion}
\subsection{Enhanced star formation activities in USS1558}
\label{SS:Enhanced_SFR}
This work has investigated the relationship between the protocluster environment of USS1558 at $z=2.53$ and the physical properties of their star-forming protocluster members. In particular, we focused on the star formation activity, the metal enrichment and the gas reservoir of our objects, three quantities that are entangled with each other and, together with the environment, regulate the evolution of the galaxies' ISM (\citealt{Peng10,Peng14}). Sect.\,\ref{SS:SFR} showed that the majority of our spectroscopic targets display enhanced SFR with respect to the field and to other protocluster samples at similar cosmic epochs, with this enhancement being especially prominent in the low-mass regime ($\mathrm{\log M_*/M_\odot<10.0}$). This effect is of special interest for environmental studies, as most works below $z=1.5$ report an anti-correlation between the star formation activity of galaxies and the number density of objects in their surroundings (e.g., \citealt{Balogh98}; \citealt{Lewis02}; \citealt{Quadri11}; \citealt{Muzzin12}; \citealt{Old20}; \citealt{Reeves21}). The reversal of the star formation-density relation would point towards drastic changes in the physical properties of overdense regions in the high-z universe (e.g. protoclusters) compared to their lower redshift counterparts (clusters), thus pointing towards the joint evolution of galaxy populations and the large scale structures they inhabit over cosmic time. Several works have found evidence of this reversal by inspecting large high-z samples covering a wide range of local densities (\citealt{Lemaux22}; \citealt{Shi23}) and by focusing on particularly overdense regions such as protoclusters (\citealt{Hayashi16}; \citealt{Wang16}; \citealt{Shimakawa18a}; \citealt{Shi20}; \citealt{Monson21}). However, this picture is not yet unanimous as many others find no significant differences between protoclusters and field in terms of SFR (e.g., see \citealt{Koyama13}; \citealt{Shimakawa18b}; \citealt{PerezMartinez23} in PKS1138, or \citealt{Sattari21} in the CC2.2 protocluster at $z=2.2$). 

Gravitational interactions between protocluster members are one of the possible mechanisms to explain higher star formation activity with respect to the field. These events are expected to be more frequent in high-z protoclusters due to the higher number density of objects in protoclusters, their yet low relative velocities which favor close encounters (\citealt{Hine16}), and their gas-rich nature (\citealt{Gottlober01}; \citealt{Genel14}). Thus we would expect that the densest protocluster regions host a larger fraction of merging starbursting galaxies. Luminous dusty star-forming galaxies (DSFG) fit well into this description and have been usually found in protoclusters (e.g., \citealt{Dannerbauer14}; \citealt{Zeballos18}; \citealt{Harikane19}; \citealt{Long20}; \citealt{Zhang22}; \citealt{Calvi23}), albeit these objects often are a minority among the protocluster galaxy populations and tend to be representative of the high end of the stellar-mass distribution. However, not all DSFGs are produced through merging events, with several works reporting remarkably disk-like morphologies and a high degree of rotational support for less extreme DSFGs in the field at $z\gtrsim2$ (e.g., \citealt{Hodge15,Hodge16,Hodge19}; \citealt{Fujimoto18}; \citealt{Rizzo21}; \citealt{Gillman23}).

The environmental analysis presented in Fig.\,\ref{F:Density_SFR} yields a flat SFR enhancement trend across low-to-high density regimes within USS1558, arguing against gravitational interactions as the sole mechanism behind the SFR enhancement. Thus, we must consider possible environmental effects related to the growth of the large-scale structure and the accretion of fresh gas from the cosmic web as candidates to explain this effect. Simulations predict that cold gas streams along the filaments of the cosmic web supply the required fuel to feed and sustain the elevated SFRs of high-z massive galaxies (e.g., \citealt{Keres05}; \citealt{Dekel06}; \citealt{Genel08}). Beyond a certain mass limit, however, the supply of cold gas in an otherwise hot medium would progressively become inefficient until it is shut down completely as we expect for massive virialized galaxy clusters (\citealt{Dekel09a}). In fact, cold accretion not only helps massive galaxies to sustain their star-forming activities (\citealt{Daddi07}; \citealt{Bethermin14}) but could also provide an extra supply of gas to lower mass satellite galaxies, thus enhancing their star formation. Such an effect would be widespread across the whole protocluster as it acts as a node of the cosmic web, and would offer a suitable explanation for the SFR enhancement we report for HAEs within USS1558. Recently, \cite{Wang22} also suggested a similar accretion scenario to explain the enhancement of star formation in the low mass regime ($\mathrm{\log M_*/M_\odot<10.0}$) within their sample while \cite{Valentino15} found a similar enhancement at slightly higher masses albeit with limited statistics. Conversely, other spectroscopic works found no such signs of SFR enhancement (i.e., \citealt{Sattari21}; \citealt{PerezMartinez23}) suggesting that the specific evolutionary mechanisms at play may significantly differ between protoclusters at the cosmic noon. Future studies particularly focusing on the star formation of very low-mass galaxies in protoclusters will be able to explore this scenario in greater depth.

\subsection{Metal deficit for low-mass protocluster galaxies}

The metallicity analysis of USS1558 protocluster HAEs (Sect.\,\ref{SS:MZR}) yielded a metallicity deficit for low mass galaxies ($\mathrm{\log M_*/M_\odot<10.0}$) of approximately 0.1\,dex with respect to both our coeval field MZR and our previous results in PKS1138 at $z=2.16$ at the same stellar mass bin (Fig.\,\ref{F:MOH}). On the other hand, this discrepancy disappears at higher masses ($\mathrm{\log M_*/M_\odot>10.0}$). The ISM metal enrichment is a byproduct of sustained star formation over a period of time. In the general field, the fundamental metallicity relation (FMR, \citealt{Mannucci10}, \citealt{Sanders18,Sanders21}) predicts that star formation and gas-phase metallicity are anticorrelated quantities, as the gas consumption required to produce metals from successive generations of stars will eventually exhaust the gas reservoir of their host galaxy. Thus, it is expected that vigorous star-forming galaxies present relatively low metallicity values with additional processes such as mergers contributing to this scenario, particularly at $z>2$ (\citealt{Bustamante18}; \citealt{Rodriguez19}; \citealt{Horstman21}; \citealt{Calabro22}). Similarly, enhanced cold gas accretion as proposed in Sect.\,\ref{SS:Enhanced_SFR} would further foster these trends. The extra supply of pristine gas from the cosmic web will naturally dilute the ISM of satellite galaxies while promoting their star formation, thus accelerating the process of mass assembly in these objects. In this context, the FMR is able to qualitatively explain the results we have obtained for USS1558 in terms of SFR and metal enrichment, with low-mass metal-poor galaxies forming stars at a greater pace than the coeval field. 

Contradicting results on the metal enrichment of cosmic noon protocluster galaxies have been reported in recent years (e.g., \citealt{Kulas13}; \citealt{Valentino15}; \citealt{Kacprzak15}; \citealt{Sattari21}; \citealt{Chartab21}; \citealt{Wang22}; \citealt{PerezMartinez23}). Selection biases of different kinds may surely play a role in this puzzling situation. However, the cosmic noon is as crucial for the evolution of galaxies as it is for the development of the large-scale structures that host them. Thus, diverse stages of protoclusters mass assembly may be an important factor driving the onset, strength, and type of environmental effects in overdense regions. If the transition between different gas accretion regimes as a consequence of the dark matter halo growth (\citealt{Dekel06,Dekel09a}) is the dominant factor driving these changes, we would expect that protoclusters made of several clumpy lower-mass substructures such as USS1558 are first born as sites of vigorous star formation and low metallicity due to the surplus of fresh gas, then evolve towards less active and more metal-rich structures as the halo growth shock heats the IGM and suppress the gas inflows, which will be shut down once the protocluster achieves virialization. It is during this last phase, that the overdense environment (now a full-fledged cluster of galaxies) deprives galaxies of external gas supply and actively removes their existing gas reservoir through ram-pressure stripping (see \citealt{Boselli22} for a review). Interestingly, individual [N{\sc{ii}}] detections within our sample, which trace its metal-rich end, tend to lie in slightly more overdense regions than the non-detections, associated with lower metallicity values (Fig\,\ref{F:Density_OH}). This may hint at different accretion regimes between the outskirts and the denser areas of the four protocluster clumps where the mass assembly has been going on for a longer time and the gas inflows may be starting to lose efficiency. However, the KS-test yields no statistical significance between the spatial distribution of both groups. Thus, a larger (and more complete) sample would be needed to explore this possibility in an unbiased way.

\subsection{Molecular gas fractions and mass-loading factors}
Finally, we have used the gas regulator model (\citealt{Peng14}) to investigate the relation between the metal enrichment of our galaxies and the status of their molecular gas reservoir ($\mathrm{f_{gas}}$). 
USS1558 protocluster galaxies occupy a similar locus than the field samples at $z=2.3$ (\citealt{Sanders23}) and $z=1.5$ (\citealt{Seko16}) reaching up to $\lambda=2$. However, our analysis is limited by the small number of galaxies with both metallicity and $\mathrm{M_{Mol}}$ estimates. Field galaxies at $z=3.3$ display higher mass-loading factors ($\lambda\approx3$, \citealt{Suzuki21}) suggesting the presence of higher outflow activity. On the other hand, the main sequence star-forming galaxies from PKS1138 (\citealt{PerezMartinez23}) lie at even lower $\lambda$ values, hinting at the suppression of outflows in these objects. The authors speculate that this effect may be the result of the increased external pressure of the IGM in this massive cluster, forcing the galaxies to recycle their gas as proposed by \cite{Kulas13}, thus yielding higher metallicity values at a fixed molecular gas fraction. However, statistical limitations prevent them to draw firm conclusions, as is also our case for USS1558. A direct comparison between the distribution of USS1558 and PKS1138 protocluster galaxies in Fig.\,\ref{F:massload} hints at USS1558 being less metal enriched at intermediate gas fractions, thus confirming its less evolved nature compared with PKS1138. Nevertheless, larger number statistics are required to quantitatively explore possible environmental imprints into the gas metallicity gas fraction plane.

\section{Conclusions}
\label{S:Conclusions}

In this work, we have analyzed new Keck/MOSFIRE NIR spectroscopy to measure the star-formation and gas-phase metallicity of a sample of HAEs belonging to the USS1558 protocluster at $z=2.53$. Archival spectrophotometric observations encompassing the rest-frame UV to submillimeter regime complemented our analysis thus providing stellar and molecular gas masses and fractions. Finally, we have examined the physical properties of 23 protocluster members as a function of environment and discussed the implications of these results on different scenarios of environmentally driven galaxy evolution during the early stages of protocluster assembly. In the following paragraphs, we summarize the main conclusions of this work:
\begin{enumerate}
\item We have spectroscopically confirmed the protocluster membership of 23 narrow-band selected HAEs out of an initial sample composed of 27 objects, thus yielding a success rate of $\sim88\%$ and including eight new sources unconfirmed by previous spectroscopic campaigns (\citealt{Shimakawa15}; \citealt{Tadaki19}). In addition, we report two foreground galaxies at $z\approx3.6$ and two non-detections. 
\item HAEs in USS1558 at $z=2.53$ display enhanced SFRs with respect to the main sequence of star formation fixed for that cosmic epoch (\citealt{Speagle14}). Furthermore, this enhancement is statistically significant when compared with protocluster HAEs belonging to PKS1138 at $z=2.16$ (\citealt{PerezMartinez23}) and with the coeval field sample of KMOS3D galaxies (\citealt{Wisnioski19}). Galaxies with enhanced SFRs tend to display low stellar masses ($\mathrm{\log M_*/M_\odot<10.0}$) and are homogeneously distributed across different density regimes within the protocluster. This suggests that USS1558 is experiencing a phase of vigorous star formation possibly driven by protocluster-wide effects such as increased accretion of pristine gas, and disfavors alternative channels such as galaxy-galaxy interactions which are expected to predominately occur within the densest areas of the protocluster.
\item We extract individual and stacked [N{\sc{ii}}] metallicity values for USS1558 protocluster members. The 10 individual metallicity measurements represent the metal-rich end of our distribution and tend to lie in slightly more overdense areas than [N{\sc{ii}}] non-detected objects, which trace the low-metallicity end of our distribution. The stacking analysis yielded similar metallicity values to those found in PKS1138 and the field for $\mathrm{10.0<\log M_*/M_\odot<10.8}$, but lower metallicity values at $\mathrm{\log M_*/M_\odot<10.0}$: $\mathrm{\Delta MZR(USS1558)=-0.14\pm0.08\,dex}$ vs $\mathrm{\Delta MZR(PKS1138)=0.00\pm0.04\,dex}$. We propose this metallicity deficit together with the SFR enhancement at low stellar masses may suggest that this regime is dominated by young galaxies still in the process of forming the bulk of their mass and whose gas reservoir may be fed by environmentally enhanced gas inflows.
\item We obtain molecular gas fraction for a subsample of 5 protocluster members by making use of the CO(3-2) measurements from \cite{Tadaki19}. After exploring the gas-phase metallicity gas fraction relation, and in the context of the gas regulator model (\citealt{Peng14}), we find that HAEs in USS1558 are distributed along mass-loading factor tracks ($\mathrm{\lambda\approx0.5-2}$) comparable with those of field samples at $1.5<z<2.3$ in contrast with previous studies at $z\approx3$ ($\mathrm{\lambda\approx3}$) arguing for higher outflow activity (\citealt{Suzuki21}), but also contrary to the environmentally suppressed outflows scenario proposed in \cite{PerezMartinez23} for the PKS1138 protocluster sample ($\mathrm{\lambda\approx0-1}$). However, a larger sample of deeper metallicity and gas fraction estimates is required to further explore this scenario.
\end{enumerate}

\section*{Acknowledgements}

We thank the anonymous referee for his/her constructive feedback, which has contributed to improving this manuscript. Part of the data presented herein were obtained at the W. M. Keck Observatory, which is operated as a scientific partnership among the California Institute of Technology, the University of California, and the National Aeronautics and Space Administration. The Observatory was made possible by the generous financial support of the W. M. Keck Foundation. The observations were carried out within the framework of Subaru-Keck/Subaru-Gemini time exchange program which is operated by the National Astronomical Observatory of Japan. We are honored and grateful for the opportunity of observing the Universe from Maunakea, which has the cultural, historical, and natural significance in Hawaii. This research is based in part on data collected at the Subaru Telescope, which is operated by the National Astronomical Observatory of Japan (NAOJ). This research is based in part on observations made with the NASA/ESA Hubble Space Telescope obtained from the Space Telescope Science Institute, which is operated by the Association of Universities for Research in Astronomy, Inc., under NASA contract NAS 5–26555. This paper makes use of the following ALMA data: ADS/JAO.ALMA\#2015.1.00395.S. ALMA is a partnership of ESO (representing its member states), NSF (USA) and NINS (Japan), together with NRC (Canada), MOST and ASIAA (Taiwan), and KASI (Republic of Korea), in cooperation with the Republic of Chile. The Joint ALMA Observatory is operated by ESO, AUI/NRAO and NAOJ. The Australia Telescope Compact Array is part of the Australia Telescope National Facility (https://ror.org/05qajvd42) which is funded by the Australian Government for operation as a National Facility managed by CSIRO. We acknowledge the Gomeroi people as the Traditional Owners of the Observatory site. This research made use of Astropy,\footnote{http://www.astropy.org} a community-developed core Python package for Astronomy \citep{Astropy13, Astropy18}. JMPM is grateful for the support from T. Kodama and Tohoku University in the development of this work. JMPM also thanks H. Dannerbauer for providing access to the COALAS datasets for the PKS1138 comparison sample. TK acknowledges the support of Grant-in-Aid for Scientific Research (A) (KAKENHI \#18H03717). MO acknowledges the support by JSPS KAKENHI Grant Number JP21K03622. Kavli IPMU is supported by the World Premier International Research Center Initiative (WPI), MEXT, Japan.

\section*{Data Availability}

The spectroscopic data used within this article belong to the Subaru-Keck exchange program ID: S20A-075 (PI: T. Kodama) and it is publicly available throughout the Keck Observatory Archive (KOA). The ground-based photometric data used in this work belongs to the Subaru programs S10B-028 and S15A-047 (PI: T. Kodama) and it is publicly available throughout the Subaru Mitaka Okayama Kiso Archive (SMOKA) system. The HST photometry used in this work belongs to the program GO-13291 (PI: M. Hayashi) and it is publicly available through the Hubble Legacy Archive. The ALMA CO(3-2) quoted in this work and used to derive molecular gas masses for the USS1558 sample belongs to the ALMA program 2015.1.00395.S (PI: T. Kodama). The ATCA CO(1-0) quoted in this work and used to derive molecular gas masses for the PKS1138 sample belongs to the ATCA large program COALAS (ID: C3181, PI: H. Dannerbauer). Enhanced data products from these datasets may be shared on reasonable request.



\bibliographystyle{mnras}
\bibliography{references.bib} 




\appendix

\section{SED fitting results}

\begin{figure*}
 \centering
 \begin{multicols}{4}
      \includegraphics[width=\linewidth]{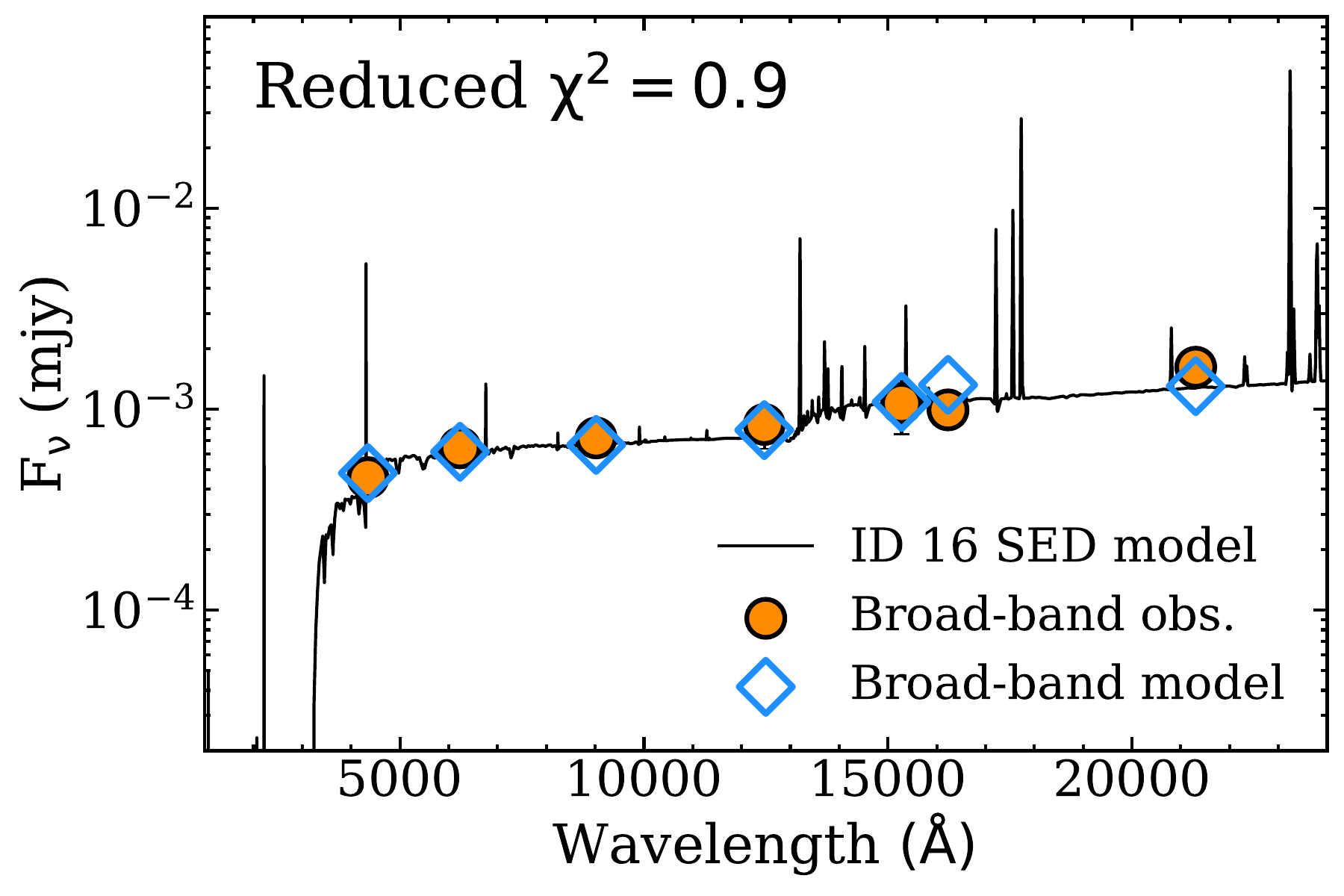}\par
      \includegraphics[width=\linewidth]{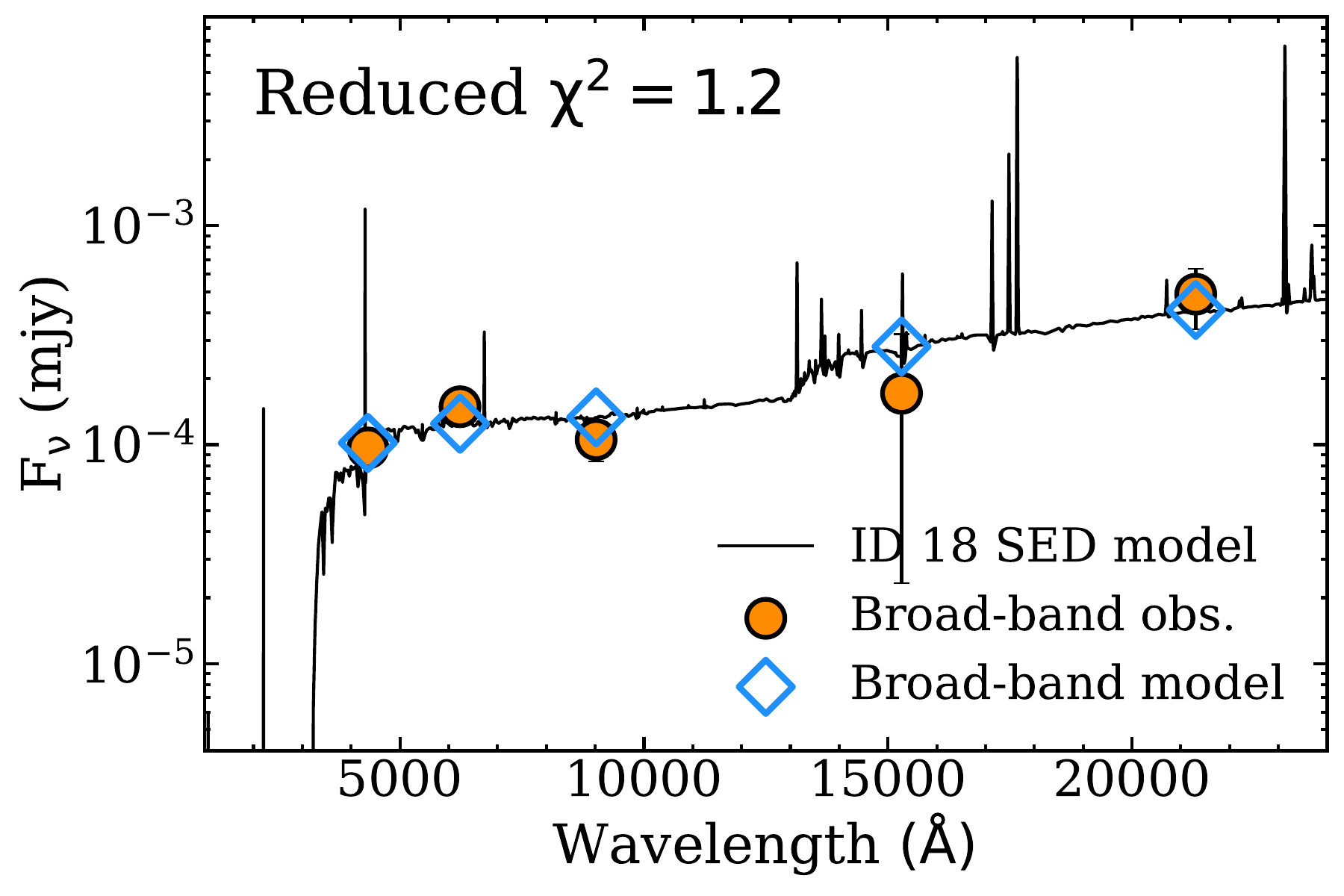}\par
      \includegraphics[width=\linewidth]{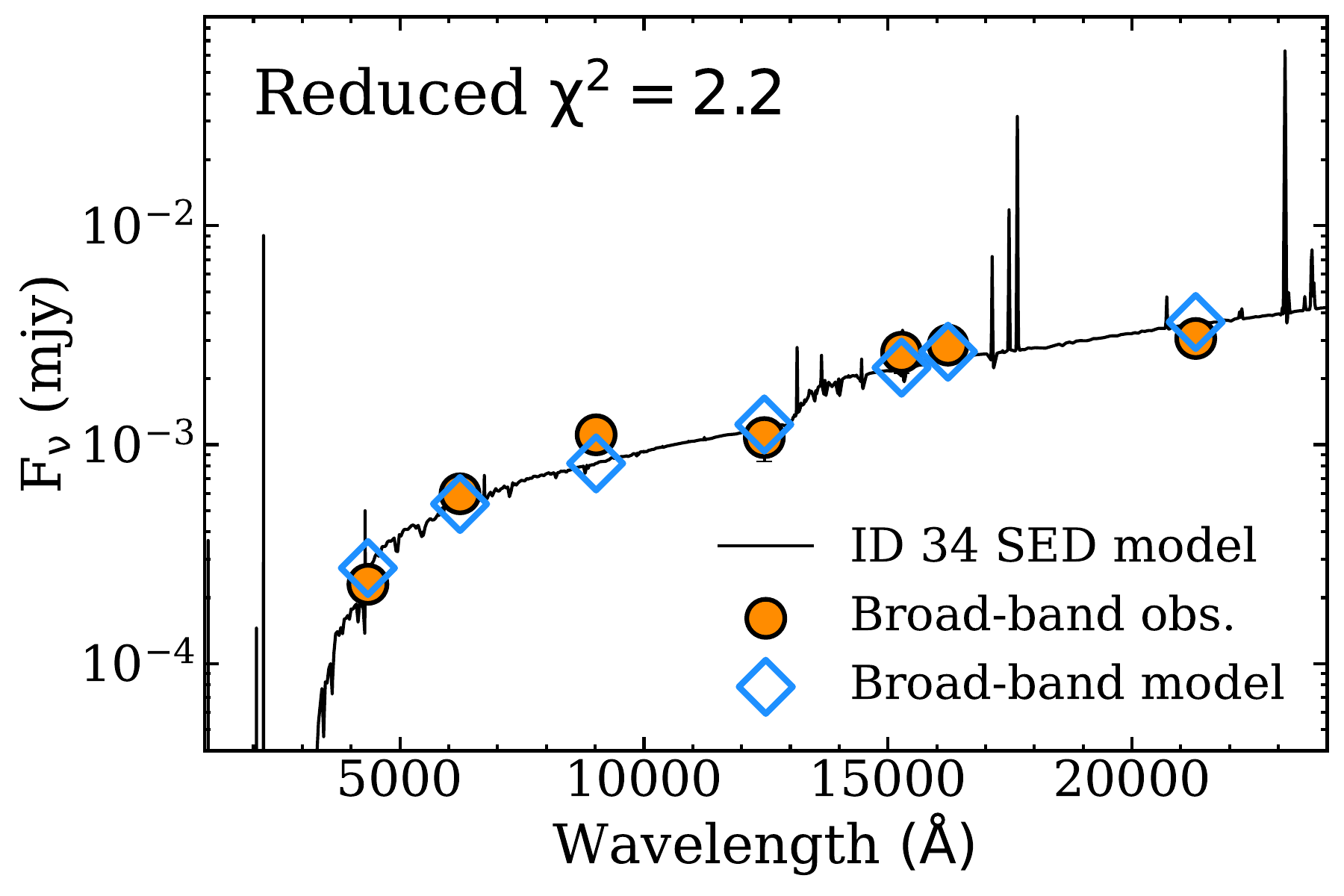}\par
      \includegraphics[width=\linewidth]{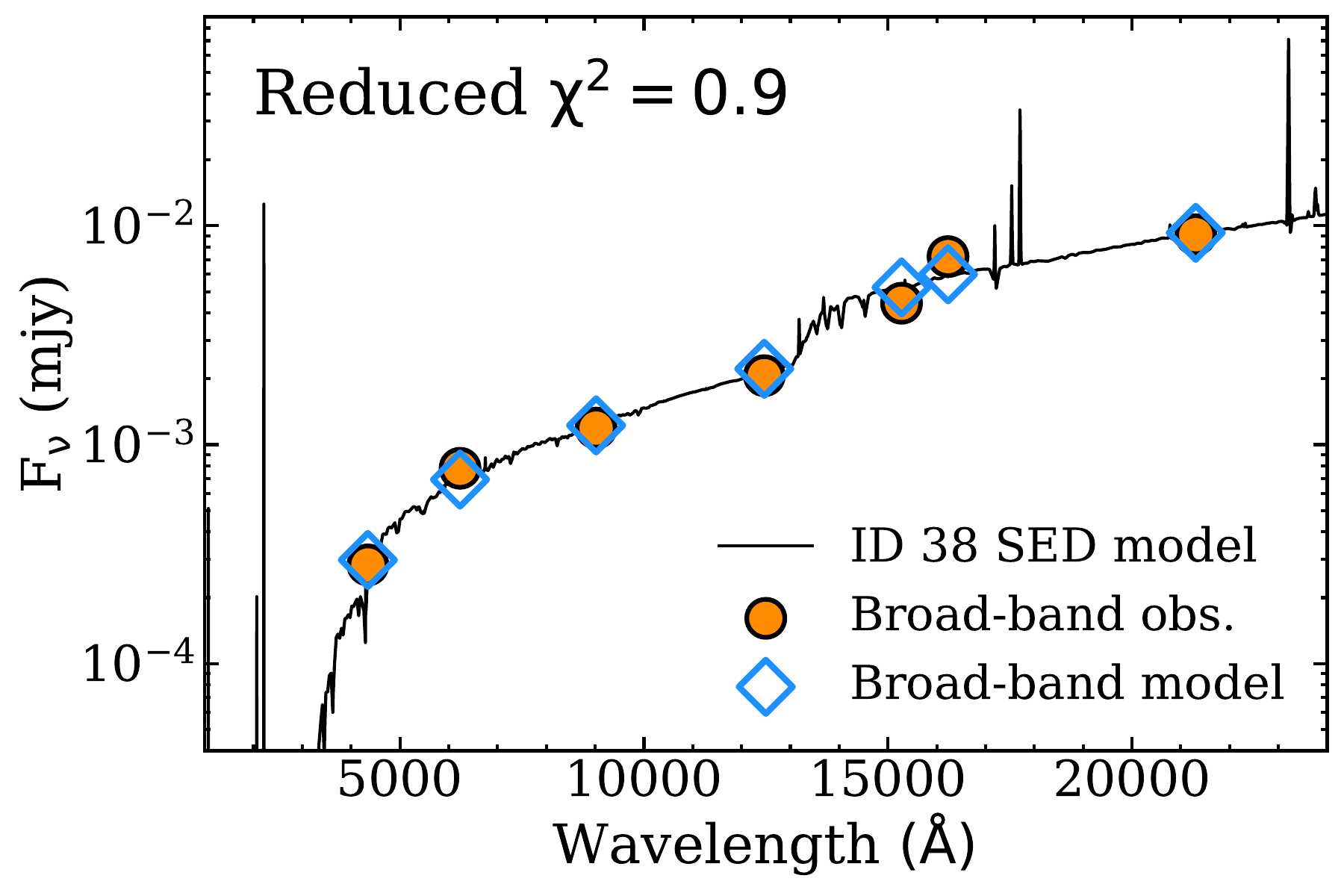}\par
      \includegraphics[width=\linewidth]{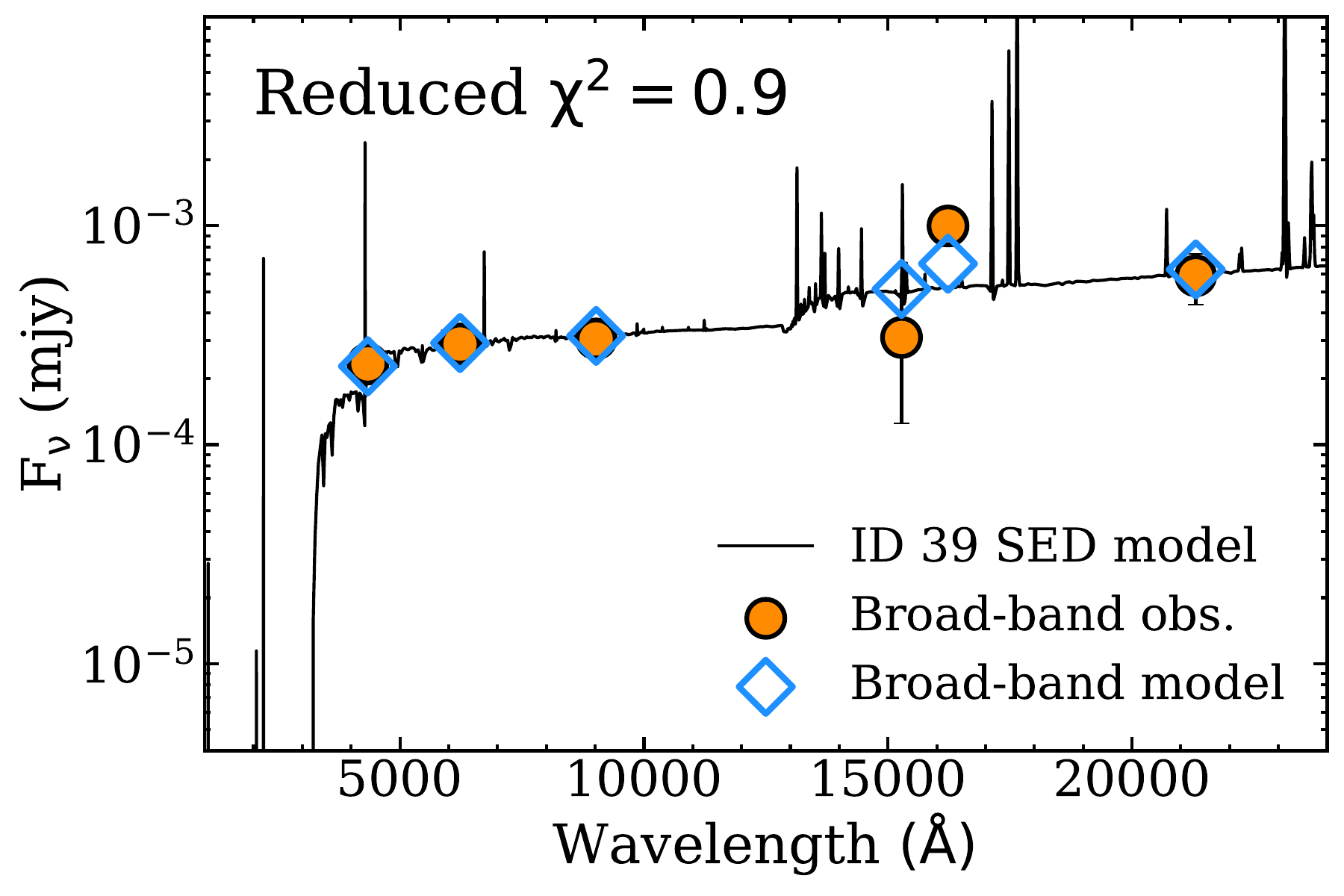}\par
      \includegraphics[width=\linewidth]{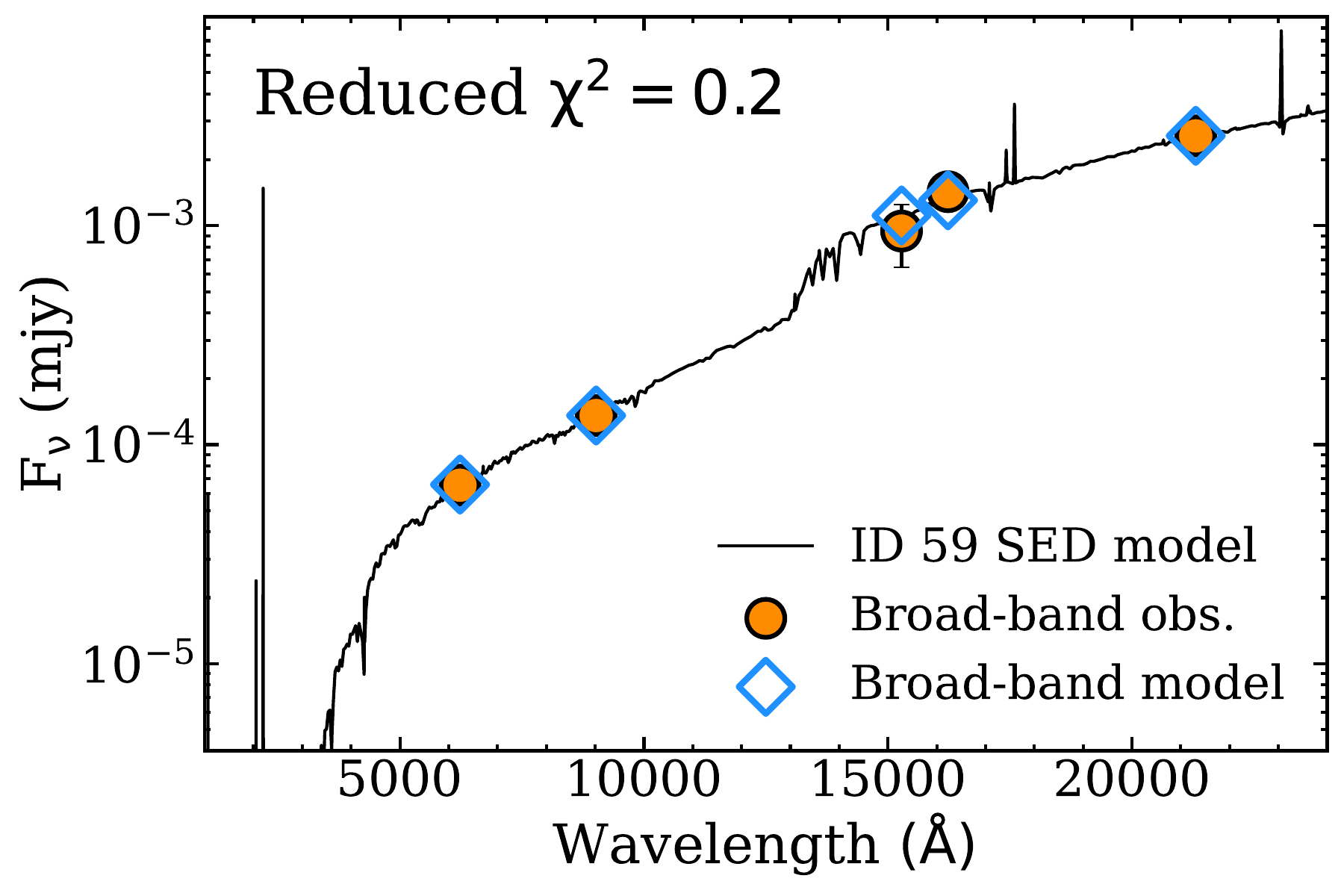}\par
      \includegraphics[width=\linewidth]{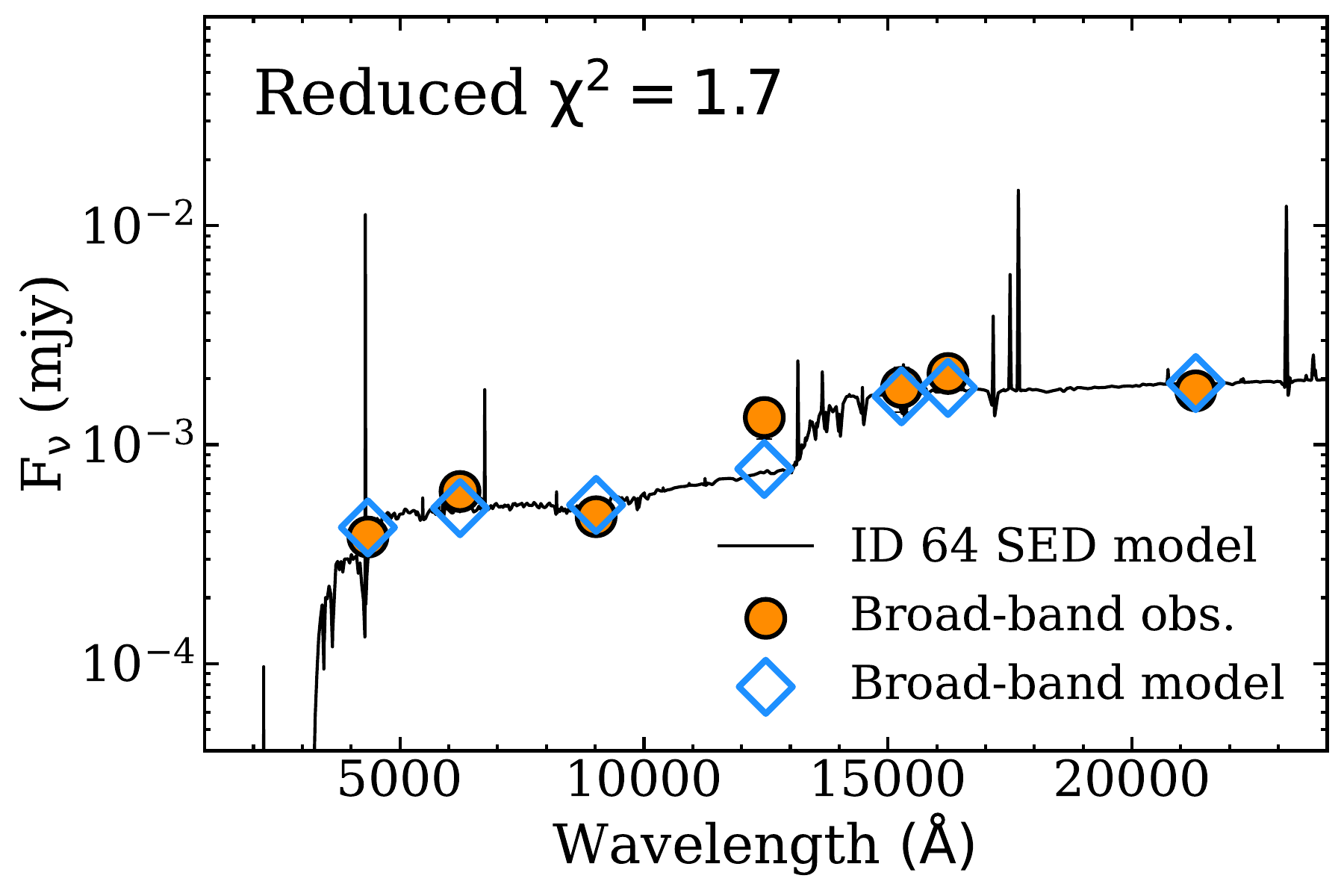}\par
      \includegraphics[width=\linewidth]{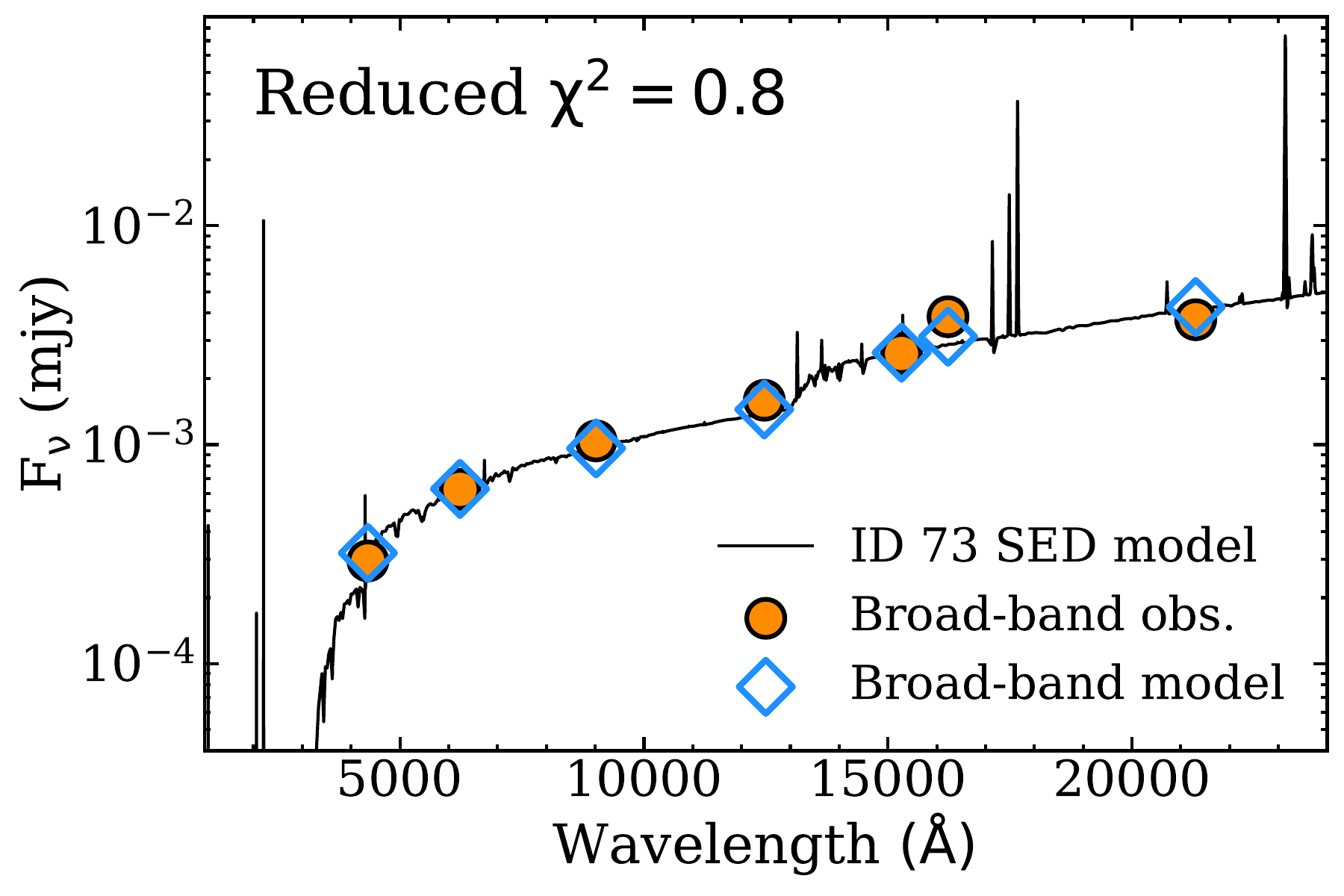}\par
      \includegraphics[width=\linewidth]{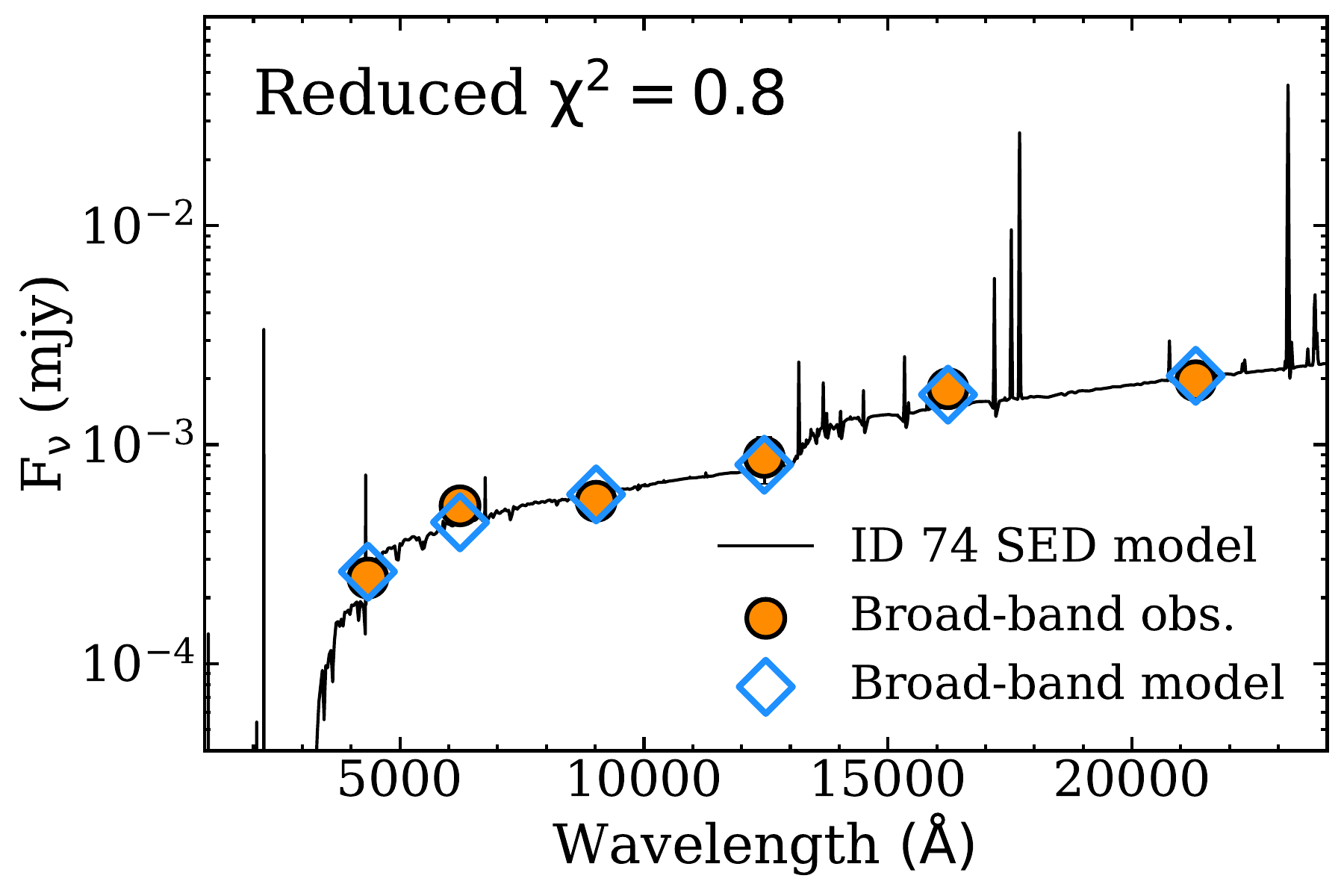}\par
      \includegraphics[width=\linewidth]{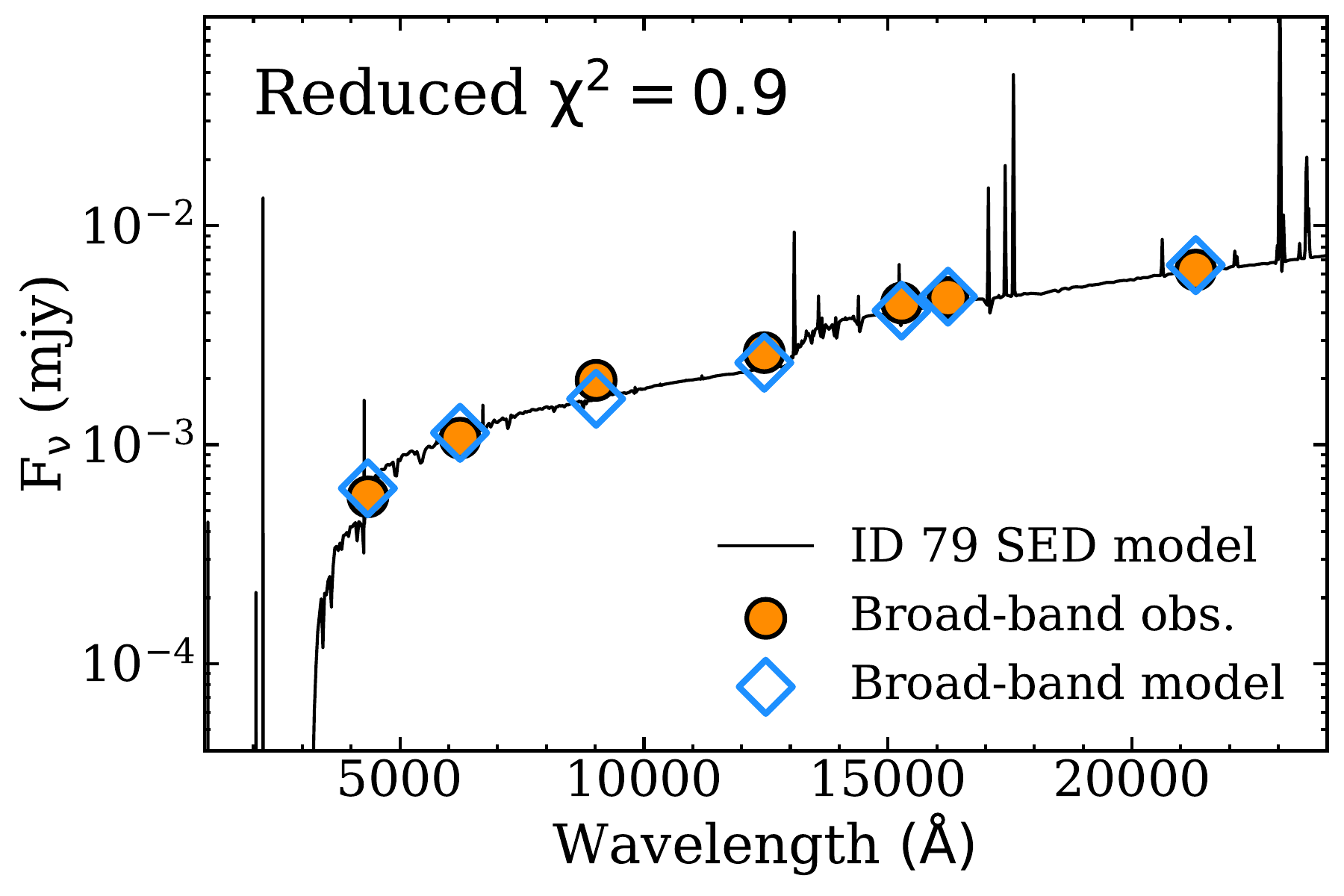}\par
      \includegraphics[width=\linewidth]{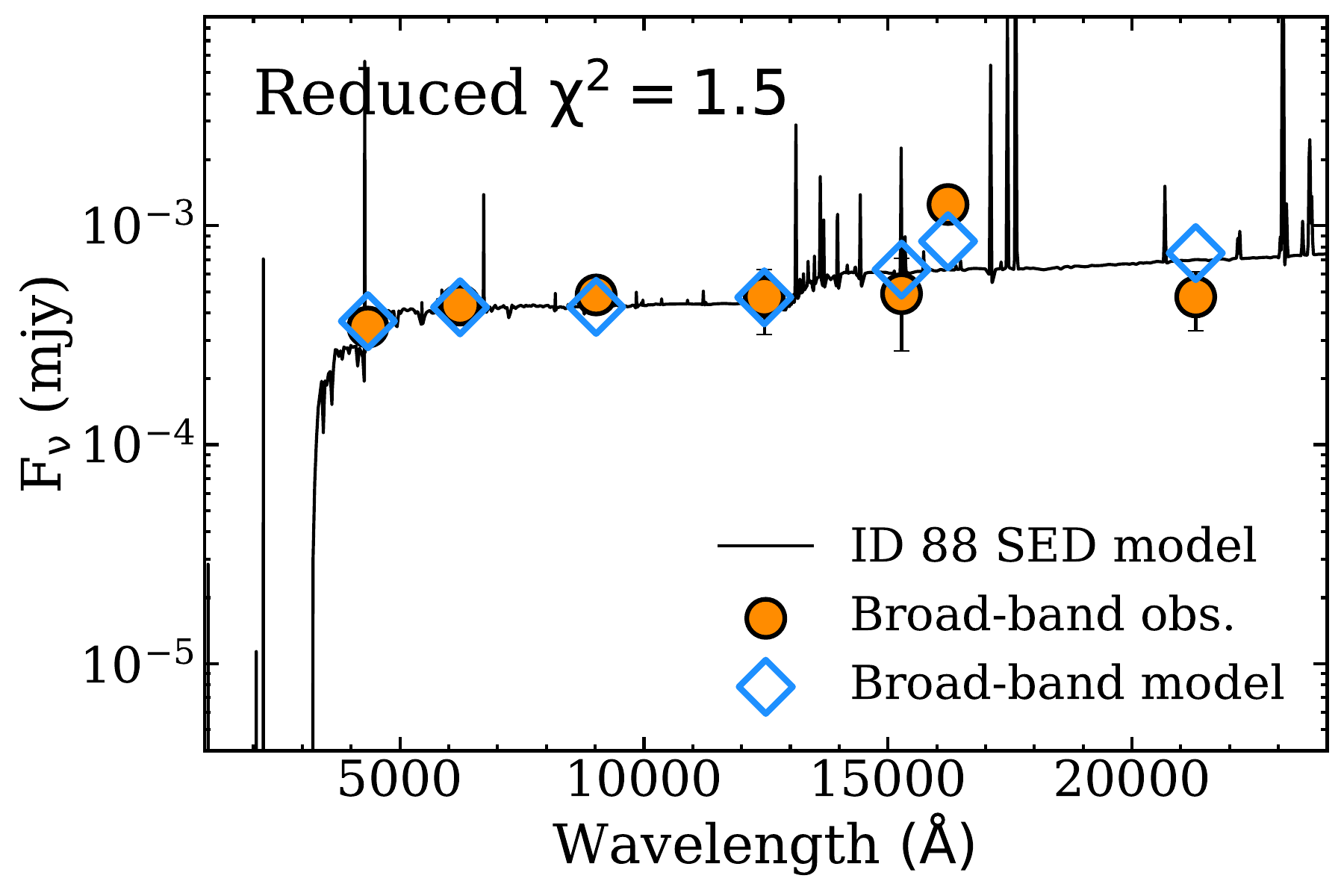}\par
      \includegraphics[width=\linewidth]{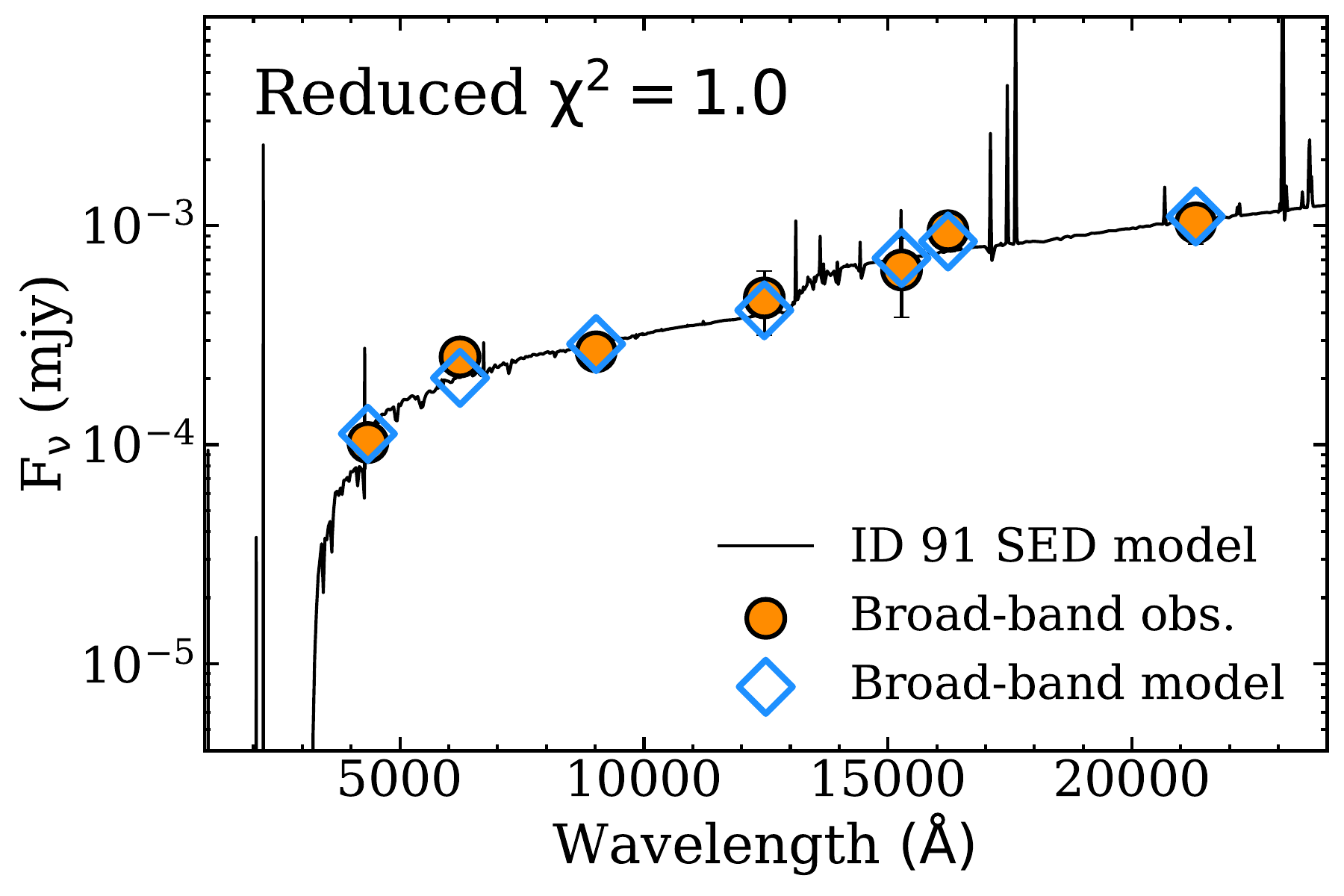}\par
      \includegraphics[width=\linewidth]{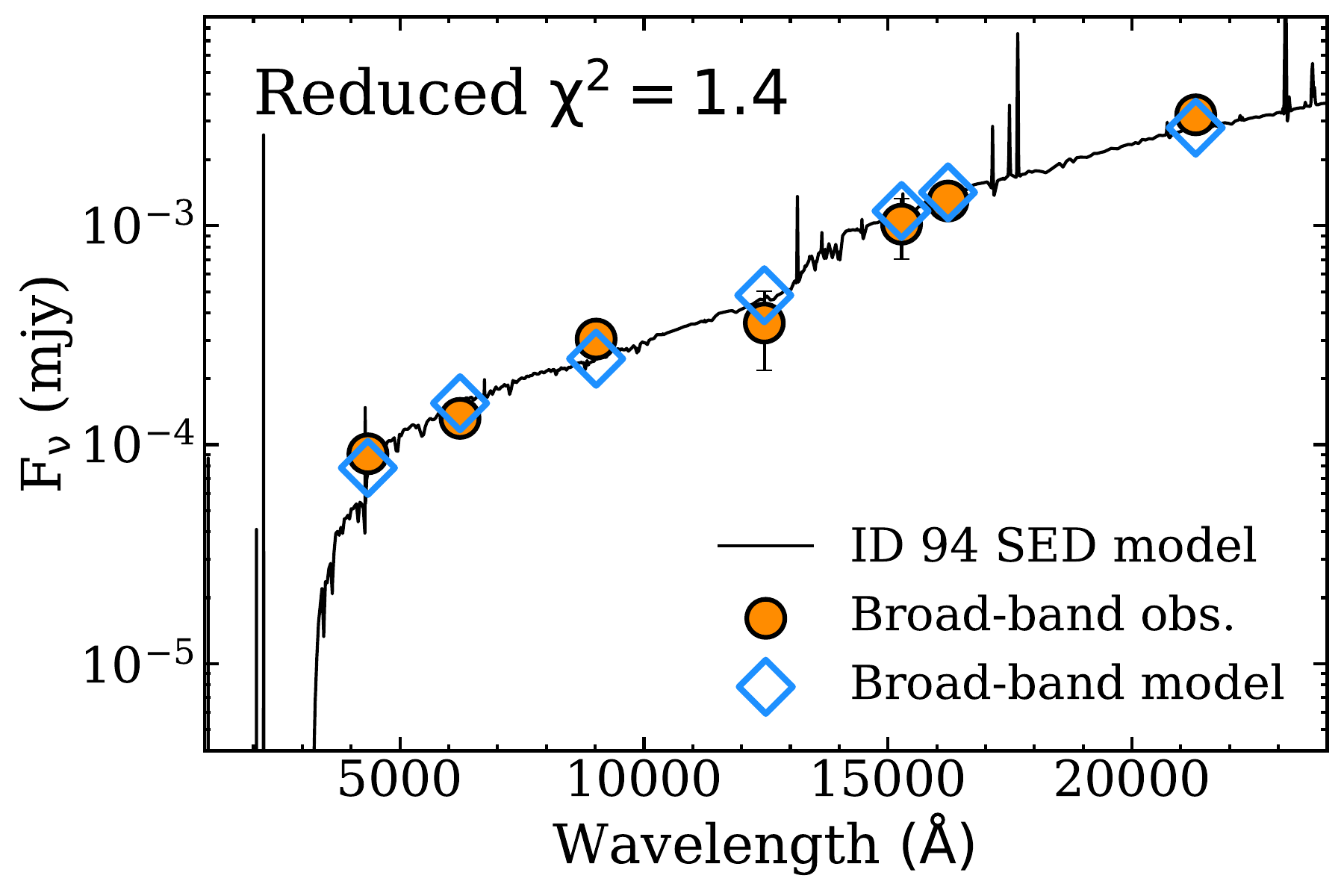}\par
      \includegraphics[width=\linewidth]{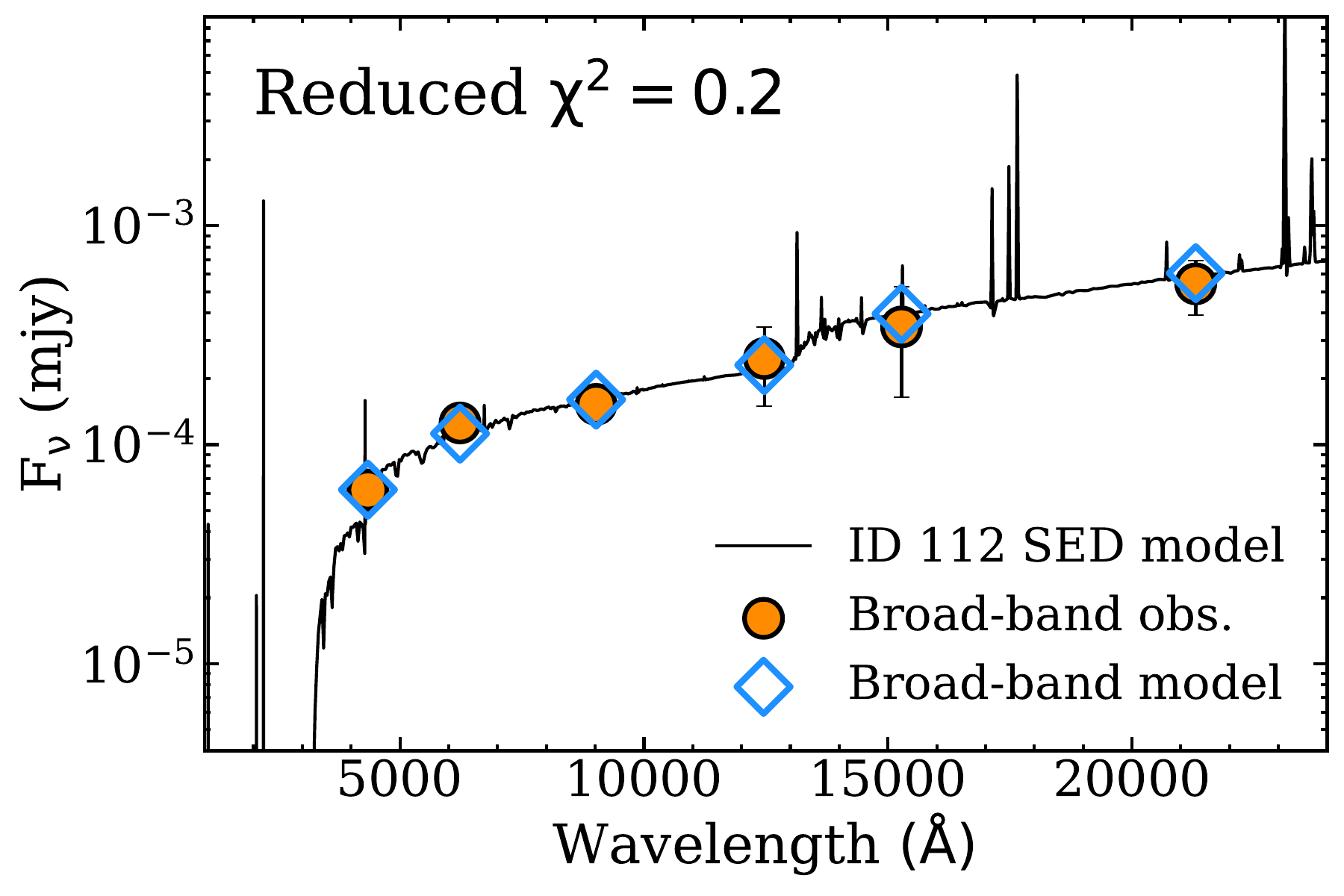}\par
      \includegraphics[width=\linewidth]{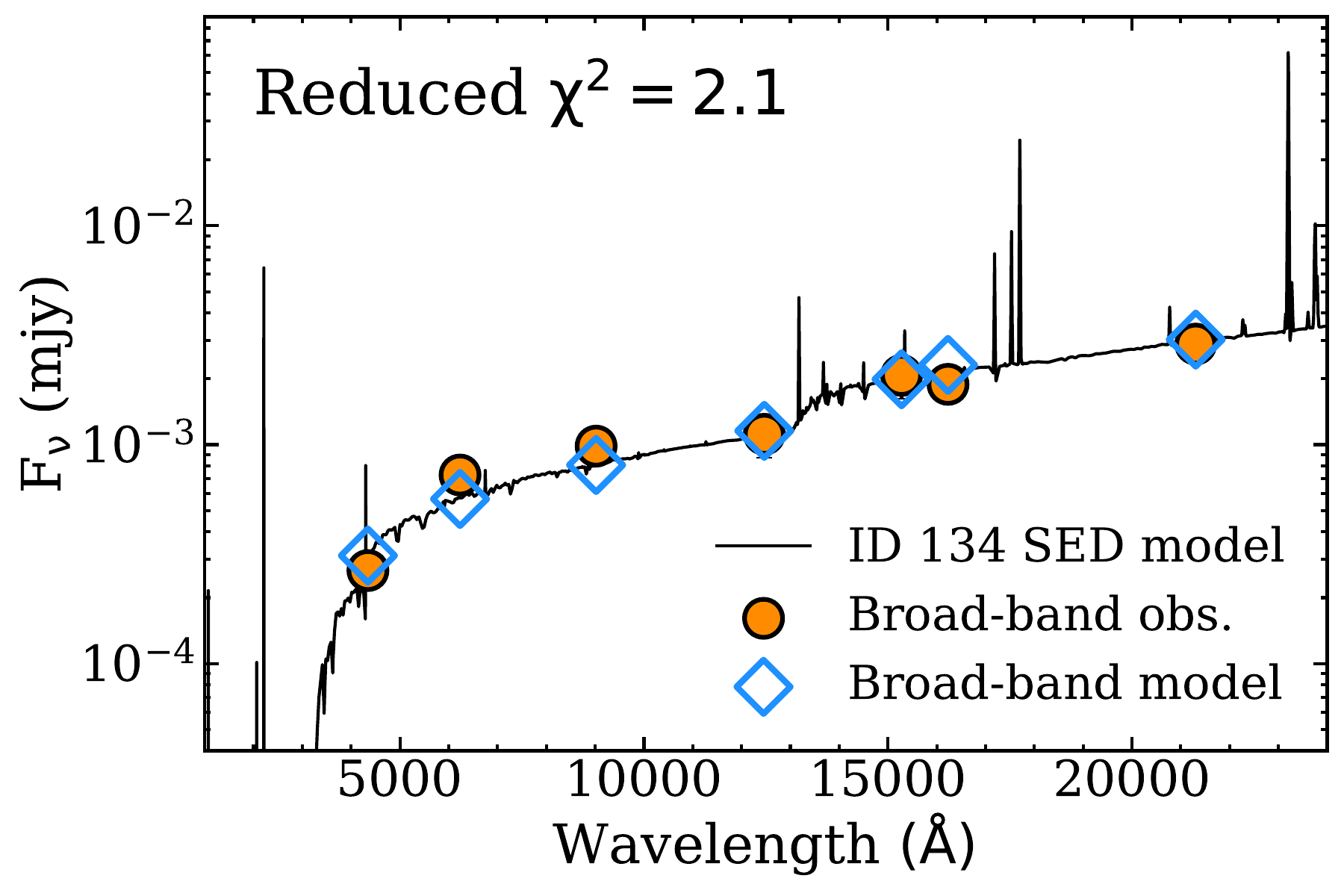}\par
      \includegraphics[width=\linewidth]{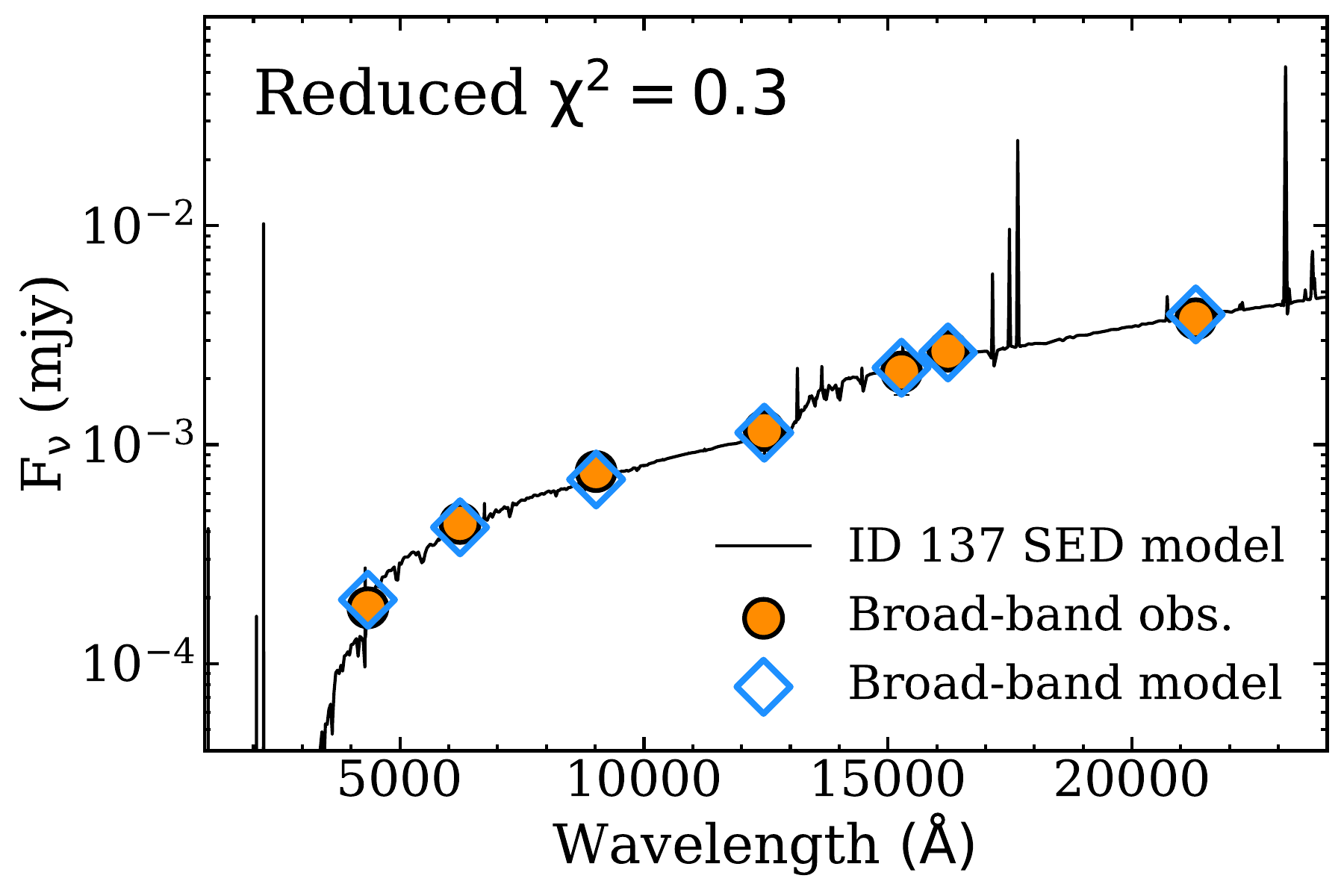}\par
      \includegraphics[width=\linewidth]{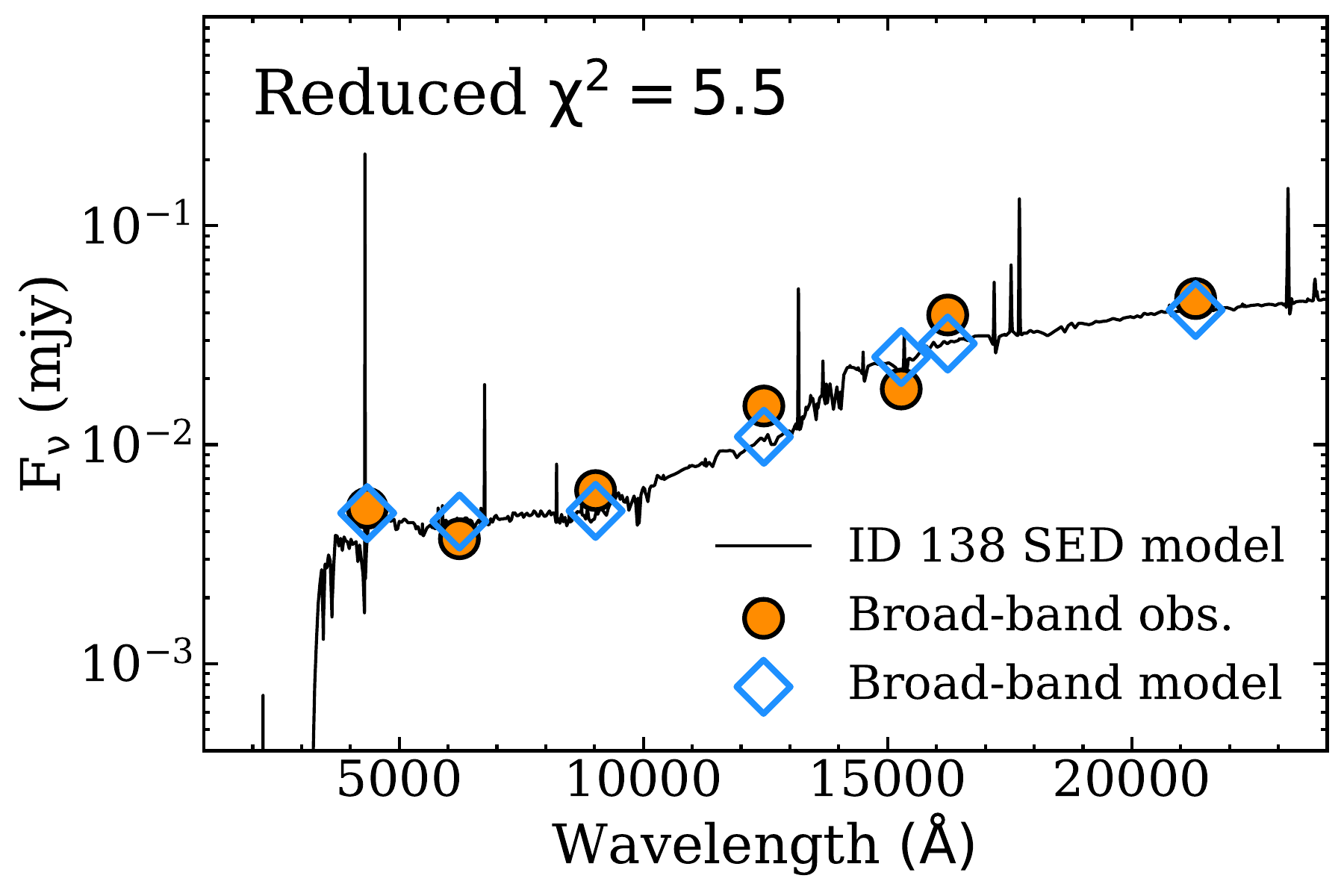}\par
      \includegraphics[width=\linewidth]{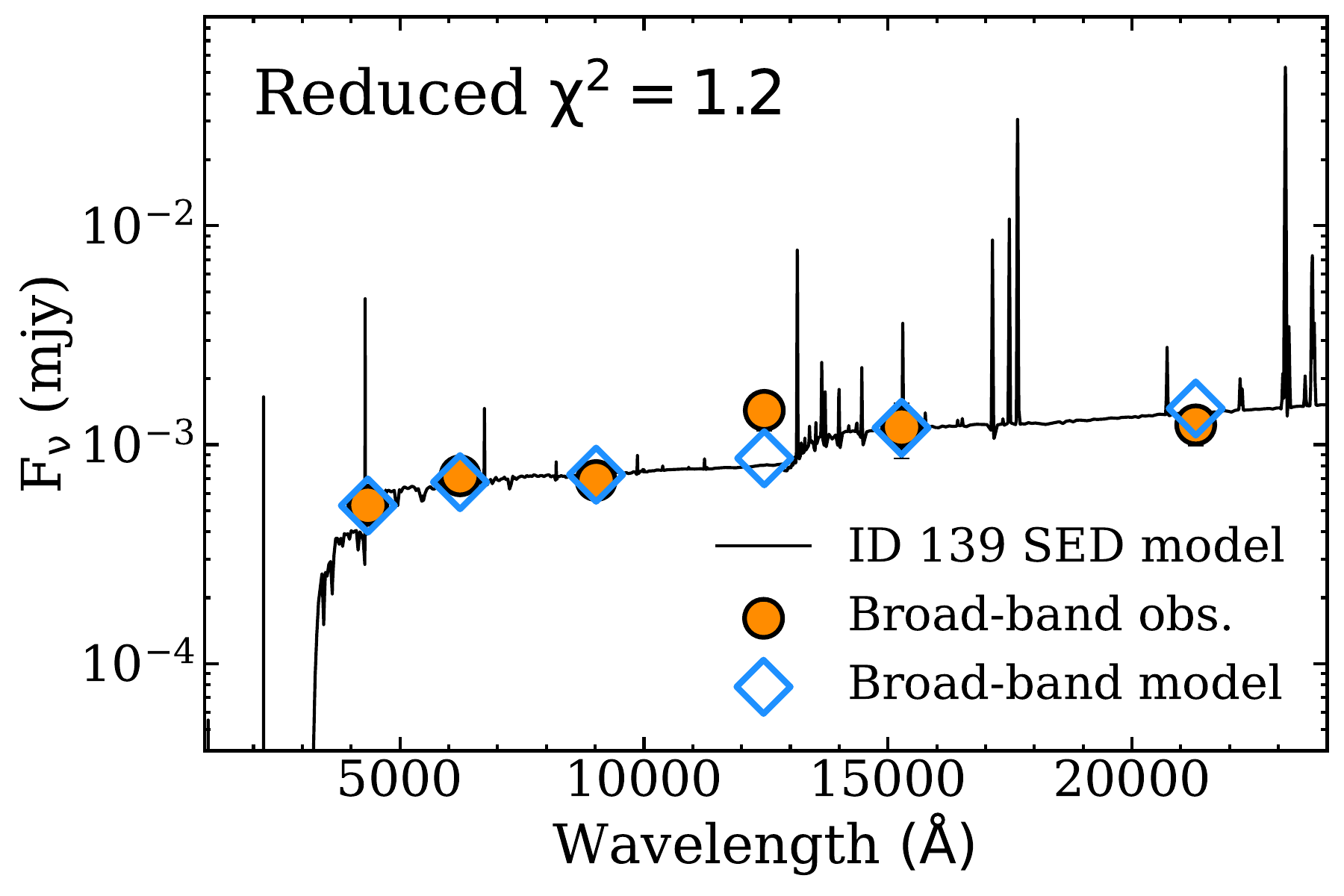}\par
      \includegraphics[width=\linewidth]{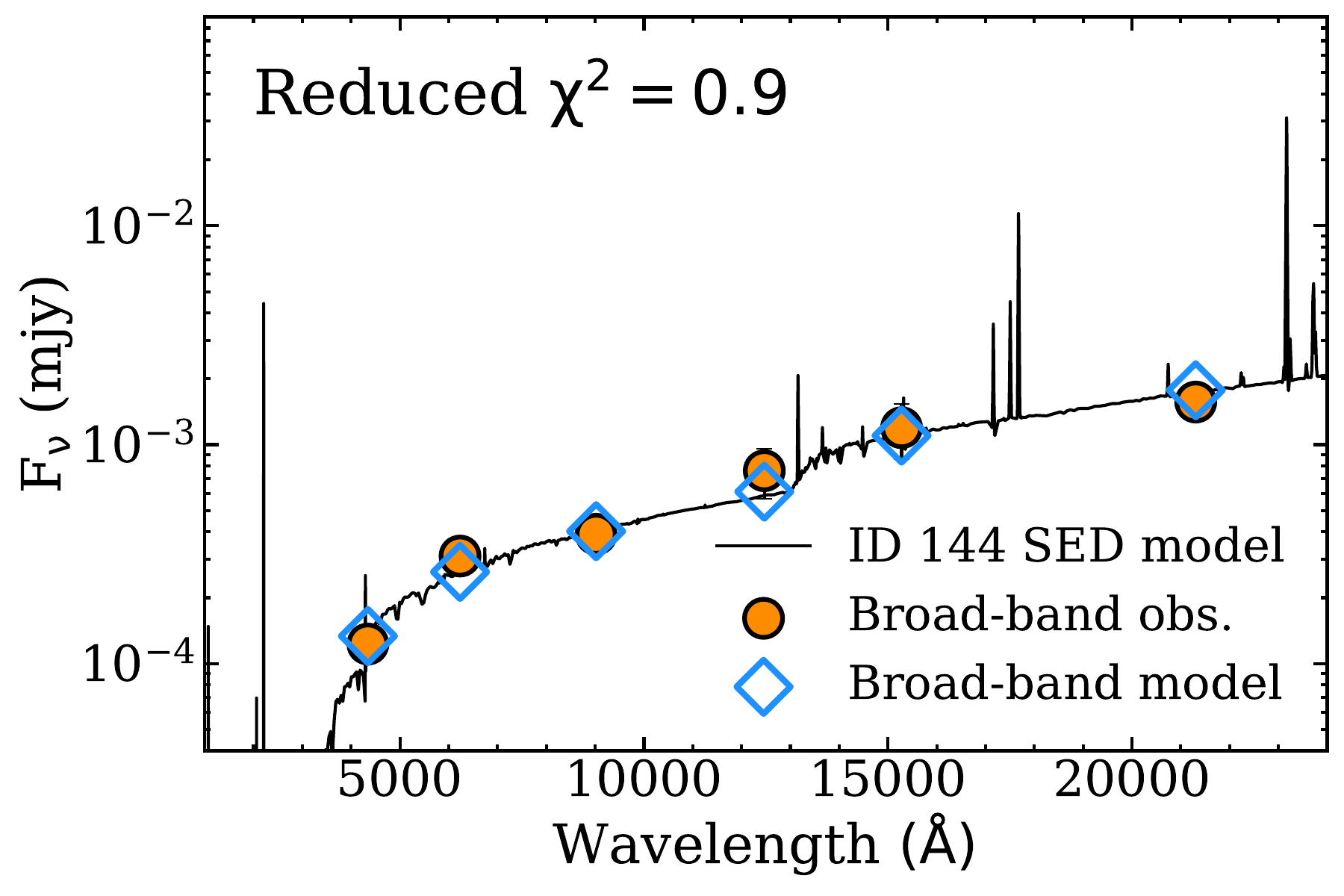}\par
      \includegraphics[width=\linewidth]{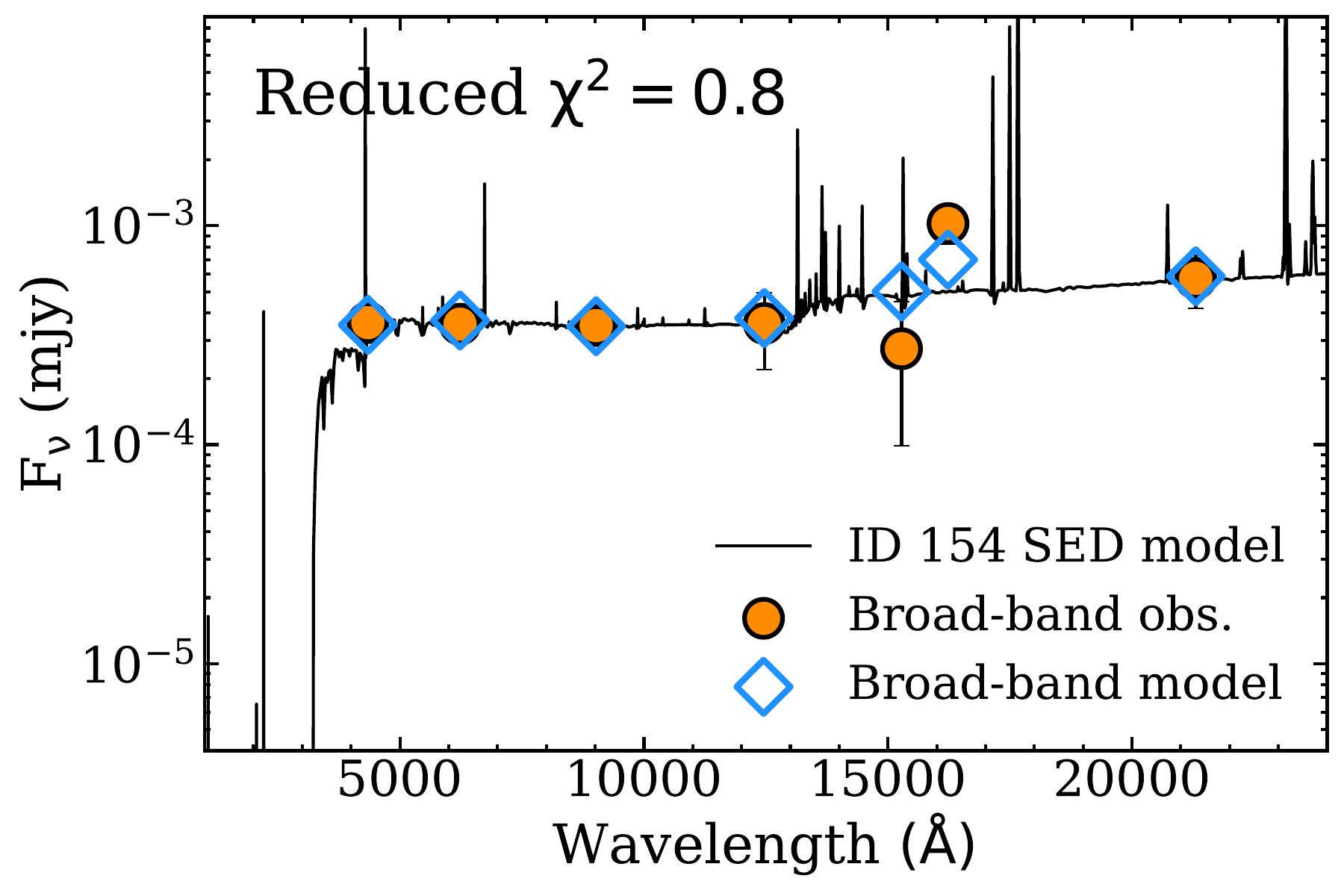}\par
      \includegraphics[width=\linewidth]{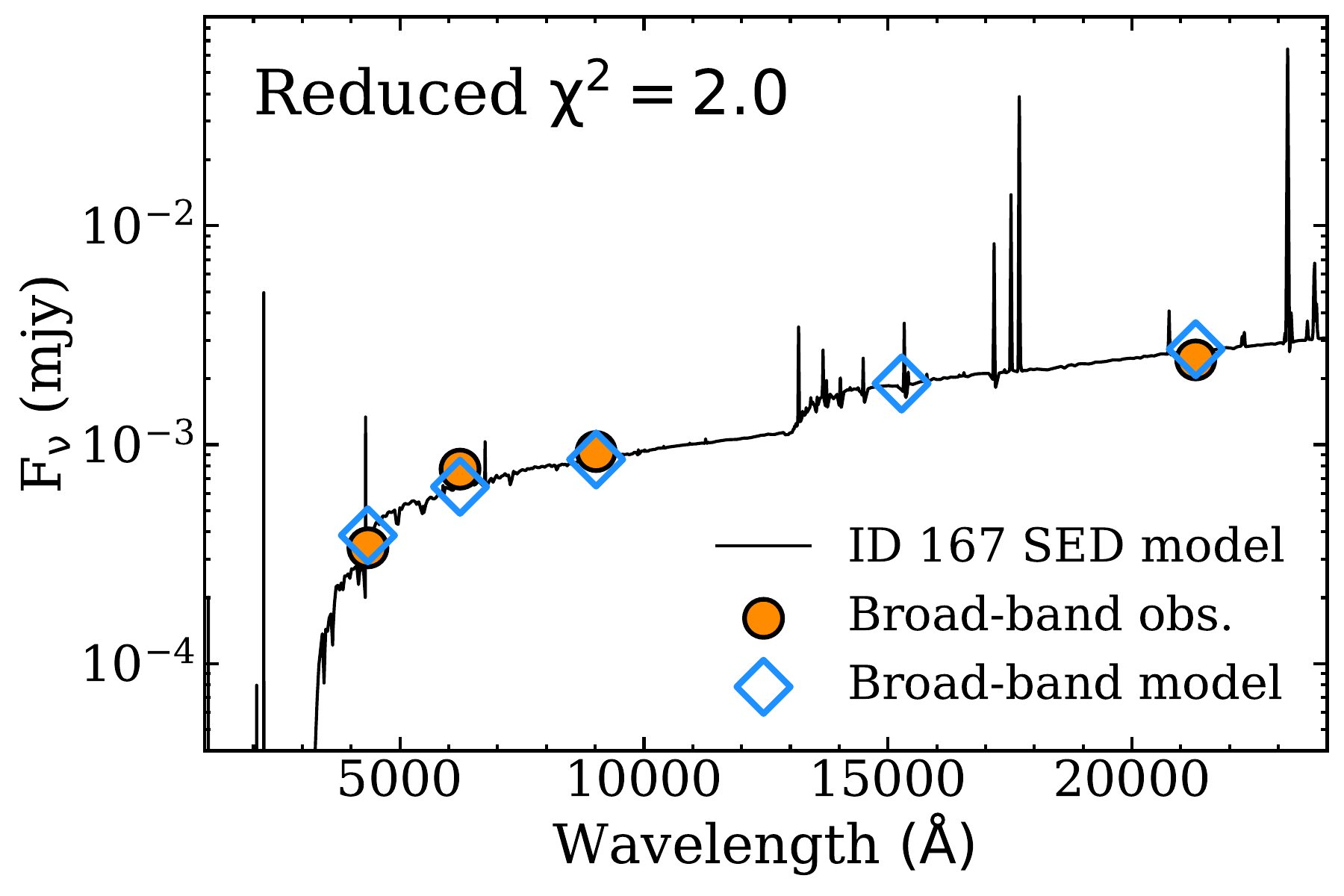}\par
      \includegraphics[width=\linewidth]{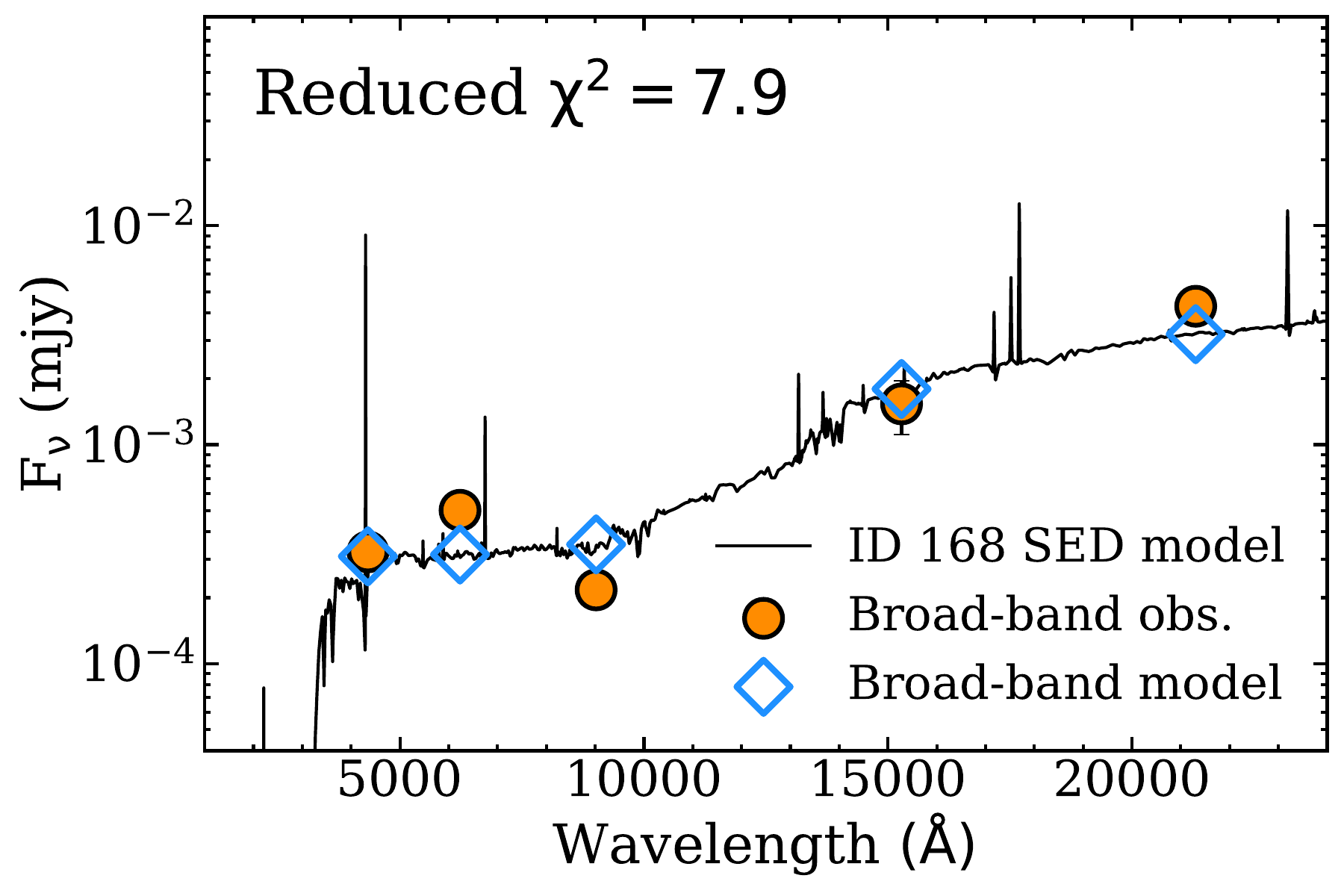}\par
      \includegraphics[width=\linewidth]{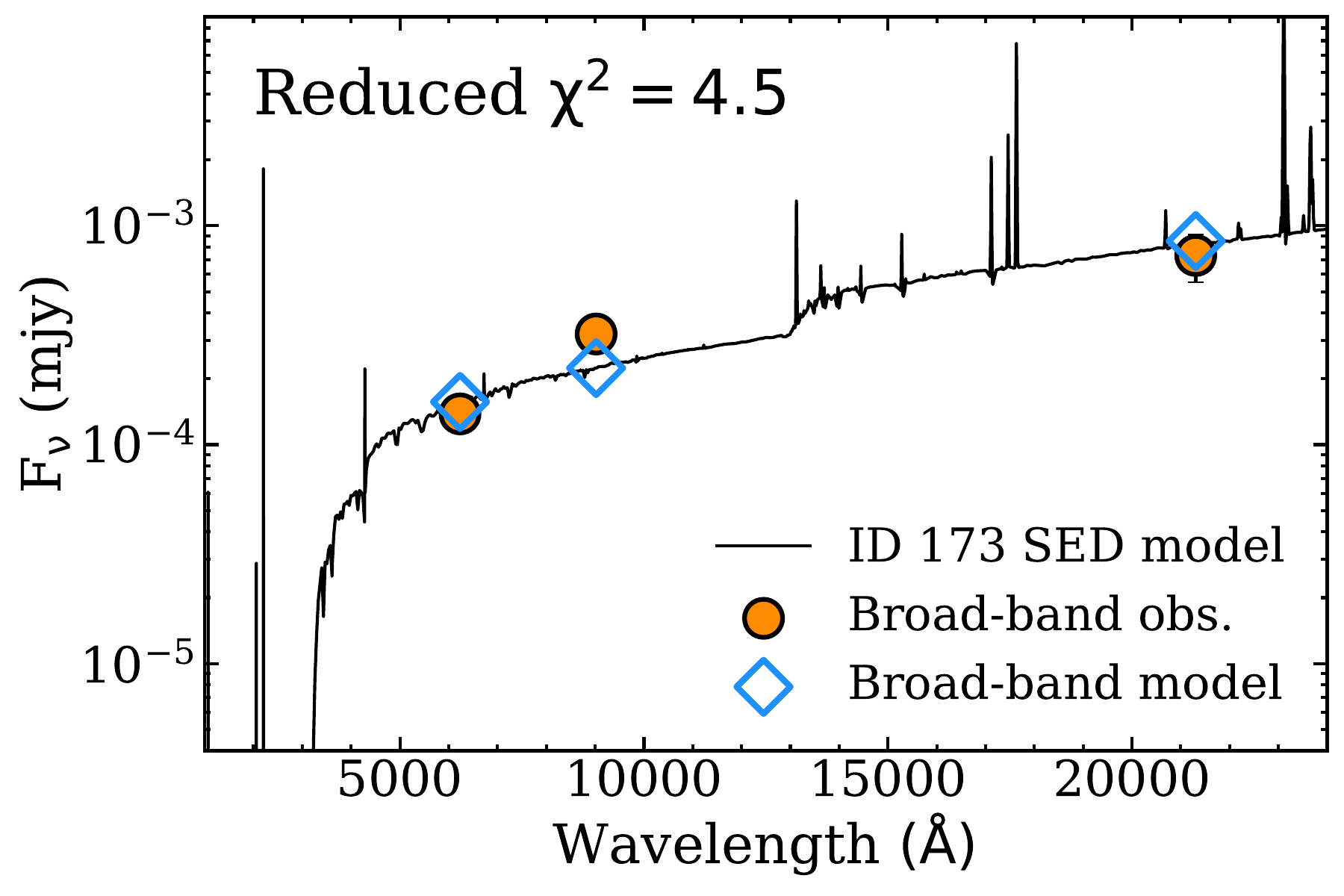}\par
      \end{multicols}
      \caption{The black solid line for each panel displays the Cigale SED model. Orange circles and blue-edged diamonds respectively show the broad-band observed and modeled fluxes for each protocluster member within our sample.}
         \label{F:SED}
\end{figure*}

\section{Summary Table}
\renewcommand{\arraystretch}{1.5}
\begin{table*}
\caption{Physical properties of our USS1558 protocluster sample. Coordinates (RA and DEC) are shown in degrees. The redshifts (z) are based on the H$\alpha$ measurements of this work. The molecular masses have been computed following the metallicity-dependent calibration by \protect\cite{Tacconi18} and making use of the ALMA CO(3-2) information published by \protect\cite{Tadaki19}. (see Sect.\,\ref{SS:MOL}). $\mathrm{A_V}$ represents the extinction extracted from the SED modelling assuming the \protect\cite{Calzetti2000} extinction law (see Sect.\,\ref{SS:SFR_method}).}

\begin{threeparttable}
\begin{tabular}{ccccccccccccccc}
\hline
\noalign{\vskip 0.0cm}
ID & RA  & DEC   & z  &  $\mathrm{\log\,M_*/M_\odot}$  & SFR & $\mathrm{12+\log(O/H)}$ & $\mathrm{\log\,M_{mol}/M_\odot}$ & $\mathrm{A_V}$ & $\Sigma_5$ \\
   & (J2000) & (J2000) & & & ($\mathrm{M_\odot/yr}$) & & & $\mathrm{(mag)}$ & ($\mathrm{Mpc^{-2}}$) \\
\noalign{\vskip 0.0cm}
\hline 
\hline 
\noalign{\vskip 0.0cm}
16 & 240.2826584 & -0.5308889 & 2.5410 & $9.82^{+0.16}_{-0.26}$   & $42\pm9$   & $8.46\pm0.13$ & -              & $0.82\pm0.21$ & 29.92  \\ 
18 & 240.2831295 & -0.5255959 & 2.5247 & $9.35^{+0.20}_{-0.37}$   & $10\pm4$   & -             & -              & $0.71\pm0.24$ & 40.27  \\ 
34 & 240.2915811 & -0.5269186 & 2.5251 & $10.51^{+0.14}_{-0.20}$  & $63\pm29$  & -             & -              & $1.86\pm0.22$ & 104.31 \\ 
38\tnote{a} & 240.2924994 & -0.5211665 & 2.5237 & $11.12^{+0.12}_{-0.17}$ & $189\pm5$ & -      & $<10.45$       & $2.11\pm0.24$ & 135.18 \\ 
39 & 240.2931181 & -0.5239965 & 2.5244 & $9.49 ^{+0.15}_{-0.23}$  & $34\pm8$   & -             & -              & $0.71\pm0.22$ & 135.51 \\ 
59 & 240.2976694 & -0.5220065 & 2.5134 & $10.88^{+0.14}_{-0.20}$  & $266\pm85$ & $8.66\pm0.13$ & $11.10\pm0.06$ & $2.66\pm0.48$ & 350.95 \\ 
64 & 240.2987622 & -0.5199375 & 2.5294 & $10.11^{+0.12}_{-0.16}$  & $35\pm15$  & $<8.60$       & $10.76\pm0.12$ & $0.74\pm0.23$ & 550.95 \\ 
73 & 240.3020857 & -0.5148570 & 2.5260 & $10.57^{+0.13}_{-0.18}$  & $311\pm45$ & $8.57\pm0.10$ & $10.73\pm0.09$ & $1.78\pm0.22$ & 82.71  \\ 
74 & 240.3019612 & -0.4821418 & 2.5344 & $10.18^{+0.13}_{-0.19}$  & $115\pm27$ & $8.45\pm0.13$ & -              & $1.37\pm0.23$ & 17.79  \\ 
79 & 240.3039954 & -0.5211177 & 2.5092 & $10.72^{+0.14}_{-0.21}$  & $179\pm33$ & $8.45\pm0.10$ & $<10.78$       & $1.67\pm0.23$ & 243.28 \\ 
88 & 240.3056859 & -0.5060703 & 2.5185 & $9.42^{+0.14}_{-0.20}$   & $43\pm5$   & $<8.30$       & -              & $0.63\pm0.19$& 121.26 \\ 
91 & 240.3079703 & -0.5108609 & 2.5178 & $9.96^{+0.14}_{-0.21}$   & $82\pm22$  & -             & -              & $1.53\pm0.23$ & 201.22 \\ 
94 & 240.3080466 & -0.4968055 & 2.5268 & $10.70^{+0.18}_{-0.33}$  & $192\pm29$ & $8.54\pm0.10$ & -              & $1.87\pm0.32$ & 25.5   \\ 
112 & 240.3143353 & -0.5083207 & 2.5246 & $9.68^{+0.16}_{-0.25}$  & $72\pm13$  & $8.47\pm0.12$ & -              & $1.39\pm0.26$ & 30.25  \\ 
134 & 240.3249030 & -0.4846260 & 2.5352 & $10.41^{+0.15}_{-0.23}$ & $102\pm33$ & -             & -              & $1.70\pm0.22$ & 89.5   \\ 
137 & 240.3247461 & -0.4756063 & 2.5267 & $10.64^{+0.14}_{-0.21}$ & $94\pm28$  & $<8.57$       & $10.66\pm0.11$ & $2.00\pm0.24$ & 53.03  \\ 
138\tnote{a,b,c} & 240.3222771 & -0.4795241 & 2.5345 & $11.83^{+0.11}_{-0.15}$ & $90\pm15$ & $8.59\pm0.06$ & -  & $0.26\pm0.25$ & 84.18  \\ 
139 & 240.3266295 & -0.4931553 & 2.5262 & $9.77^{+0.14}_{-0.20}$  & $43\pm9$   & -             & -              & $0.73\pm0.20$ & 15.69  \\ 
144 & 240.3290271 & -0.4616180 & 2.5332 & $10.20^{+0.13}_{-0.20}$ & $55\pm13$  & -             & -              & $1.73\pm0.23$ & 48.61  \\ 
154 & 240.3312181 & -0.4995278 & 2.5278 & $9.36^{+0.16}_{-0.25}$  & $30\pm4$   & -             & -              & $0.39\pm0.20$ & 9.05   \\ 
167 & 240.3382885 & -0.4621250 & 2.5332 & $10.28^{+0.14}_{-0.20}$ & $115\pm19$ & $8.55\pm0.09$ & -              & $1.45\pm0.22$ & 13.22  \\ 
168\tnote{a} & 240.3392922 & -0.4609448 & 2.5358 & $10.62^{+0.13}_{-0.18}$ & $19\pm4$ & $8.82\pm0.09$ & -       & $0.20\pm0.17$ & 10.75  \\ 
173 & 240.3479611 & -0.4806875 & 2.5213 & $9.84^{+0.16}_{-0.24}$  & $74\pm24$  & -             & -              & $1.47\pm0.32$ & 5.78   \\ \hline
\noalign{\vskip 0.0cm}
\end{tabular}
\begin{tablenotes}
\item[a] These objects are labeled as AGN candidates due to displaying $\log$([N{\sc{ii}}]/H$\alpha$)$\geq-0.35$ or broad H$\alpha$ profiles ($\sigma>700$ km\,s$^{-1}$).
\item[b] Confirmed X-ray emission (\citealt{Macuga19}).
\item[c] Radio galaxy 4C-00.62 (\citealt{Kajisawa06}).
\end{tablenotes}
\end{threeparttable}
\label{T:BigTable}
\end{table*}


\bsp	
\label{lastpage}
\end{document}